\documentclass[preprint]{aastex}

\shorttitle{Highly-Ionized High-Velocity Gas}
\shortauthors{Sembach et al.}
\slugcomment{Submitted to the {\it ApJ Supplement Series} (25-July-2002)}

\begin{document}

\newcommand{\kms}{\,km\,s$^{-1}$}     

\title{Highly-Ionized High-Velocity 
Gas In the Vicinity of the Galaxy}

\author{K.R.~Sembach\altaffilmark{1}, 
        B.P.~Wakker\altaffilmark{2},
	B.D.~Savage\altaffilmark{2}, 
	P.~Richter\altaffilmark{2,3},
	M.~Meade\altaffilmark{2},
	J.M.~Shull\altaffilmark{4},
	E.B.~Jenkins\altaffilmark{5},
	G.~Sonneborn\altaffilmark{6}, and 
	H.W. Moos\altaffilmark{7}}
\altaffiltext{1}{Space Telescope Science Institute, 3700 San Martin Dr.,
	Baltimore, MD  21218~~({\it sembach@stsci.edu}}
\altaffiltext{2}{Department of Astronomy, University of Wisconsin, 475 N.
	Charter St., Madison, WI  53706}
\altaffiltext{3}{Current address: Observatorio Astrofisico di Arcetri, Largo 
	E. Fermi 5, Florence, Italy}
\altaffiltext{4}{Center for Astrophysics and Space Astronomy, Department of 
	Astrophysical and Planetary Sciences, University of Colorado, Boulder,
	CO  80309}
\altaffiltext{5}{Princeton University Observatory, Peyton Hall, 
	Princeton, NJ  08544}
\altaffiltext{6}{Laboratory for Astronomy and Solar Physics, Code 681, 
	NASA Goddard Space Flight Center, Greenbelt, MD 20771}
\altaffiltext{7}{Department of Physics \& Astronomy, The Johns Hopkins 
	University, 3400 N. Charles St., Baltimore, MD  21218}

\begin{abstract}

We report the results of a $FUSE$ study of
high velocity \ion{O}{6} absorption along complete sight lines through 
the Galactic halo in directions toward 100 extragalactic objects
and 2 halo stars.
The high velocity \ion{O}{6} traces a variety of phenomena, including
tidal interactions with the Magellanic Clouds, accretion of gas, outflowing
material from the Galactic disk, warm/hot gas interactions in a highly 
extended
Galactic corona, and intergalactic gas in the Local Group.
We identify 85 
high velocity \ion{O}{6} features at $\ge3\sigma$ confidence at velocities of
$-500 < v_{LSR} <  +500$ \kms.
There are an additional 6  
confirmed or very likely ($>90$\% confidence) features plus 2 tentative 
detections 
between $v_{LSR} = +500$ and +1200 \kms; these very high velocity \ion{O}{6}
features trace intergalactic gas beyond the Local Group.
The 85 \ion{O}{6} features have velocity centroids ranging 
from $-372 \lesssim \bar{v}_{LSR} \lesssim -90$ \kms\ 
to $+93 \lesssim \bar{v}_{LSR} \lesssim +385$ \kms, 
line widths b~ $\sim 16-81$ \kms\ with an average of 
$\langle$b$\rangle$ = $40\pm14$ \kms, and
an average 
\ion{O}{6} column 
density $\langle \log N \rangle = 13.95\pm0.34$ with  a median value of
13.97. 
Values of b greater than the $17.6$ \kms\
thermal width expected for \ion{O}{6}
at $T \sim 3\times10^5$\,K indicate that 
additional non-thermal broadening mechanisms are common.
The \ion{O}{6} $\lambda1031.926$ absorption is detected at $\ge 3\sigma$ 
confidence along 59 of the 102 sight lines surveyed.  
The high velocity \ion{O}{6} detections indicate
that $\sim60$\% of the sky (and perhaps as much as $\sim85$\%, depending on 
data quality considerations)  is covered by high velocity H$^+$ associated with
the \ion{O}{6}.  
$N({\rm H}^+)\gtrsim4\times10^{16}$ cm$^{-2}$ 
if the high velocity hot gas has a metallicity similar
to that of the Magellanic Stream.  About 30\% of the sky is covered by 
the hot, high velocity H$^+$ at a level of 
$N({\rm H}^+)\gtrsim4\times10^{17}$ cm$^{-2}$, which is 
similar to the detection rate found for \ion{H}{1} 21\,cm emission 
produced by warm
neutral gas at a comparable 
column density level.  
Some of the high velocity \ion{O}{6} is associated with known 
\ion{H}{1} structures (the Magellanic Stream,
Complex~A, Complex~C, the Outer Spiral Arm, 
and several discrete \ion{H}{1} HVCs).
Some of the high velocity \ion{O}{6} features have no counterpart 
in \ion{H}{1} 21\,cm emission, including  discrete absorption features and
positive velocity absorption wings extending from $\sim100$ to $\sim300$ \kms\
that blend with lower velocity absorption
produced by the Galactic thick disk/halo. The discrete 
features may typify clouds located in the Local Group, while 
the \ion{O}{6} absorption wings may be tidal debris or material expelled
from the Galactic disk.
Most of the \ion{O}{6} features have velocities incompatible 
with those of the Galactic halo, even if the halo has decoupled from 
the underlying Galactic disk. 
The reduction in the dispersion about the mean of the high velocity \ion{O}{6}
centroids when the velocities are converted from the
LSR to the GSR and LGSR reference frames is necessary (but not 
conclusive) evidence that some of the clouds are located outside the 
Galaxy.  
Most of the \ion{O}{6} cannot be produced by photoionization, even
if the gas is irradiated by extragalactic ultraviolet background radiation.
Several observational quantities indicate that collisions in hot gas
are the primary
ionization mechanism responsible for the production of the \ion{O}{6}.  These
include the ratios of \ion{O}{6} column densities to those of other highly
ionized species (\ion{C}{4}, \ion{N}{5}) and the strong correlation between
$N$(\ion{O}{6}) and \ion{O}{6} line width.
 Consideration of the possible sources of collisional ionization 
favors production of some of the \ion{O}{6} at the boundaries between 
cool/warm 
clouds of gas and a highly extended ($R \gtrsim 70$ kpc),
 hot ($T > 10^6$\,K), low-density ($n \lesssim 
10^{-4}-10^{-5}$ cm$^{-3}$) Galactic corona or Local Group medium.
 The existence of a hot, highly extended Galactic corona or Local Group medium
and the prevalence of high velocity \ion{O}{6} are consistent with predictions
of current galaxy formation scenarios.
Distinguishing between the various phenomena producing high velocity \ion{O}{6}
in and near the Galaxy will require continuing studies of the distances, 
kinematics,  elemental abundances,  and physical states of the different 
types of high velocity \ion{O}{6} found in this study.
Descriptions of galaxy evolution will need to account for the highly ionized
gas, and future X-ray studies of hot 
gas in the Local 
Group  will need to consider  carefully the relationship of the 
X-ray absorption/emission to the complex high velocity 
absorption observed in \ion{O}{6}.

\end{abstract}


\section{INTRODUCTION}

Understanding galaxy formation and evolution requires observational 
information about hot, highly ionized gas in and near galaxies.   Numerical 
simulations of 
cosmological structure formation in the presence of cold dark matter 
indicate that a significant fraction of the baryonic material at low 
redshift should be shock-heated to temperatures of $10^5-10^7$\,K as the 
gas collapses and forms clusters and groups of galaxies (e.g., 
Cen \& Ostriker 1999; Dav\'e et al. 1999, 2001).  This reservoir of hot gas is 
detectable through ultraviolet absorption-line measurements (Tripp,
Savage, \& Jenkins 2000;
Savage et al. 2002a), but a complete picture of how the 
hot gas in galaxies and the intergalactic medium (IGM) are related does not
yet exist because a variety of internal and external processes affect the 
heating and distribution of the interstellar gas in and
around galaxies.  In addition to 
the galaxy formation process, the accretion of satellite 
galaxies, tidal interactions, star formation,
galactic winds, and galaxy-IGM interactions may all
contribute to the production of hot gas.  To some 
extent, all of these activities operate in and around our own
galaxy, so studying the hot gas in the immediate environment of the Milky Way
is a logical step in assessing the relevance and roles of 
these processes locally.

One of the primary science 
objectives of the {\it Far Ultraviolet Spectroscopic 
Explorer} ($FUSE$) mission is to determine the  properties 
of hot, highly ionized gas in the low-redshift universe.  A key component 
of this research has been the study of \ion{O}{6} absorption along
many sight lines through the Galactic halo.  Such observations
were not possible until $FUSE$ was launched because previous observatories
lacked either the spectral resolution or sensitivity to study the velocity 
structure of the \ion{O}{6} absorption toward distant background sources.  
Among the most interesting
results to date is the detection of \ion{O}{6} in high velocity clouds (HVCs)
(Sembach et al. 2000; Murphy et al. 2000).
For decades astronomers have studied the neutral (\ion{H}{1})
content of the clouds and have debated the origin and location of the high 
velocity gas.  While it is
generally accepted that no single model can account for all of the observed 
properties of the \ion{H}{1} HVCs (see Wakker \& 
van~Woerden 1997 for a review), there is no unanimity on the fundamental 
properties of the different types of HVCs. With the 
re-introduction of the idea that some of the \ion{H}{1} HVCs
 could be located outside of the Milky Way if they are embedded
in halos of dark matter (Blitz et al. 1999), the possible range of locations
and cloud properties has widened.
Determining the ionization and hot gas content of the high velocity gas bears
directly on the locations of the clouds and their interactions with other
components of the gaseous interstellar and intergalactic media.

The \ion{O}{6} $\lambda\lambda1031.926, 1037.617$ doublet lines in the 
far-ultraviolet spectral region are the best  lines to use for 
kinematical investigations of hot ($T \gtrsim 10^5-10^6$\,K) gas in the 
low-redshift universe.  Oxygen is the most abundant element heavier than 
helium, and the \ion{O}{6} lines have large oscillator strengths 
($f_{1032} = 0.133$, $f_{1038} = 0.0661$; Morton 1991).  X-ray spectroscopy
of higher ionization lines (e.g., \ion{O}{7}, \ion{O}{8}) is possible with
{\it XMM-Newton} and the {\it Chandra X-ray Observatory} for a small number 
of sources, but the spectral resolution 
(R~$\equiv \lambda/\Delta\lambda \lesssim 1000$)
is modest compared to that afforded by $FUSE$ (R~$\sim 15,000$).
Lower ionization lines observable at high spectral resolution at 
ultraviolet wavelengths are generally
either much weaker than the \ion{O}{6} lines (e.g., 
\ion{N}{5}~$\lambda\lambda1238.821, 1242.804$) or 
are better tracers of collisionally ionized 
gas at temperatures $T \lesssim 10^5$\,K (e.g., 
\ion{C}{4} $\lambda\lambda1548.195, 1550.770$, \ion{C}{3} $\lambda977.020$,
\ion{Si}{4} $\lambda\lambda1393.755, 1402.770$, \ion{Si}{3} $\lambda1206.500$).
This latter set of ions is also considerably more susceptible to 
photoionization than \ion{O}{6} since the photoionization cross sections of 
the ions are large and the ionization potentials are less than 54\,eV, the
energy of the \ion{He}{2} ionization edge.   \ion{O}{6} can be produced
by photoionization under special conditions involving a hard radiation
field and a very low gas density, but as we will show later, this does not
appear to be a viable production mechanism for {\it most} of the \ion{O}{6}
observed in and near the Milky Way.

This article is one in a series of three papers devoted to $FUSE$ 
observations of \ion{O}{6} absorption along 
complete paths through the Galactic halo in the directions of quasars,
active galactic nuclei (AGNs), and BL Lac objects.  
The other two papers include 
a catalog with basic measurements and illustrations of all of the \ion{O}{6} 
profiles obtained in the survey (Wakker et al. 2002) and 
a companion study of the \ion{O}{6} absorption associated with the thick disk 
of the Milky Way (Savage et al. 2002b).  Here, we concentrate 
on the properties of the \ion{O}{6} absorption at high velocities with respect
to those expected for gas participating in differential Galactic rotation
(generally, ``high velocity'' 
refers to $100 < |v_{LSR}| < 400$ \kms\ in this paper).
High velocity \ion{O}{6} absorption associated with the Magellanic Clouds is 
discussed elsewhere (Friedman et al. 2000; Danforth et al. 2002; Hoopes et al. 
2001, 2002; Howk et al. 2002b).  

This paper is organized as follows.  In \S2 we describe the observations and 
data reduction.  We present the \ion{O}{6} HVC measurements in \S3 and 
describe the general types of high velocity gas seen in \S4.  Section 5 
contains information about the column densities, sky covering fractions,
velocities, and line widths 
of the \ion{O}{6} absorption.  In \S6 we discuss the high
velocity \ion{O}{6} features associated with previously identified high 
velocity clouds, while \S7 specifies the few sight lines that show high 
velocity \ion{H}{1} 21\,cm emission with no corresponding \ion{O}{6} 
absorption.  Section~8 highlights new high velocity gas detected only in
\ion{O}{6} absorption.  In \S9 we consider how
the high velocity gas is ionized.  
Section~10 describes the general kinematical 
behavior of the high velocity gas.
Section 11 contains a discussion of the high velocity
\ion{O}{6} features and the implications of this work 
for understanding hot gas in the 
low-redshift universe.  We conclude  with a summary of results in \S12.

\section{OBSERVATIONS AND DATA REDUCTION}

Full details of the $FUSE$ observations and data reduction for the
objects surveyed for high velocity \ion{O}{6} are given by Wakker et al.
(2002). 
We summarize some of the key points here.  
The objects included in this study form an incomplete flux-limited
sample of extragalactic objects drawn primarily from the $FUSE$ Science Team 
\ion{O}{6}, D/H, and AGN programs conducted in the first two years of the 
mission (December 1999 -- November 2001). 
Publicly available Guest Investigator (GI)
observations have been included when it is clear that the science goals of our
study and those of the GI do not conflict.   Most of the objects are either
quasars or low-redshift Type~1 Seyfert galaxy nuclei, which have relatively 
smooth continua in the wavelength region of interest.  A few additional 
background sources,  such as bright regions of low-redshift starburst 
galaxies and two high latitude stars located 
$>5$ kiloparsecs from the Galactic disk,  have been included as 
well.  Neither stellar sight line included in the sample (vZ\,1128, 
PG\,0832+675) shows 
evidence of high velocity \ion{O}{6} absorption.  Of the more than
two hundred objects considered, only about half had data suitable 
for studies of the \ion{O}{6} lines.  
Total exposure times ranged from $\sim$ 5 ksec to $\sim600$
ksec per object, with  a median of $\sim16$ ksec per object.  

$FUSE$ contains four co-aligned spectrographs that cover the 905--1187\,\AA\
spectral region.  The optics (a mirror and a holographically-ruled 
diffraction grating) in two of the channels are coated with aluminum
and lithium-fluoride (LiF) for 
maximum throughput above 1000\,\AA.  The other two channels contain
optics with SiC coatings for maximum throughput below 1000\,\AA.
Each object was centered in the $30\arcsec \times 30\arcsec$ (LWRS) 
aperture of the LiF1 channel, which is also the channel used for 
guiding.  In the remaining channels, the objects sometimes drifted 
around in the LWRS apertures.  These drifts do not affect our study 
since each photon event recorded by the micro-channel plate detectors is 
time-tagged and can be corrected for these motions.
Descriptions of the $FUSE$ instrumentation and 
on-orbit performance are given by Moos et al. (2000, 2002) and Sahnow
(2000a, 2000b).  
Additional information can be found in the {\it FUSE Observer's 
Guide}.\footnotemark\
\footnotetext{On-line at 
{\tt http://fuse.pha.jhu.edu/support/guide/guide.html.}}

All of the data were reduced in a homogeneous fashion with the standard 
$FUSE$ pipeline software ({\tt CALFUSE} v1.8.7) available  
from the Johns Hopkins University.\footnotemark\
\footnotetext{For a subset of the observations, we compared the 
{\tt CALFUSE} v1.8.7 processed data to the data processed with the 
latest (v2.0.5) available 
software.  We found no significant differences in the 
reduced datasets in the spectral regions considered in this study, except 
for $\sim\pm30$ \kms\ differences in the velocity calibration.  See 
Wakker et al. (2002) for additional information about the zero-point velocity 
scale differences
between the two processing versions for objects in our sample.}
  The \ion{O}{6} lines are covered by 
all four $FUSE$ channels, but we have opted to use data from only the two
most sensitive channels (LiF1, LiF2), which have optics coated with Al+LiF.
These two channels account for roughly 76\% of the $FUSE$ effective area
at the wavelengths of the \ion{O}{6} lines.  For the higher signal-to-noise
ratio  ($S/N$) 
observations, 
we used only the data from the LiF1 channel and checked this
result with the LiF2 data.  However, 
we combined the two independent measurements for lower $S/N$ observations
 to increase the detection confidence (see
Wakker et al. 2002).  To maximize $S/N$,
we used data obtained during both orbital day and 
night since the \ion{O}{6} lines are in spectral regions unaffected by
terrestrial airglow emission.
The fully sampled 
$FUSE$  data have a FWHM velocity resolution of $\sim20-25$ \kms\
($R \sim 12,000-15,000$).

We set the zero point of the $FUSE$ wavelength scale for each observation
by referencing the observed \ion{Ar}{1} $\lambda1048.220$ and \ion{Si}{2}
$\lambda1020.699$ absorption features to the \ion{H}{1} 21\,cm emission
observed in each direction [see Wakker et al. (2002)
for a description of the method
and information about the \ion{H}{1} data used in these comparisons].  
This comparison was done for the low velocity portions of the profiles.   
H$_2$ absorption features in the 1020--1040\,\AA\ spectral region provided
an additional consistency check on the wavelength scales for some sight
lines.  We estimate a wavelength  uncertainty of $\sim0.03$\,\AA\ 
($\sim10$ \kms) for the 
fully reduced data.  None of the conclusions of this study are affected by 
this zero-point uncertainty.
Unless otherwise specified, all velocities in this paper
are referred to the Local Standard of Rest (LSR) reference frame, which
has a standard solar motion of 19.5 \kms\ in the direction $l_{std} = 56\degr,
b_{std} = 23\degr$ (Mihalas \& Binney 1981).

Figure~1 contains the fully reduced $FUSE$ spectra between 1015\,\AA\ and 
1040\,\AA\ for three sight lines through the Galactic halo (NGC\,7469,
PG\,1116+215, and PG\,1259+593).   These sight 
lines exhibit varying degrees of complexity in their observed \ion{O}{6}
absorption and in the amount of blending with other atomic and molecular 
lines in the bandpass.  All three
sight lines contain high velocity \ion{O}{6} absorption that is observable 
in the 1031.926\,\AA\ line.
Toward NGC\,7469 and PG\,1259+593, the high velocity
 \ion{O}{6} $\lambda1037.617$ absorption  is blended 
with either \ion{C}{2}$^*$ $\lambda1037.012$ or with nearby
molecular hydrogen lines, whose wavelengths are indicated in the top panel.
Similar illustrations of high velocity \ion{O}{6} 
absorption observed toward AGNs and QSOs by $FUSE$ in this spectral region 
are given by Oegerle et al. (2000), Savage et al. (2000, 2002a, 2002b), 
Sembach et al. (2000, 2001b), and Heckman et al. (2001).

\section{MEASUREMENTS}

In our study of high velocity \ion{O}{6}, we draw upon the 
measurements and detailed sight line comments presented in the $FUSE$ 
\ion{O}{6} catalog paper (Wakker et al. 2002)
for the 102 objects with sufficient data 
quality to study the absorption.  The catalog contains information
about line strengths (equivalent widths), velocities, and contamination 
by other absorption features.  The catalog includes illustrations of the 
observed absorption profiles and some basic statistical information 
about the properties of the absorption and data quality.
We present a complete list of the high velocity \ion{O}{6} features 
in Table~1, sorted by Galactic longitude. 
The table entries include 
the name of the background continuum source observed, 
the Galactic coordinates ($l, b$) of the background source, 
a data quality factor (Q), an
``identification'' (ID) for the high velocity \ion{O}{6} absorption based on
the correspondence with Galactic structures or high velocity gas observed in 
\ion{H}{1} 21\,cm emission,  the velocity range spanned by the \ion{O}{6}
($v_{min},v_{max}$),  
the average velocity and width of the \ion{O}{6} absorption ($\bar{v}$ and b), 
the logarithmic column density, and the significance 
of the \ion{O}{6} $\lambda1031.926$ absorption-line detection 
($W_\lambda/\sigma_W$, where $\sigma_W$ is the 
error in the equivalent width due to statistical noise fluctuations
and continuum placement uncertainties).  The data quality
factor is an indicator of the continuum $S/N$ (per 20 \kms\ resolution element)
in the spectrum near the \ion{O}{6} lines as follows:  $Q = 1~(S/N = 3-5)$,
$Q = 2~(S/N = 5-9)$, $Q = 3~(S/N = 9-14)$, $Q = 4~(S/N >14)$. 

\subsection{Identifying the High Velocity \ion{O}{6} Absorption}

The derived properties of the high velocity gas
depend in part upon the adopted separation of the high velocity  
absorption from lower velocity absorption arising 
in the nearby Galactic disk and halo.  In regions of the sky 
where there
is little high velocity gas observed in other species (e.g., \ion{H}{1},
\ion{C}{2}, \ion{C}{4}), the nearby \ion{O}{6} absorption is generally 
found at $-100 \le v_{LSR} \le +100$ \kms, which we adopt as the typical 
velocity range of the Galactic
thick disk and halo \ion{O}{6} absorption (see Savage et al. 2002b).
We define absorption outside this velocity range as high
velocity gas, with a few minor 
exceptions to account for directions where there 
is reason to believe that the thick disk absorption extends to higher
velocities.  For example, differential Galactic rotation may broaden the 
profiles in some directions, particularly low latitude sight lines in the 
$l = 90\degr$ and $l = 270\degr$ directions, as well as at longitudes in
the general direction of the Galactic center.  In a few cases we contracted 
the thick disk limits and allowed the high velocity gas integration to extend 
to lower velocities by  $\sim10-20$ \kms\ to account for absorption associated
with known high velocity features (e.g., Mrk~817, which exhibits \ion{O}{6} 
absorption in Complex~C at velocities $v_{LSR} < -90$ \kms). For 5 lines of 
sight, the separation is based on \ion{O}{6} velocity structure.  We list the 
integration limits ($v_{min}, v_{max}$) for the high velocity gas in
Table~1.

We examined the $FUSE$ spectra for absorption produced by the stronger
member of the \ion{O}{6} $\lambda\lambda1031.926,1037.617$ 
doublet because the higher 
velocity portions of the weaker line ($\lambda1037.617$) 
are usually severely blended with other interstellar lines.  We searched 
for absorption at velocities $ -1200 < v_{LSR} <  +1200$ \kms, and
with few exceptions all of the identified \ion{O}{6} absorption is confined
to the $|v_{LSR}| < 400$ \kms\ velocity range.  The few possible cases
where higher velocity features may be present are described in Appendix~A
of the \ion{O}{6} catalog (Wakker et al. 2002) and are not considered 
further here.

The identification of high velocity \ion{O}{6} $\lambda1031.926$ 
requires careful consideration of the absorption produced by nearby lines
of other atomic and molecular species (see Sembach 1999 for model 
templates of the absorption in the $FUSE$ bandpass).  
At negative velocities, the primary 
lines that blend with \ion{O}{6} $\lambda1031.926$ are 
the H$_2$ (6--0) P(3) line at 1031.191\,\AA\ (--214~\kms\ with respect to the 
rest wavelength of the \ion{O}{6} line) and the \ion{Cl}{1}
lines  at 1031.885\,\AA\, (--303~\kms) and 1031.507\,\AA\ (--122~\kms).
At positive velocities, the primary contaminant is the H$_2$ (6--0) R(4) line 
at 1032.349\,\AA\ (+123 \kms).  Of these, the H$_2$ lines are the most 
important, since even small H$_2$ column densities in the $J = 3$ or $J = 4$ 
rotational levels produce detectable absorption lines; 20\,m\AA\ features 
correspond to $N_J$ as low as $(1-2)\times10^{14}$ cm$^{-2}$ (see, e.g.,
Figure~1, top panel).  We carefully
modeled the strengths of the H$_2$ $J=3$ and $J=4$ 
lines in the \ion{O}{6} wavelength region using other H$_2$ lines in the same 
rotational levels at other wavelengths (see Wakker et al. 2002
for a description of 
the procedure).  A weak  
transition of \ion{Co}{2} at 1031.542\,\AA\ (--112 \kms) produces negligible 
absorption along the low-density sight lines studied in this 
work.\footnotemark  
\footnotetext{For $N$(H\,{\sc i}) = $10^{21}$ cm$^{-2}$, a cosmic
(meteoritic) abundance ratio (Co/H) = $8.13\times10^{-8}$
(Anders \& Grevesse 1989), and a typical interstellar Co depletion 
$\delta_{Co} \leq 0.03$ (Federman et al. 1993), we expect 
$W_{1032}$(Co\,{\sc ii})~$\ll 1$\,m\AA.} The 
deuterated hydrogen HD (6--0) R(0) $\lambda1031.912$ line, which occurs at low 
velocities (--4 \kms\ with respect to \ion{O}{6} $\lambda1031.926$), can be 
important for studies of the interstellar medium (ISM) in the 
Galactic disk and halo but has not yet 
been observed in absorption at high velocities.  HD is detected toward only 
one object in our sample
(NGC\,7469, at low velocity) and is not important 
for our study of the high velocity gas along this sight line, though it must
be accounted for in our study of the halo gas in this direction
(Savage et al. 2002b).

Another possible source of confusion with the high velocity \ion{O}{6} 
$\lambda1031.926$ absorption is 
redshifted \ion{H}{1} and  metal-line absorption in the intergalactic 
medium along the sight line or in the immediate environment of the QSO/AGN.
The most important lines to search for within $\pm500$ \kms\ of the 
\ion{O}{6} $\lambda1031.926$ line 
are \ion{H}{1} Ly$\beta$ ($z \sim 0.0044-0.0077$) and  \ion{H}{1} 
Ly$\gamma$ ($z \sim 0.0593-0.0628$), but we also found instances of 
\ion{H}{1} Ly$\delta$ ($z \sim 0.0847-0.0883$), \ion{H}{1} Ly$\epsilon$ 
($z \sim 0.0985-0.1022$), \ion{C}{3} $\lambda977.020$ ($z \sim 0.054-0.0580$),
\ion{Si}{2} $\lambda1020.699$ ($z \sim 0.0093-0.0127$), and 
\ion{S}{3} $\lambda1012.502$ ($z \sim 0.0175-0.0209$).
Whenever possible, we have checked existing {\it Hubble Space Telescope (HST)}
data for corresponding Ly$\alpha$ absorption or $FUSE$ data for 
other \ion{H}{1} Lyman series lines to confirm or refute such identifications.
The middle and bottom panels of Figure~1 contain several examples of 
intervening absorption lines arising in the IGM.
We refer the reader to Wakker et al. (2002)
for comments about IGM absorption features
near the \ion{O}{6} lines.

\subsection{Column Densities}

We derived the column densities listed in Table~1 from the apparent optical 
depths of the 
\ion{O}{6} $\lambda1031.926$ absorption lines 
under the assumption that there is no unresolved saturated 
structure in the profiles.  The apparent column density as a function
of velocity, $v$ (\kms), is given by 
$N_a(v) = (m_e c) (\pi e^2)^{-1}\,\tau_a(v)\,(f\lambda)^{-1} 
= 3.768\times10^{14}\,\tau_a(v)\,(f\lambda)^{-1}$ 
[cm$^{-2}$~(km~s$^{-1}$)$^{-1}$], 
where the wavelength $\lambda$ is given in \AA ngstroms, and the 
optical depth is related to the 
estimated continuum intensity and the
observed line intensity through the relation
$\tau_a(v)$ = {\rm ln}[$I_c(v)/I(v)$].  The $N_a(v)$ profiles represent valid, 
instrumentally-blurred versions of the actual column density profiles when 
no unresolved saturated structures are present or when the absorption is
weak ($\tau_{max} \lesssim 1-3$).
In such cases, $N = \int{N(v) dv} = \int{N_a(v)} dv$; 
see Savage \& Sembach (1991) for
a discussion of the applicability of the apparent optical depth
method and uncertainties involved in calculating column densities from 
$N_a(v)$ profiles. At the spectral resolution afforded by $FUSE$, 
the \ion{O}{6} lines are fully resolved for gas at temperatures  
$T\gtrsim1.4\times10^5$\,K. At lower temperatures, the lines may still be 
resolved if the turbulent motions in the cloud are modest.  For example,
if b$_{turb} \approx 8$ \kms, the lines will be resolved for 
$T \gtrsim 8\times10^4$\,K.
At lower temperatures, the abundance of \ion{O}{6} in collisionally ionized
gas is expected to be less than a few percent of the total oxygen abundance, 
even in non-equilibrium cooling situations (Shapiro \& Moore 1976).

We report two errors for each of the column densities in Table~1.  The 
first error ($\sigma_{sc}$) is a quadrature addition of the uncertainties
associated with continuum placement and statistical (Poisson) noise
propagated through the line profile following the procedures outlined by
Sembach \& Savage (1992).  This error accounts for most of the uncertainty 
involved in making an objective estimate of the line strength.
The second error ($\sigma_{sys}$) is a quadrature addition
of systematic uncertainties accounting for fixed-pattern noise, choice 
of velocity integration limits, and removal of H$_2$ features.  This error,
which often dominates the total error, accounts for some of the more 
subjective decisions that must be made when measuring the line strength.
To be conservative, we have made generous allowances for the various 
components of $\sigma_{sys}$.  The 
fixed-pattern noise component of the systematic error accounts for an
intensity uncertainty of $\sim10$\% integrated over the width of one 
resolution element.
Fixed-pattern noise structures can mimic weak, high velocity 
absorption lines.  This pattern can vary 
from one observation to the next, depending upon the positions of the spectra
on the microchannel plates.  Whenever possible, we checked the data from 
multiple instrument channels (LiF1, LiF2) to 
determine the impact of fixed-pattern noise on the observed absorption 
features.\footnotemark\  
The velocity integration component of the systematic error accounts for 
changes of $\pm15$ \kms\ in the values of $v_{min}$ and $v_{max}$.  The 
H$_2$ decontamination component accounts for modest uncertainties in the 
profile shapes and strengths of the model H$_2$ lines used to deblend the 
absorption ($\Delta\,v_0 = 10$ \kms, $\Delta$\,b~$ = 10$ \kms, 
and $\Delta\,[I/I_c] = 0.2$).
Wakker et al. (2002) describe all of these sources of error and the methods 
used to calculate them.
\footnotetext{Only two 
of the objects considered in this study (Ton\,S210, HE\,0226-4110) 
were observed with special focal-plane image motions 
to reduce the fixed-pattern noise in the spectra.}  

We estimated $3\sigma$ upper limits to the column densities for the 43 
sight lines for which we find no evidence of high velocity \ion{O}{6}
absorption in the following manner.  We scaled the equivalent 
width error ($\sigma_{W}$) obtained for the Galactic 
thick disk/halo component of the 
absorption found by Wakker et al. (2002) 
to a value appropriate for an integration range 
of 100 \kms.  We then  calculated the corresponding $3\sigma$ column density 
limit assuming the 
linear curve-of-growth relation between $W_\lambda$~and~$N$:  
$W_\lambda = (\pi e^2) (m_e c^2)^{-1} N f \lambda^2$.  For 
\ion{O}{6} $\lambda1031.926$, the $3\sigma$ column density limit is 
$N {\rm(cm^{-2})} < 2.4\times10^{12}~\acute{\sigma_{W}} {\rm (m\AA)}$, where 
$\acute{\sigma_{W}}$ is the scaled equivalent width error.
Table~2 contains the Galactic coordinates ($l,b$), data quality ($Q$), 
scaled equivalent width error ($\acute{\sigma_{W}}$), and column 
density limit for each of the 43 sight lines.  Notes appended to the entries
indicate whether the listed limits may be affected by other absorption
features.

The values of $N$(\ion{O}{6}) listed in Table~1 
should be free of saturation effects if the \ion{O}{6} is collisionally
ionized.  If the gas is photoionized, the temperature may be as low as
$\sim10^4$\,K. Consider the following conservative example.
A single isolated Gaussian \ion{O}{6} $\lambda1031.926$ feature that arises
in $10^4$\,K gas and has {\it no} additional broadening beyond its thermal
width (b~$\approx 3.2$ \kms)  must have an intrinsic central optical depth 
$\tau_{max}(v)\approx1$ before a 0.10 dex saturation correction is required 
for a $FUSE$ 
observation of the $\lambda1031.926$ line.
When observed at the resolution of $FUSE$, this line has a 
maximum apparent optical
depth $\tau_{a,max}(v)\approx0.21$ and a maximum apparent column density per unit velocity $N_{a,max}(v)\approx6\times10^{11}$ cm$^{-2}$ 
(\kms)$^{-1}$.
  For comparison, an isolated
\ion{O}{6} $\lambda1031.926$
line with $\tau_{max}(v)\approx1$ and 
b~$\approx 15$ \kms\ observed by $FUSE$ would have 
$\tau_{a,max}(v)\approx0.75$ and 
$N_{a,max}(v)\approx2\times10^{12}$ cm$^{-2}$ (\kms)$^{-1}$.
Since all but a few of the 
observed \ion{O}{6} features have line widths well in excess of the 
instrumental width, cold narrow components can contribute only a fraction
of the total observed columns, and therefore the effects of these components 
on the resulting $N_a(v)$ profiles is diminished by  
broader components (i.e., their central optical
depths would have to be much greater than unity to impact the total
column density by 0.1 dex.)

Ideally, one would like to check explicitly whether unresolved saturated
structure is present in the observed \ion{O}{6} absorption profiles by 
comparing the values of $N_a(v)$ for both lines of the doublet.  In
practice, this is difficult to do  for high velocity \ion{O}{6} since the 
high velocity portions of the 1037.617\,\AA\ line are often blended with other 
Galactic absorption features.  The comparison at high velocities 
is possible in two cases (ESO\,572-G34 and PG\,1116+215), and no unresolved
saturated structure is found.
For sight lines where it is possible to 
examine the low velocity portions of the profiles, the $\lambda1031.926$ and 
$\lambda1037.617$ results are usually in good agreement, implying that 
saturation corrections for absorption in the thick disk/halo of the 
Galaxy are not required for most of the observations.
Wakker et al. (2002) illustrate the $N_a(v)$ profiles for the two \ion{O}{6}
lines for all of the objects surveyed.  In 17 of the 22 cases where a  
comparison of $N_{a,\lambda1037}/N_{a,\lambda1032}$ can be made, the 
ratio is unity to within the $1\sigma$ uncertainties.  In the remaining
5 cases, the values of $N_{a,\lambda1037}/N_{a,\lambda1032}$ range from
$\sim1.1$ to 1.6, with 4 of the 5 cases having saturation corrections
of less than 0.25 dex (see Savage et al. 2002b).  
Since the high velocity \ion{O}{6} 
absorption features are generally weaker than the thick disk/halo features,
saturation corrections for the high velocity gas are small
enough that they do not affect the scientific conclusions of this study.

We show examples of the \ion{O}{6} apparent column density profiles 
for the  Mrk~509, ESO\,572-G34, and PG\,1116+215 sight lines in Figure~2.  
Toward
Mrk\,509, the thick disk/halo \ion{O}{6} $\lambda1031.926$ 
absorption at low velocities is several times stronger 
than the high velocity absorption.   The $\lambda1037.617$ 
absorption is blended with other features at high velocities, so a direct
comparison of the $N_a(v)$ profiles at $v_{LSR} < -100$ \kms\ is not
possible. However, the good agreement in the $N_a(v)$ 
profiles between --50 and +100 \kms\  suggests that the weaker, high 
velocity absorption seen in the $\lambda1031.926$ line 
should not be affected by unresolved saturated structure.
Toward both ESO\,572-G34 and PG\,1116+215, the high velocity and 
thick disk/halo absorption features have comparable  strengths and widths. 
The good agreement in the 
$N_a(v)$ profiles at velocities between --50 and +250 \kms\
shows that the data are unaffected by unresolved
saturated structure.  Thus, the $N_a(v)$ profiles for these sight lines are 
also valid instrumentally-smeared representations of the true 
values of $N(v)$.  Since the high velocity \ion{O}{6} 
line widths and column densities
are correlated (see \S 9.3), we conclude that saturation effects do not 
strongly influence the \ion{O}{6} column densities derived for the high 
velocity gas.

\subsection{Velocity Centroids and Line Widths}

For each sight line where high velocity \ion{O}{6} is observed, we calculated
the centroid velocity of the absorption by integrating 
the first moment of the apparent optical depth profiles,
$\bar{v} = {\int{v\,\tau_a(v) dv}/\int{\tau_a(v) dv}}$, over the 
velocity range ($v_{min}$ to $v_{max}$) used for the column density 
calculation.  We calculated the velocity widths by integrating the 
second moments of the apparent optical depth profiles, 
b~=~$\sqrt{2\int{(v-\bar{v})^2 \tau_a(v)dv}/\int{\tau_a(v)dv}}$,
over the same velocity range.  The factor of $\sqrt{2}$ in this equation
scales the derived width to an approximate Doppler spread parameter;
for a single Gaussian component this relationship is exact: 
b~$=\sqrt{2} \sigma$.
Most of the high velocity \ion{O}{6} 
features observed are weak [$\tau_a(v) < 1$].  Therefore, the derived 
value of the line width can be sensitive to the  integration limits
and to noise fluctuations near the ends of the integration range.  
It is best to examine the values of b for the sample in a statistical sense 
rather than regarding these as robust measures of the high velocity widths for 
individual sight lines.  We list values of $\bar{v}$ and b in
Table~1.

\section{GENERAL GROUPS OF HIGH VELOCITY FEATURES}

In the top panel of Figure~3, we identify the 102 sight lines in 
our sample.  
In this Hammer-Aitoff 
projection, the Galactic
anti-center is at the center of the plot, and Galactic longitude increases 
from right to left.  The name of each object listed in Table~1
appears next to each point.  
The absence of data points at low Galactic latitudes 
($-15\degr\lesssim b \lesssim15\degr$) is the result of 
selecting objects sufficiently far away from the Galactic plane
to avoid significant
extinction of far-ultraviolet radiation.\footnotemark\  
Several other regions of the figure
are also sparsely populated, including the Galactic center region for 
$0\degr \lesssim b \lesssim 70\degr$ and the anti-center region for 
$-60\degr \lesssim b \lesssim 0\degr$.
\footnotetext{A reddening of E(B--V) = 0.10 magnitudes 
corresponds to a flux reduction
factor of 4 at 1032\,\AA\ (i.e., A$_\lambda$ = 1.5 mag) 
for the standard Galactic interstellar 
extinction curve with a total to selective extinction ratio 
R$_{\rm V}$ = 3.1 (Cardelli, Clayton, \& Mathis 1989).}

Along 59 of the 102 sight lines we 
detect high velocity \ion{O}{6} $\lambda1031.926$ absorption with integrated 
(total) values of $W_\lambda / \sigma_{W} \ge 3.0$.
Whenever possible, we have identified the high velocity \ion{O}{6} 
absorption features listed in Table~1 with corresponding high velocity
features observed in \ion{H}{1} 21\,cm emission.  \ion{H}{1} data of 
varying quality exist for all of the sight lines considered here.  The
available single-dish radio data are shown by Wakker et al. (2002). 
The sources of \ion{H}{1} data include the Leiden-Dwingeloo
survey (Hartmann \& Burton 1997), NRAO 140-foot telescope observations
(Murphy et al. 1996), the Villa Elisa survey (Arnal et al. 2000;
Morras et al. 2000), and Effelsberg 100\,m telescope
observations (see Wakker et al. 2001).

For some sight lines, the correspondence
in the velocities and celestial coordinates of the \ion{O}{6} and 
\ion{H}{1} allow for a straight-forward association to be made, but for many
features the association with \ion{H}{1} emission is much less
certain.  Even in cases where there is obvious \ion{H}{1} emission at the 
velocities of the \ion{O}{6} absorption, one must keep in mind that the 
exact relationship between the \ion{H}{1} and \ion{O}{6} can be complicated
by ionization differences (neutral versus ionized) and beam size 
effects (typically 9\arcmin--36\arcmin\ for the \ion{H}{1} emission versus 
a micro-arcsecond beam size for the \ion{O}{6} absorption). 
With these caveats in mind, we split the sample of high velocity \ion{O}{6}
features into several categories, which are listed in Table~3.
We summarize the number of 
features in each category, the Galactic coordinates 
spanned by the \ion{O}{6} features observed in each category (not necessarily
the full range spanned by the \ion{H}{1}), and the mean
 \ion{O}{6} column densities
($\langle \log N \rangle \pm \sigma_{\langle \log N \rangle}$), 
central velocities ($\langle \bar{v} \rangle$), 
and velocity widths ($\langle {\rm b}\rangle$).
Several of these classifications contain high velocity \ion{O}{6}
features with relatively unambiguous \ion{H}{1} counterparts 
(Complex~A, Complex~C, the Outer Arm, and the Magellanic Stream). 
Other classifications contain high velocity \ion{O}{6} features that
do not have 
obvious \ion{H}{1} counterparts, and these are labeled as either 
Local Group clouds or as ``Other''.  We discuss the above 
categories of features and selection criteria in more detail in \S\S6 and 8.

We show the locations and  various categories of \ion{O}{6} features on the 
sky in  the bottom panel of Figure~3.  
Each category is represented by a different symbol.
We include the 43 null detections listed in Table~2 
to indicate directions that have been searched for \ion{O}{6} but do not have 
any detectable high velocity gas at a level consistent with the data quality
of the observations.  Most of these null detections (34/43) occur
in the northern Galactic sky, and several occur within a few degrees of 
sight lines where high velocity \ion{O}{6} is present. 

\section{RESULTS FOR THE SAMPLE OF HIGH VELOCITY \ion{O}{6} ABSORPTION 
FEATURES}

Eighty-five
high velocity \ion{O}{6} absorption features are present along 59 of 
the 102 complete halo sight lines having sufficient $FUSE$ data 
quality to detect high velocity \ion{O}{6} $\lambda1031.926$ absorption
at a significance of 
$\gtrsim3\sigma$.  Most of these detections are for absorption features
with equivalent widths $\ge 30$\,m\AA.  Six detections have 
$W_\lambda < 30$\,m\AA.  We include a weak feature toward PG\,1259+593 
with $W_\lambda \approx 14$\,m\AA\ even though the formal detection
confidence is only 
$\sim2.3\sigma$ since the data quality is excellent and the 
feature is likely real.
In this section, we provide information about the column densities,
sky covering factors, and velocities of the \ion{O}{6}
absorption.

\subsection{Column Densities}

We plot a histogram of the logarithmic 
column densities of the high velocity \ion{O}{6}
features in Figure~4, where we have binned the data into 0.10 dex intervals.   
The minimum and maximum (logarithmic)
column densities are 13.06 and 14.59, respectively, with an average value 
of  $\langle\log N \rangle = 13.95\pm0.34$ and a median of 13.97.  
(The value of $\log \langle N \rangle= 14.06\pm^{0.22}_{0.71}$ is similar.)
The average high velocity \ion{O}{6} column 
density is a factor of $\sim2.7$ times (0.42 dex) lower than the value of 
$\langle \log N \rangle =14.38\pm0.18$ 
reported for the Galactic thick disk/halo \ion{O}{6} 
by Savage et al. (2002b).  The 
thick disk/halo distribution of column densities, which is indicated in 
Figure~4 with a dashed line, is reasonably well represented as a 
Gaussian distribution.  However, the observed velocity distribution for the
high velocity \ion{O}{6} is skewed
and considerably broader because it is possible to trace lower column
density features at velocities where blending with other \ion{O}{6} 
absorption is minimal.  Low column density features exist in the 
thick disk/halo distribution as well (see Jenkins 1978a), 
but they are presumably
blended with stronger \ion{O}{6} features at similar velocities.  
The lowest \ion{O}{6} column density observed for the Milky Way 
thick disk/halo ($\log N = 13.88$ toward Mrk\,1095,
not counting upper limits; Savage et al. 2002b) 
is comparable to the mean of the high velocity sample.

There are 17 sight lines for which the high velocity gas dominates the 
{\it total} \ion{O}{6} absorption;
these include ESO\,265-G23, HE\,1115-1735, 
MRC\,2251-178, Mrk\,304, Mrk\,335, Mrk\,926, Mrk\,1502, NGC\,588,
NGC\,595, NGC\,1705, NGC\,7469,
NGC\,7714, PG\,0052+251, PG\,1626+554, PG\,2349-014, PHL\,1811, and 
UGC\,12163.  
In these directions the total \ion{O}{6}
column densities of the high velocity gas are 
$\sim0.31\pm0.18$ dex higher than those of the Galactic thick disk/halo.

We plot the total high velocity \ion{O}{6} column
density as a function of thick disk/halo \ion{O}{6} column density in the 
top panel of 
Figure~5.  This plot shows roughly an order of magnitude scatter in the 
high velocity gas columns at all values of $N$(\ion{O}{6})$_{\rm MW}$
encountered along these sight lines.
Within this scatter there is an apparent anti-correlation of 
$N$(\ion{O}{6})$_{\rm HV}$ and $N$(\ion{O}{6})$_{\rm MW}$.  For the 53/59 
sight lines where good measures of both quantities are possible (excluding 
upper limits), we find a linear correlation coefficient of --0.26 and 
a slope $\Delta \log N_{\rm HV} / \Delta \log N_{\rm MW} \approx -0.40$
(indicated by the dashed line).
The Spearman rank correlation probability that the two populations are
uncorrelated is only 5\%, which indicates that the two populations are 
anti-correlated at about the $2\sigma$ level. 
If the six upper limits on $\log N_{\rm MW}$ are included in the analysis 
as measured values, the significance of the all-sky anti-correlation increases 
to $2.8\sigma$, with an even higher significance if the actual values are 
less than the measured limits.  (If the actual values  are 
a factor of two less than the limits, then the significance of the 
anti-correlation climbs to $3\sigma$.)  

A weak anti-correlation between $\log N_{\rm MW}$ and 
$\log N_{\rm HV}$ may be expected if some of the high velocity gas is 
lower velocity disk/halo gas that has been displaced to high velocities by 
energetic processes occurring within the Galaxy (e.g., supernovae).  We 
estimate the approximate fraction of gas that would need to be displaced 
to cause the observed anti-correlation by considering a subset of the data
points shown in the figure.  If the 
high velocity data points with $\log N$(\ion{O}{6})$_{\rm HV} < 14$ are removed
from this plot, the correlation disappears.  For these 22 data points, 
$\langle$ $N$(\ion{O}{6})$_{\rm HV}$/$N$(\ion{O}{6})$_{\rm MW+HV}$ $\rangle$ 
= $0.40\pm0.21$.  Thus,
an average displacement of $\lesssim40$\% of the total \ion{O}{6} along these 
sight lines would be sufficient to produce the observed trend. Sixteen of
the 22 sight lines occur in the northern sky, where the thick disk/halo
\ion{O}{6} distribution is enhanced relative to the southern sky.

Selection effects likely influence the trend seen in the top panel of 
Figure~5.  The enhancement of thick disk/halo \ion{O}{6} in the 
northern sky (Savage et al. 2002b) together with the low values of 
$N$(\ion{O}{6})$_{\rm MW}$ along southern sight lines containing 
large columns 
of high velocity \ion{O}{6} at longitudes $30\degr < l < 130\degr$
({\tt MS}, {\tt MSe}, {\tt LG} categories - see Table~3) 
results in the general trend seen. 
The anti-correlation is weakened
if data points in the northern and southern Galactic sky are considered 
separately.  In the north, the data points have a distribution that shows an
anti-correlation at $<1\sigma$ confidence.  In the south, the anti-correlation
has a significance of $\sim1.5\sigma$ if limits are included, and 
$\sim1.0\sigma$ if the limits are omitted.  Thus, we conclude that much of 
the observed anti-correlation may simply be due to differences in the 
Milky Way and HVC distributions in the two hemispheres.

In the bottom panel of Figure~5 we plot the integrated high velocity
\ion{O}{6} column density against the integrated 0.25 keV X-ray intensity
along the 59 sight lines where high velocity
\ion{O}{6} is detected.  The soft X-ray data points are from
the ROSAT all-sky survey (Snowden et al. 1997) and are R1+R2 band
averages of the eight 12\arcmin\ pixels ($\sim1150$ arcmin$^2$ total) 
surrounding each sight line, ignoring the pixel directly along the 
sight line to avoid the X-ray emission from the background AGN or QSO being
observed.
The anti-correlation of $N$(\ion{O}{6})$_{\rm HV}$ with $I$(X) is slightly
stronger here than in the top panel of Figure~5.  The trend 
largely disappears if sight lines with  $\log N$(\ion{O}{6})$_{\rm HV} < 14$
are omitted from consideration.  
Interestingly, a  similar comparison of $I$(X) with 
$N$(\ion{O}{6})$_{\rm MW}$ by Savage et al. (2002b) reveals no correlation
between $N$(\ion{O}{6})$_{\rm MW}$ and $I$(X), which if taken at face value
seems to indicate that the X-ray and high velocity \ion{O}{6} are coupled
more strongly.  This too may be the result of selection effects.
The interpretation of this plot is
complicated by various problems encountered when comparing infinitesimal point 
source absorption-line observations with low resolution (both spatial
and spectral) X-ray emission observations.  Besides sampling different 
paths through the ISM, additional factors such as strong spectral attenuation
in the X-ray band by intervening hydrogen, small scale structure in the 
ISM, and strong X-ray signals from foreground ISM gas
make it difficult to quantify the X-ray emission
associated explicitly with the high velocity gas.  Future high-resolution
X-ray observations will be needed to determine if this trend holds up
under more careful scrutiny.

\subsection{Sky Distribution -- Column Densities}

The absorption-line measurements for the individual sight lines in our study 
sample gas along very small solid angles.  However, it is possible to 
construct a rough picture of the \ion{O}{6} distribution on the sky when
the sight lines are considered together.  
Figure~6 contains all-sky plots of the distribution of high velocity 
\ion{O}{6} column 
densities.  In the top panel, the 
column densities are coded for each sight line 
according to symbol size, with filled 
circles representing detections and  downward-pointing
triangles indicating upper limits.  In cases where multiple features are 
present along a sight line, the filled circle is split, and the values
for the individual features are coded accordingly.  In the bottom panel
of Figure~6,
we plot a small filled circle, to indicate the precise location of each 
sight line, surrounded by a color-coded
region (12\degr\ in radius) indicating the high velocity \ion{O}{6} 
column densities of the features listed in Table~1.  In
some directions, the region consists of several colors since multiple
high velocity features are present.  For sight lines where no high
velocity gas is detected (see Table~2), we color-code the column densities of 
the upper limits with small ($2.5\degr$ radius) circles.
Each sight line can be identified by
cross-referencing its position with the object names in Figure~3.
Large \ion{O}{6}  column densities for the high velocity gas 
occur in the southern 
Galactic hemisphere in directions associated with either the Magellanic
Stream (e.g., Fairall~9, NGC~1705) or possible extragalactic clouds (e.g.,
Mrk~304, Mrk~926).  In a few cases (e.g., ESO\,265-G23, ESO\,572-G34), 
large  column densities also occur in the northern Galactic sky.  

A comparable map of the sky distribution of Galactic thick disk/halo \ion{O}{6}
column densities is shown and discussed by Savage et al. (2002b).  
The \ion{O}{6} catalog paper 
contains maps of the high velocity \ion{O}{6} column densities in 
narrow velocity channels.  These maps allow one to examine the 
column densities as a function of velocity and position as well as 
the relationship of the high velocity \ion{O}{6} to lower velocity \ion{O}{6}. 

\subsection{Sky Covering Factor}
To make a meaningful estimate of the sky covering factor of the high
velocity gas, it is important to note that the data quality varies from
sight line to sight line. If we consider only the 
highest quality ($Q=4$) sight lines in the \ion{O}{6}
sample, we find that 22 of 26 
sight lines have {\it total} high velocity \ion{O}{6} 
equivalent widths $\Sigma W_{1032} \ge 30$\,m\AA\ at a detection confidence
[$W_{1032} / \sigma_{W}]_{tot} \ge 3.0$.  The remaining four $Q=4$
sight lines
(NGC\,1068, PG\,0804+761,VII\,Zw\,118, and vZ\,1128) have equivalent
width errors that would allow high confidence detections of 30\,m\AA\ 
features if they were present (see Table~2).  The $Q=4$ sight lines are 
scattered across the sky and are not unique to one particular region.  
At lower data qualities
($Q=1-3$), the percentage of sight lines with $\Sigma W_{1032} \ge 30$\,m\AA\
detected at high confidence drops to about 50\%, with few of the non-detection
sight lines having sufficient data quality to rule out weak absorptions.
Fifty-nine of the 102 sight lines in the sample have total high velocity 
\ion{O}{6} 
equivalent widths $\Sigma W_{1032} > 30$\,m\AA\ at a detection confidence 
[$W_{1032} / \sigma_{W}]_{tot} \ge 3.0$.   Thus, a key result of our survey 
is that a large fraction (at least 58\% and perhaps as many as 85\%) of 
the sight lines contain high velocity \ion{O}{6} with $\Sigma W_{1032} > 
30$\,m\AA.

At higher equivalent width thresholds, $\Sigma W_{1032} \gtrsim 100$\,m\AA,
the effect of data quality on the resulting detection rate is diminished
considerably.  We find that $\sim42$\% of the $Q=4$ sight lines have 
$\Sigma W_{1032} \gtrsim 100$\,m\AA\ at $3\sigma$ confidence, compared to 
$\sim38$\% for the $Q=1-3$ sight lines. There is no significant difference
in the detection rate within the individual data quality intervals at this
threshold.  We list the numbers and percentages of sight lines
containing high velocity \ion{O}{6} for various equivalent width 
thresholds and data qualities in Table~4. 
Values for equivalent width thresholds of less than 100\,m\AA\ for 
$Q=1-3$ sight lines are likely 
incomplete and should probably be treated as lower limits.
These numbers are useful for 
comparisons with detection rates for other \ion{O}{6} absorption systems, such as 
the weak IGM \ion{O}{6} absorbers (see Savage et al. 2002a) or the 
Galactic thick disk/halo.
Positive detections of \ion{O}{6} associated with the thick disk/halo occur
for 91 of the 
102 sight lines in our sample, and the eleven 
non-detections occur for sight lines where the data quality is marginal 
(Savage et al. 2002b).

We provide a more detailed estimate of the sky covering fraction of high 
velocity \ion{O}{6} in Table~5, where we list the numbers and percentages
of sight lines in our sample having total 
high velocity \ion{O}{6} column densities, $\Sigma N$(\ion{O}{6})$_{\rm HV}$,
greater than a 
column density threshold, $N_0$.  We list values for the entire sample as 
well as for several slices in Galactic longitude and latitude 
($0\degr < l < 180\degr$, $180\degr < l < 360\degr$; $b < 0\degr$, $b> 0\degr$).  
Note that values of the covering fraction
for $N_0 \lesssim 10^{14}$ cm$^{-2}$ are subject to data 
quality biases of the type discussed above and may underestimate the true
covering fractions.  For 
the entire sample, we find that $\sim60$\% of the sight lines surveyed
(and likely more if the data quality could be improved)
contain high velocity \ion{O}{6} with $N \ge 10^{13}$~cm$^{-2}$, 
and $\sim30$\% have $N \ge 10^{14}$ cm$^{-2}$.  These two \ion{O}{6}
column density thresholds correspond roughly to $N$(H$^+$) $\gtrsim
10^{16} Z^{-1}$ and $\gtrsim 10^{17} Z^{-1}$, respectively, 
where $Z$ is the metallicity of the hot gas
(on a linear scale where $Z=1$ is solar), and the ionized hydrogen 
column density of the hot gas is 
$N$(H$^+$)~=~$N$(\ion{O}{6})~(O/H)$_\odot^{-1}$~${\rm f}_{O\,VI}^{-1}~Z^{-1}$.  
In making these estimates of H$^+$ in the \ion{O}{6}-bearing gas, 
we adopt a reference abundance 
(O/H)$_\odot = 5.45\times10^{-4}$ (Holweger 2001) and an \ion{O}{6}
ionization fraction f$_{\rm O\,VI}~\equiv$~(\ion{O}{6}/O)~$\lesssim~0.2$,
which is the maximum value expected under conditions of either 
collisional ionization or photoionization (see \S9 and Tripp \& Savage
2000).  

Assuming that the high velocity \ion{O}{6} covering factors listed in 
Table~5 provide a fair representation of the entire sky, we find that 
roughly 30--35\% of the sky is covered by the hot, high velocity
 H$^+$ at a level of 
$\gtrsim 4\times10^{17}$ cm$^{-2}$ if the gas has a typical
metallicity comparable to that of the Magellanic Stream, 
$Z_{MS} \sim 0.2-0.3$ (Lu et al. 1998; Gibson et al. 2000; 
Sembach et al. 2001a).  (At this column density, data quality issues 
do not seriously affect this estimate).  
This $N$(H$^+$) estimate would increase by a factor of 
$\sim2$ if the metallicity is closer to the value of $Z \sim 0.1$ found
for Complex~C in the directions of 
Mrk~290 (Wakker et al. 1999) and PG\,1259+593 (Richter et al. 2001). 
 The covering factor of the hot, high velocity H$^+$ inferred from 
the high velocity \ion{O}{6} 
measurements is similar to
the high velocity \ion{H}{1} covering factor at a comparable column density
level.  
In a sensitive survey of \ion{H}{1} 21\,cm emission in 860 directions, Lockman
et al. (2002) find that $\sim37$\% of the sky is covered by 
high velocity \ion{H}{1} at a $4\sigma$ detection threshold of 
$\sim8\times10^{17}$ cm$^{-2}$.

We note that the detection rate of high velocity gas in the southern 
Galactic hemisphere appears to be slightly higher than for northern 
latitudes.  For the full sample, we find 
high velocity detection rates of 75\% (27/36) and 36\% (13/36) 
for $N_0 = 10^{13}$ and 10$^{14}$ cm$^{-2}$ and $b < 0\degr$, 
compared to 48\% (32/66) and 27\% (18/66) for $b > 0\degr$.  The 
values for the two hemispheres differ by about $1.5\sigma$ if the 
sight lines are drawn from a random population.  Small number
statistics, data quality, 
and selection effects can play important roles in such
comparisons.  For example, a relatively large percentage of the northern 
sight lines (9/66) lie in the general direction of Complex~C,
where high velocity \ion{O}{6} is prevalent (see \S 6.1).
Therefore, it is desirable to diversify and increase both the sample size
and data quality
(preferably by factors of 2 or more) to test whether the trend we note
in the current sample is a general feature of the high velocity sky or 
a subtle selection effect.

\subsection{Velocity Centroids}

The top panel of Figure~7 contains a histogram of the high velocity 
\ion{O}{6} feature centroids in the LSR reference frame.  
These centroids are the 
profile-weighted velocities (first moments) derived by integrating the 
apparent optical depth profiles over the velocity ranges listed in Table~1.  
The most extreme 
velocity centroids are found for the sight lines toward  
NGC~588 ($v_{LSR} \approx -372$ \kms), NGC\,4670 ($v_{LSR} \approx +363$ \kms),
and Mrk\,478  ($v_{LSR} \approx +385$ \kms).  
The average negative velocity centroid is $-202\pm77$
\kms, and the average positive velocity centroid is $187\pm70$ \kms.  
The distribution of velocities is relatively symmetric about $v_{LSR}=0$ \kms. 
 The histogram is sampled in 20 \kms\ velocity bins.
The dashed line in the figure shows a Gaussian fit to the observed 
centroids.    
The  Gaussian has a width $\sigma_{\langle \bar{v} \rangle} \approx 207$ \kms\ 
(b $\approx 293$ \kms; FWHM $\approx 488$ \kms) and is centered at 
$\sim -33$ \kms.  Both $\langle \bar{v} \rangle$ and $\sigma_{\langle \bar{v} \rangle}$ are uncertain because of the absence
of points at $|v_{LSR}| \lesssim 100$ \kms. 

For comparison with the distribution of high velocity centroids in the LSR
reference frame, we also
show the distributions in the Galactic Standard of Rest (GSR) and the 
Local Group Standard of Rest (LGSR) in the middle and bottom panels of 
Figure~7.  Velocities in these reference frames are related to the LSR 
velocity through the equations
 $v_{GSR} = v_{LSR} + 220$\,(\kms) $\sin(l) \cos(b)$ and 
 $v_{LGSR} = v_{GSR} -62$\,(\kms) $\cos(l) \cos(b) + 40$\,(\kms) $\sin(l) \cos(b) - 35$\,(\kms)~$\sin(b)$ (\mbox{Karachentsev} \& Makarov 1996).   
In both cases the 
dispersion in the high velocity gas
centroids is smaller than in the LSR reference frame
($\sigma_{\langle \bar{v} \rangle} \approx 144$ \kms\ in the GSR, and $\sigma_{\langle \bar{v} \rangle} \approx 128$ \kms\ in the 
LGSR).  

In Table~6 we list
the means and standard deviations of the high velocity feature 
centroids  for the four Galactic regions with either 
$b<0\degr$ or $b>0\degr$, and $l<180\degr$ or 
$l > 180\degr$.
Values of $\langle \bar{v} \rangle \pm \sigma_{\langle \bar{v} \rangle}$
are given for the LSR, GSR, and LGSR reference frames.  The numbers of 
high velocity features in each Galactic region  are
listed at the bottom of the table. In most quadrants and in the total sample
the mean 
centroid velocity is smaller in the GSR than in the LSR.  In all but 
one quadrant
($l>180\degr, b>0\degr$) the dispersions in the centroids are also smaller
in the GSR than in the LSR, suggesting that this is a better reference 
frame for describing the velocities.  In other words,  the high velocity 
\ion{O}{6} in our sample has an observed distribution 
of bulk velocities that is more evenly and tightly clustered
about the velocity of the Galactic center than about the velocity expected
for gas participating in differential Galactic rotation at the Galactocentric
radius of the Sun (with the important caveat that the HVC features 
were selected in a manner to avoid LSR velocities 
$|v_{LSR}| \lesssim 100$ \kms).
A similar result, which applies to the velocity distribution of 
neutral hydrogen clouds and is subject to many of the same
kinds of selection effects, has been used to support 
arguments for an extragalactic location for some of the \ion{H}{1}
HVCs (e.g., Blitz et al. 1999).  

Features designated {\tt EPn} dominate the high velocity \ion{O}{6} absorption
in the discrepant quadrant for which the standard deviation of the values of 
$\bar{v}$ is larger in the GSR than in the LSR ($l > 180\degr, b > 0\degr$).
This group of clouds is not better represented
in the GSR reference frame and is distinct from the other categories in 
this regard.  As noted in the previous section, some of these features
may trace thick disk/halo gas accelerated to high velocities.

We illustrate the velocities of the \ion{O}{6} features in the LSR reference
frame in Figure~8, where we show the dependence of the velocity on both
Galactic longitude (top panel) and Galactic latitude (bottom panel).  
The vertical bars on each point 
indicate the velocity widths (b-values) of the observed features, not the 
errors in $\bar{v}$.  At $l > 180\degr$, all but one feature   
have positive LSR velocities; the exception is the {\tt LG} feature toward
Ton\,S210, 
which is located near the south Galactic pole.  
At $l < 180\degr$, $\bar{v} > 0$ \kms\
for only 8 features, while $\bar{v} < 0$ \kms\ for 43 features.  This
strong segregation of velocities in Galactic longitude is also apparent
in the latitude plot in the bottom panel of Figure~8.

We provide a similar plot for the 
longitude dependence in the GSR reference frame in Figure~9.   Despite
showing a somewhat tighter clustering of data points about $v_{GSR} = 0$ 
\kms, there is still considerable scatter in the velocity centroids as 
a function of longitude 
($\sigma_{\langle \bar{v} \rangle} \sim 144$ \kms\ for the whole sample). The
dashed lines in the figure bound the range of velocities expected for 
Galactic disk sight lines with $|b| < 20\degr$ and $|v_{LSR}| \lesssim 100$ \kms.  
As expected, few points fall in this zone because of sample selection 
effects.  In
the GSR reference frame, the general velocity distribution of data points
clusters more tightly about zero velocity than in the LSR reference frame.  
The few 
sight lines that fall in the range of velocities expected for the thick 
disk/halo include Ton\,S210 ($l \approx 225\degr$) and several sight lines
in the general direction of Complex~C ($l \sim 80-110\degr$; 
Mrk\,817, Mrk,876, PG\,1351+640, PG\,1415+451, and PG\,1626+554).  Both 
Complex~C and the compact HVC toward Ton\,S210 are likely extragalactic 
clouds that will be discussed further in \S6.

\subsection{Sky Distribution -- Velocities}

In Figure~10 we plot the distribution of \ion{O}{6} centroid velocities
projected onto the sky for both the LSR and GSR reference frames.    
The projection is the same as that in Figure~6, with the Galactic 
anti-center at the center of the plots.  In each direction we plot a small 
filled circle to indicate the precise location of the 
sight line, with a corresponding  a color-coded
region  indicating velocity if high velocity \ion{O}{6} absorption
is detected.  In
some directions, the region contains several colors since multiple
high velocity features are present within the 12\degr\ radius of the region.  
Each sight line can be identified by
cross-referencing its position with the object names in Figure~3.
The color coding of the velocities 
(blue/green = negative velocities, orange/red = positive velocities) is
the same in both panels.  A  comparison of the two panels shows that the color
shading in the GSR panel is generally lighter, indicating that the velocities
are smaller in this reference frame, as discussed above.

We show a similar plot for the \ion{H}{1} high velocity sky observed 
in 21\,cm emission in Figure~11a, where we have coded the
velocities in the same color system, but at the finer spatial resolution 
afforded by the \ion{H}{1} data.  Figures~11b and 11c display
the same data in azimuthal equal-area projections looking up toward
the north Galactic pole and down toward the south Galactic pole.
Several large structures or groups of 
\ion{H}{1} high velocity clouds are visible
in Figure~11, including:  1) the Magellanic Stream, which passes 
through the south Galactic pole and extends up to $b \sim -30\degr$,
with positive velocities for 
$l \gtrsim 180\degr$ and 
negative velocities for $l \lesssim 180\degr$; 
2) high velocity cloud Complex C, which covers a large portion of the 
northern Galactic sky between $l=30\degr$ and $l = 150\degr$ and has 
velocities of roughly --100 to --170 \kms; 3) the extreme positive 
velocity clouds in the northern sky ({\tt EPn}), which are located in the general
region $180\degr \lesssim l \lesssim 330\degr, b \approx 30\degr$; and 4)
the Galactic center negative velocity (GCN) clouds located near $l \sim45\degr,
-60\degr \lesssim b \lesssim -30\degr$. 

In Figures~11a-c, we have indicated the 
positions and velocities of the high velocity \ion{O}{6} features with
filled circles having the same color coding scale as the \ion{H}{1} data.  
If multiple high velocity features are present, the 
circle is subdivided into multiple colored regions.  Small open circles 
(or ``X'' marks for the two stellar sight lines) denote 
the directions where no high velocity \ion{O}{6} is detected (see 
Table~2).  Several key points about these figures are worth highlighting:
\medskip
\\
\noindent
1) There is excellent velocity correspondence between the \ion{H}{1}
and \ion{O}{6} in Complex~C, with sight lines passing near to Complex~C,
but not through the high velocity \ion{H}{1},
showing no \ion{O}{6} absorption (see \S6.1). 
\medskip
\\
\noindent
2) Toward Complex~A, there is a close pair of sight lines exhibiting an
\ion{O}{6} detection (Mrk\,106) and a non-detection (Mrk\,116) (see \S6.2).
\medskip
\\
\noindent
3) The high velocity component toward NGC\,3310 ($l = 156.6\degr, 
b = +54.1\degr$) has a velocity similar to that of Complexes A and C, even 
though there is no \ion{H}{1} 21\,cm emission detected along the sight line 
at these velocities.
\medskip
\\
\noindent
4) The H\,1821+643 sight line contains \ion{O}{6} absorption at the 
velocities of the Outer Arm as well as at more negative velocities
(see \S6.4).  The progression of velocities between the Outer Arm
and Complex~C is relatively smooth.
\medskip
\\
\noindent
5) In the $l < 180\degr, b < 0$ quadrant of the sky there are often two 
negative velocity components, with the highest negative velocity components
occurring in the Magellanic Stream ({\tt MS})
or the extension of the Magellanic Stream ({\tt MSe})
at velocities typical of the Stream (see \S6.3).  Components with $\bar{v}
\sim -120$ \kms\ concentrate to longitudes less than those of the Stream,
whereas components at $l \sim120\degr-140\degr$ have velocities 
typical of the Stream.
\medskip
\\
\noindent
6) There is good velocity correspondence between the \ion{H}{1}
and \ion{O}{6} velocities in the positive velocity portion of the 
Magellanic Stream at $l > 180\degr$ (see \S6.3).  The \ion{O}{6}
features off the main axis of the 
Stream at these longitudes 
have  velocities similar to those of the \ion{H}{1} near the Stream.
\medskip
\\
\noindent
7) In the $l > 180\degr, b > 0$ quadrant of the sky there are positive 
velocity \ion{O}{6} and \ion{H}{1} features, sometimes at similar velocities.
These are some of the {\tt EPn} features listed in Tables~1 and 3.
In some cases, the \ion{O}{6} features have substantially higher velocities
than the \ion{H}{1}.
The --277 \kms\ feature toward ESO\,265-G23 ($l = 285.9\degr, b=+16.6\degr$)
has a velocity and location close to that of \ion{H}{1} in the leading arm of 
the Magellanic Stream identified by Putman et al. (1998);  \ion{H}{1} at
similar velocities is seen $\sim1\degr$ away.  
\medskip
\\
\noindent
8) At $l \sim 180\degr, b > 0\degr$ there is high velocity \ion{O}{6} with 
$\bar{v} \sim +150$ \kms.  Some of these features are broad 
absorption wings extending from the lower velocity absorption produced by
the Galactic thick disk/halo (see \S8.1).
\medskip
\\
\noindent
9) There may be high velocity \ion{H}{1} near the +143 \kms\ \ion{O}{6}
feature toward PKS\,0405-12 ($l = 204.9\degr, b = -41.8\degr$.
\medskip
\\
\noindent
10) High velocity \ion{O}{6} features toward  Mrk\,478 ($l=59.2\degr,
b=+65.0\degr$, $\bar{v} \approx +385$ \kms), 
NGC\,4670 ($l=212.7\degr, b = +88.6\degr$, $\bar{v} \approx +363$ \kms), 
and Ton\,S180 
($l=139.0\degr, b=-85.1\degr$, $\bar{v} \approx +251$ \kms) stand out as 
having particularly unusual
velocities compared to those of other \ion{O}{6} features in similar regions 
of the sky. These features are classified as {\tt Oth} in Table~1, and may be 
located outside the Local Group (i.e., in the IGM).
\medskip
\\
\noindent
11) Sight lines that contain both negative and positive high velocity 
features include Mrk\,509, Ton\,S180, and several Complex~C sight lines
(Mrk\,817, Mrk\,876, PG\,1259+593, PG\,1351+640, and PG\,1626+554).
\medskip

Although not shown in Figures~10 and 11, \ion{O}{6} and \ion{H}{1} are 
also present 
at intermediate velocities ($|v_{LSR}|\sim30-80$ \kms) along many of the 
sight lines considered in this work. 
The intermediate velocity \ion{O}{6}
absorption and its relationship to the intermediate velocity \ion{H}{1} 
sky are discussed briefly by 
Savage et al. (2002b).  The kinematics, elemental 
abundances, molecular content, and locations of the 
intermediate velocity clouds (IVCs) indicate that they are part of the 
Galactic processes that circulate and distribute interstellar material in the 
thick disk and halo of the Galaxy (Houck \& Bregman 1990; Richter et al. 2001, 
2002).  The IVC population appears in most cases to be distinct from 
higher velocity phenomena traced in \ion{H}{1} (see Wakker 2001), 
and it is likely that the \ion{O}{6} IVC population is not 
strongly connected with the high velocity phenomena we observe in \ion{O}{6}
absorption.  There is no general correlation in the amount of intermediate 
and high velocity \ion{H}{1} or \ion{O}{6}
observed along the sight lines in our sample.

Table~7 contains a general breakdown of the number of sight lines
containing negative and positive high velocity \ion{O}{6} features.
For several cuts in Galactic longitude and latitude we list the number of 
sight lines containing at least one high velocity \ion{O}{6} component with a 
velocity centroid, $\bar{v}$, exceeding a 
threshold velocity, $v_0$. The table quantifies the graphical
results shown in Figures~8 and 10, namely that the $l<180\degr$ region
of the sky contains predominantly negative velocities, while the $l>180\degr$
region of the sky contains predominantly positive velocities.  Combined,
roughly 58\% (59/102) of the sight lines have \ion{O}{6} features with 
$|\bar{v}| > 90$ \kms, and 28\% (29/102) have \ion{O}{6} features with 
$|\bar{v}| > 200$.  A small percentage of sight lines, 11\% (11/102), has 
$|\bar{v}| > 300$, and of these all but 3 (Mrk\,478, NGC\,1705, and 
NGC\,4670) are located in the 
southern Galactic hemisphere at $45\degr < l < 135\degr$.  Data quality
affects these percentages since weak features ($W_{1032} 
\lesssim 100$\,m\AA) are below the detection limit for some sight lines.
Thus, the percentages enclosed within parentheses 
in Table~7 are best treated as lower limits.  For 
comparison, we also tabulate the all-sky percentages
for the highest quality ($Q=4$) sight lines within square brackets 
in Table~7.  As expected, the $Q=4$ sight line
percentages are higher than those for the entire sample.

\subsection{Line Widths}
The observed line widths  of the high velocity \ion{O}{6} features range 
from b~$\sim16$ to $\sim81$ \kms.  We show the distributions of b-values for
the high velocity gas and thick disk/halo absorption in 
Figure~12.   These b-values have not been corrected for a small amount of 
instrumental broadening (b$_{inst} \sim 12-15$ \kms).
The average high velocity feature width is 
$\langle$b$\rangle = 40\pm14$ \kms\ if all of the high velocity features 
listed in Table~1 are considered. In some cases, the 
line widths are less reliable since they depend strongly on the exact
choice of the velocity cutoff between thick disk/halo absorption and the 
high velocity absorption used in the calculation of b. Thus, one must 
be careful
about interpreting the meaning of the line width for some sight lines,
especially when the thick disk/halo and high velocity portions of the 
absorption merge without recovery to the continuum at intermediate 
velocities.  These less reliable values of b are enclosed in parentheses 
in Table~1. For the 
19 features having reliable b-values, $\langle$b$\rangle = 42\pm12$ \kms,
which is indistinguishable from the value for the entire sample.

The average high velocity line width is smaller than the value of 
$\langle$b$\rangle = 60\pm15$ \kms\ found for the low velocity 
\ion{O}{6} (Savage et al. 2002b).  This is  due in part to the 
reduced blending of the high velocity gas as well as the multi-component 
nature of the low velocity gas, which by definition occurs between
$-100 \lesssim v_{LSR} \lesssim 100$ \kms.  If the broadening is purely thermal
in nature,  the smallest Doppler widths in the high velocity sample 
(b $\sim16$ \kms)
correspond to temperatures of $\sim3\times10^5$\,K, the temperature at which
\ion{O}{6} peaks in abundance under conditions of collisional ionization 
equilibrium (Sutherland \& Dopita 1993).  These small widths are only
slightly larger than the instrumental resolution, so it is possible that the
temperatures may be lower in these few cases.
The larger line widths are likely dominated
by turbulent broadening, shear, or flows rather than thermal broadening at
high temperatures since the fractional
abundance of \ion{O}{6} decreases rapidly at  $T \gtrsim 10^6$\,K 
(i.e., b$_{thermal} \gtrsim 34$ \kms) (Shapiro \& Moore 1976; 
Sutherland \& Dopita 1993).

\section{O\,{\sc vi} ASSOCIATED WITH PREVIOUSLY IDENTIFIED HIGH VELOCITY
CLOUDS}

We detect high velocity \ion{O}{6} along every sight line in our sample
for which there is a known high velocity \ion{H}{1} cloud observed in 
21\,cm emission along the sight line.  These \ion{H}{1} HVCs include large 
complexes of gas, such as Complex~C and the Magellanic Stream, as 
well as smaller isolated clouds.  We discuss some of these high 
velocity features in this section.

\subsection{Complex C}

High velocity cloud Complex~C is a large assembly of high velocity gas in the 
northern Galactic sky between $l\sim30\degr$ and $l \sim 150\degr$.  It is 
well-known for its substantial \ion{H}{1} content and distance
(M~$>1.2\times10^6$ M$_\odot$; $d > 5$ kpc, z $>$ 3.5 kpc;
 van~Woerden et al. 1999a; Wakker 2001),
lack of molecular hydrogen (Murphy et al. 2000; Richter et al. 2001),
and substantially sub-solar abundances.  The metallicity of Complex~C is  
$Z \sim 0.1$ (Wakker et al. 1999; Richter et al. 2001), but may be as 
high as $Z \sim 0.3$ in some 
locations (Gibson et al. 2001a).  The gas in Complex~C has LSR
velocities ranging from approximately --100 to --170 \kms, 
which is sufficient to separate
the absorption from both the Milky Way thick disk/halo absorption in this
direction (extending out to $v_{LSR} \sim -100$ \kms) and intermediate 
velocity absorption features along the path to Complex~C, such as the 
Intermediate Velocity Arch ($v_{LSR} \sim -55$ \kms) (see Wakker 2001).  
Complex~C contains ionized gas as revealed through
previous observations of \ion{O}{6} absorption (Murphy et al. 2000;
Sembach et al. 2000) and H$\alpha$ emission (Tufte, Reynolds,
\& Haffner 1998; Wakker et al. 1999).

Nine of the sight lines in our sample 
pass through Complex~C and contain \ion{O}{6} absorption at velocities 
characteristic of the high velocity neutral gas.  These are 
Mrk\,279, Mrk\,290, Mrk\,501,
Mrk\,506, Mrk\,817, Mrk\,876, PG\,1259+593, PG\,1351+640, and PG\,1626+554.
We illustrate the \ion{H}{1} 21\,cm
emission in the region of sky containing Complex in Figure~13 to show the 
general morphology of the neutral gas in these directions.  Two velocity
ranges, $-200 < v_{LSR} < -150$ \kms\ (top panel) 
and $-150 < v_{LSR} < -100$ \kms\ (bottom panel), are
shown to illustrate the differences in the \ion{H}{1} structure in these
velocity ranges.  These \ion{H}{1} data are from the Leiden-Dwingeloo survey 
(Hartmann \& Burton 1997).  In the more negative
velocity map, the two sight lines
that lie in densest concentrations of \ion{H}{1} are Mrk\,876 and PG\,1351+640.
In the less negative velocity map, all of the sight lines lie in the 
direction of some \ion{H}{1} emission, and additional intermediate 
and high velocity features (IV~Arch, Complex~A, Complex~K) are identified.

The \ion{O}{6} detections toward Complex~C 
are clearly visible in the all-sky maps shown in Figures~10 and 11.
We plot the \ion{O}{6} $\lambda1031.926$ and \ion{C}{2}
$\lambda1036.337$ absorption profiles for each of the 9 Complex~C
sight lines in Figure~14.  
The \ion{O}{6} absorption traces highly ionized gas, while the \ion{C}{2}
absorption traces a combination of neutral and weakly ionized gases.
Included for each sight line is the \ion{H}{1} 21\,cm 
emission profile from Murphy et al. (unpublished, see Wakker et al. 2002) 
for comparison with 
the absorption features.  The single-dish NRAO 140-foot telescope 
21\,cm emission measurements have a beam size of 21\arcmin\ (FWHM), a 
channel spacing of $\sim1-2$ \kms, and 
a $3\sigma$ sensitivity of $\sim (1-2)\times10^{18}$ cm$^{-2}$ for a 50 \kms\
integration range.  The 21\,cm emission
provides information about the velocity structure of the primary neutral gas
concentrations along the sight lines, while the \ion{C}{2} profile reveals 
the total velocity extent of the neutral plus weakly ionized gas.  

\ion{O}{6} absorption is present at the velocities of the \ion{H}{1} 21\,cm 
emission in Complex~C.  Thus, the highly ionized and neutral gases 
appear to be kinematically related in these different directions.  
Values of $\log N$(\ion{O}{6})
range from 13.67 to 14.35, with an average of 13.93 and a standard
deviation of 0.25 dex.  The
\ion{O}{6} centroids range from a low of $\sim -110$ \kms\ (PG\,1259+593) 
to a high of $\sim-150$ \kms\ (Mrk\,279, Mrk~876,
PG\,1626+554).  
We summarize the relative velocities and column densities
of the \ion{H}{1} and \ion{O}{6} in Table~8.
The \ion{O}{6} and \ion{H}{1} velocities track each other
reasonably well, with a typical scatter of $\sim15$ \kms.
There is no obvious dependence of the \ion{O}{6}
column density on the \ion{H}{1} column density.  
For example, two sight lines (Mrk\,279 and PG\,1351+540) with 
$\log N$(\ion{O}{6})
$\approx  13.67$ have \ion{H}{1} columns that differ by 0.46 dex,
and two sight lines (Mrk\,506 and Mrk\,876) with $\log N$(\ion{O}{6})
$\approx 14.06$  have \ion{H}{1} columns that differ by 0.66 dex.  However,
there is a weak trend for the ratio of the two species to change with 
position across Complex~C, with higher values of 
$N$(\ion{H}{1})/$N$(\ion{O}{6}) at higher longitudes.

The strong \ion{C}{2} and \ion{O}{6} lines have similar negative 
velocity extents along each sight line.  The \ion{C}{2} traces very small 
amounts of \ion{H}{1}, as well as H$^+$, so it is difficult to know 
if the velocity extents of the neutral and highly ionized gas are identical,
 but this result is at least suggestive that they are similar.

Several additional sight lines in the \ion{O}{6} survey pass near 
Complex~C but do not have high velocity \ion{O}{6} detections. 
 We comment briefly on a few of these sight lines below.

\begin{quote}  Mrk\,205 ($l=125.45\degr, b = +41.67\degr$)
has marginal $FUSE$ data ($Q = 1$, $S/N \sim4$), resulting in a $3\sigma$ 
upper limit of log $N$(\ion{O}{6}) $<$ 13.79 for the high velocity gas at
Complex~C velocities.
 High-resolution Space
Telescope Imaging Spectrograph (STIS) 
data show that the sight line intercepts neutral gas at Complex~C 
velocities (Tripp et al. 2002), but the $FUSE$ data are insufficient for 
an analysis of the absorption.  
\end{quote}

\begin{quote}
 The Mrk\,487 sight line ($l = 87.84\degr, b = +49.03\degr$) has
 a \ion{C}{2} $\lambda1036.337$
absorption feature near --140 \kms.  There is no detectable \ion{H}{1}
21\,cm emission at this velocity along the sight line, but nearby Complex~C 
sight lines (e.g., Mrk\,290, PG\,1626+554) show \ion{C}{2} absorption at 
similar velocities.  No high velocity \ion{O}{6} is 
present at a level of $\log N < 14.03~(3\sigma)$, even though the sight line
 passes only $2.6\degr$ from the Mrk\,290 sight line, where high velocity 
\ion{O}{6} is detected with $\log N \approx 14.20$. This 
separation corresponds to a linear dimension of 
$\sim220\,({\rm z}~/~5\,{\rm kpc})$ pc.
\end{quote}

\begin{quote}
3C\,249.1 ($l=130.39\degr, b = +38.55\degr$)
lies a bit further off Complex~C than Mrk\,205 but is still 
only a few degrees from the edge of moderately strong \ion{H}{1} 21\,cm
emission between --150 and --100 \kms\ (see Figure~13). No high velocity 
\ion{O}{6} is present at a level of $\log N < 13.83~(3\sigma)$.
\end{quote}

\begin{quote}
The NGC\,3310 sight line ($l=156.60\degr, b = +54.06\degr$) has a high
velocity \ion{O}{6} component at velocities near those of Complex~C,
but no \ion{H}{1} 21\,cm emission is detected at this velocity or at 
lower velocities typical of Complex~A.  The \ion{O}{6} feature has a
velocity of $\sim-162$ \kms\ and is classified as {\tt Oth} in Table~1. 
\end{quote}

\begin{quote}
PG\,1444+407 ($l=69.90\degr, b = +62.72\degr$), 
PG\,1415+451 ($l=84.72\degr, b = +65.32\degr$), and 
SBS\,1415+437 ($l=81.96\degr,b=+66.20\degr$)
all lie to the north of Complex~C, off of the main \ion{H}{1} 21\,cm emission.
No high velocity \ion{O}{6} is detected at Complex~C velocities.
\end{quote}

\subsection{Complex~A}

Sight lines to two extragalactic objects in 
our sample (Mrk~106, Mrk\,116) pass through high
velocity cloud Complex~A. Complex~A is located at a distance of 
4--10~kpc (z = 2.6--6.8 kpc), 
has a mass of $\sim2.0\times10^6$~M$_\odot$,
and has \ion{H}{1} velocities of $-210 \lesssim v_{LSR} \lesssim -140$~\kms\
(van~Woerden et al. 1999b; Wakker 2001).  We detect \ion{O}{6} absorption 
with $\log N \approx 13.81$ ($W_{1032} \approx 65$ m\,\AA) between 
--150 and --100 \kms\ toward Mrk\,106.  This absorption strength is comparable
to that of the weaker detections toward Complex~C (e.g., Mrk\,279, Mrk\,501,
Mrk\,817, PG\,1351+640).  The
centroid of the absorption (--125~\kms)  is shifted by +32 \kms\ from
the centroid of the \ion{H}{1} 21\,cm Complex~A feature in this direction.
The \ion{O}{6} detection is weak, with $W_{1032} / \sigma_{W} = 3.0$.
Thus, while it is plausible that 
the absorption could be due to Complex~A, a higher quality $FUSE$ spectrum of 
this object would be desirable to confirm both the strength and  velocity 
of this feature.

We do not detect high velocity
\ion{O}{6} toward Mrk\,116, and set a $3\sigma$ limit of
$\log N < 13.49$.  At the distance of Complex~A, the $2\degr$
separation of Mrk\,116 and Mrk\,106 corresponds to a linear separation
of 140--350 pc.  One other object in our sample, PG\,0832+675, lies in front 
of Complex~A and does not exhibit any \ion{O}{6} or low ionization 
species at Complex~A velocities (see also 
Ryans et al. 1997 for a non-detection of \ion{Ca}{2} $\lambda3933.663$ 
absorption).  The lack of \ion{C}{2} absorption in the 
$FUSE$ data sets the lower limit to the distance of Complex~A.

\subsection{The Magellanic Stream}

The Magellanic Stream is a complex of high velocity gas that has been
tidally torn from the Magellanic Clouds as they interact with the Milky Way.  
The Stream is readily visible in the \ion{H}{1} maps shown in Figure~11.
It consists of gas at both positive and negative velocities, with a zero
velocity crossing near the south Galactic pole.  The \ion{H}{1} 21\,cm
emission of the Stream covers nearly
1000 square degrees on the sky (see Putman 2000).
Several sight lines in our sample either intersect the \ion{H}{1} Stream 
(Fairall\,9, NGC\,7714, NGC\,7469, PG\,2349-014)
or lie very near to it on the sky and have velocities similar to the 
Stream velocities (NGC\,1705, PKS\,0558-304).  We classify these as {\tt MS}
sight lines in Table~3.  An additional 12 objects lie in the general 
vicinity of the Magellanic Stream; we label these as possible extensions
of the Magellanic Stream ({\tt MSe}).  The association of the 
{\tt MSe} \ion{O}{6}
features with the \ion{H}{1} gas of the Magellanic Stream is less certain 
than for the {\tt MS} sight lines since there is no detectable 21\,cm emission
at Stream velocities along these sight lines.  Since the {\tt MSe} and {\tt LG} 
features in this region of the sky have similar velocities, it is possible
that some of the {\tt MSe} features may eventually be more properly identified as
Local Group clouds.

We show the 
\ion{H}{1} 21\,cm emission profiles and \ion{O}{6} and \ion{C}{2} absorption
profiles for six 
{\tt MS/MSe} sight lines in Figure~15.   In the four cases where both
high velocity \ion{H}{1} and \ion{O}{6} are detected, the \ion{O}{6} 
absorption is broader than the \ion{H}{1}.  
For the three {\tt MS} 
sight lines with negative Stream velocities, the high velocity
\ion{O}{6} absorption 
overlaps (or nearly overlaps in the case of NGC\,7714) the velocities of 
the \ion{H}{1} and extends to considerably less negative velocities.  
We tentatively classify the \ion{O}{6} feature toward NGC\,7714
($\bar{v} = -259$ \kms), as well
as the higher velocity features toward NGC\,7469 ($\bar{v} = -304$ \kms)
and PG\,2349-014 ($\bar{v} = -326$ \kms) as being associated with the 
Magellanic Stream, while the remaining lower velocity features for the 
latter two sight lines may or may not be associated with the Stream
(these are designated {\tt LG} in Table~1). The 
\ion{O}{6} column densities of the {\tt MS} components for these three 
sight lines
are similar: $\log N$  = 14.18 (NGC\,7469), 14.13 (NGC\,7714), 
and 14.00 (PG\,2349-014).

For the three primary 
{\tt MS/MSe} sight lines with positive Stream velocities, the 
situation 
is also complex.  The association
of the high velocity \ion{H}{1} and \ion{O}{6} is fairly clear for 
Fairall\,9.  However, NGC\,1705 and PKS\,0558-304 both lie near to the 
Stream \ion{H}{1}, but neither sight line intersects any obvious Stream 
\ion{H}{1} emission (see Figure~15).
The velocities of the \ion{O}{6} absorption features are similar to those
of other \ion{H}{1} clouds cataloged by Wakker \& 
van~Woerden (1991) in the neighborhood of these two directions, but 
the association of these 
\ion{O}{6} clouds with the Magellanic Stream is less clear than 
for Fairall\,9 since they lie several degrees off of the main axis of the 
Stream.  Nevertheless, small \ion{H}{1} clouds associated with the Stream
lie in the same area of the sky as these sight lines (Putman et al. 2002).
The two sight lines exhibit high velocity \ion{C}{2}
absorption at velocities similar to those of the \ion{O}{6}.  
Given the similarity of the 
positive velocity \ion{O}{6} absorption to that seen at negative velocities
in the other Stream directions, we tentatively classify the positive
velocity \ion{O}{6} features shown in Figure~15 as highly ionized Magellanic
Stream or Magellanic Stream extension gas.  These features have 
\ion{O}{6} column densities of 
$\log N$ = 14.33 (Fairall\,9), 14.31 and 13.78 (NGC\,1705), and 
13.68 (PKS\,0558-504).  We note that the only other {\tt MSe} 
sight line in this region of the sky, HE\,0226-4410, has a slightly
smaller velocity ($\bar{v} =164$ \kms) and a similar column density
($\log N = 13.98$). 

The highest velocity portions of the \ion{C}{2} $\lambda1036.337$ 
absorption along the 6 sight lines shown in Figure~15 occur at 
velocities similar to those of the high velocity \ion{O}{6} absorption.  
At positive velocities, the \ion{C}{2} absorption is completely confused with 
low velocity \ion{C}{2}$^*$ $\lambda1037.012$ absorption in the ISM 
(at +195 \kms\ with
respect to \ion{C}{2}).  As in the case of Complex~C, the relationship
between \ion{O}{6} and \ion{C}{2} suggests that the \ion{C}{2}
is probably tracing ionized gas.  The slight offset of the 
highest velocity portions of the \ion{C}{2} absorption 
relative to the \ion{H}{1} is consistent with an ionized boundary 
around the Stream.  This finding also agrees well with higher resolution
STIS data.  Sembach (2002) presents a preliminary analysis of 
the high 
velocity gas along the NGC\,1705 sight line and finds that the high velocity 
\ion{O}{6}
absorption is offset slightly from the lower ionization gas.  Furthermore,
the \ion{O}{6} toward NGC\,1705 
has a somewhat smoother velocity structure and does not 
have the distinct component structure seen in the lower ionization lines
with STIS.

\subsection{The Outer Spiral Arm}

Three low latitude sight lines in our sample lie in the general direction of 
the Outer Arm seen in \ion{H}{1} 21\,cm emission in the general
direction $49\degr \le l \le 161\degr, 4\degr \le b \le 31\degr$ at a 
Galactocentric distance of about 15.5 kpc (Kepner 1970; 
Haud 1992).  HS\,0624+6907 ($l=145.7\degr, b = +23.4\degr$) exhibits Outer
Arm \ion{H}{1} emission but has no associated \ion{O}{6} feature.
H\,1821+643 ($l=94.0\degr, b = +27.4\degr$) and 3C\,282 
($l=61.3\degr, b = +17.5\degr$) exhibit \ion{O}{6} absorption with
$\bar{v}_{LSR} \sim -119$ \kms\ and $\bar{v}_{LSR} \sim -90$ \kms, 
respectively.  These velocities are in excellent agreement with both the 
observed
and expected \ion{H}{1} velocities in the Outer Arm (--115 \kms\ toward
H\,1821+643 and --85 \kms\ toward 3C\,282) (Hulsbosch \& Wakker 1998;
Haud 1992).  These velocities are similar to those of some of the 
higher latitude Complex~C sight lines at nearby longitudes, but it is
unlikely that the two complexes of gas are related.

\subsection{Extragalactic Clouds}

Several high velocity clouds in our sample other than Complex~C and 
the Magellanic Stream are excellent candidates for being located
outside the Milky Way.  We discuss below a few of those clouds 
that have been studied by previous investigators at other wavelengths.

\subsubsection{The \ion{C}{4} HVCs Toward Mrk\,509 and PKS\,2155-304}

The Mrk\,509 sight line ($l = 35.97\degr, b = -29.86\degr$)
passes through two highly ionized HVCs detected in \ion{C}{4} 
absorption that have ionization properties consistent with  
clouds irradiated by local extragalactic background radiation
(Sembach et al. 1999).  The \ion{C}{4} absorption components are
centered at --283 \kms\ and --228 \kms.
These clouds have little associated low ionization 
absorption. There is no detectable \ion{H}{1} 21\,cm emission
directly along the sight line [$\log N$(\ion{H}{1})$ < 17.5$ ($3\sigma$)], 
although \ion{H}{1} emission near the velocities of the \ion{C}{4} HVCs
is present within a few degrees of the sight line (Sembach et al. 1995b).

Sembach et al. (2000) first  detected \ion{O}{6} absorption between --345 
and --100 \kms\ toward Mrk\,509 (see also Figure~2).  The \ion{O}{6}
absorption is prominent  at --247 and --147 \kms, with the lower velocity
component shifted slightly redward of the high velocity
\ion{C}{4} absorption.  The detection 
of \ion{O}{6} at these velocities is somewhat unexpected, since  
photoionization by the extragalactic background is not expected to produce 
the observed quantities of \ion{O}{6} unless the path lengths are 
very long and the cloud sizes very large (see \S9).  
At these velocities we find $\log N$(\ion{O}{6}) $\sim 14.36$, which is a 
factor of $\sim10$ higher than predicted by the photoionization model that 
best fits the data for \ion{C}{4} and lower ionization stages (see Sembach 
et al. 1999).  We  conclude that additional ionizing sources are 
responsible for increasing the \ion{O}{6} column density above that 
expected for the pure photoionization case.

A similar result holds for the two highly ionized HVCs toward PKS\,2155-304.
These clouds are also likely ionized by the extragalactic background 
(Sembach et al. 1999), but they contain \ion{O}{6} in quantities much
greater than those expected for a simple photoionization scenario.  One 
possible explanation for the high velocity \ion{O}{6} features observed along 
both the Mrk~509 and PKS~2155-304 sight lines is that these are 
photoionized extragalactic clouds that have begun to interact with the 
hot Galactic corona or other low density medium (see \S11).

\subsubsection{The Isolated Compact HVCs Toward Ton\,S210 and Mrk\,205}

The class of high velocity \ion{H}{1} clouds classified as ``compact and 
isolated'' based on their morphological properties is a difficult group of
clouds to study in absorption at ultraviolet wavelengths since their
sky covering factor is low ($\lesssim1$\% for those cataloged by Putman 
et al. 2002).  Various authors have suggested that the  compact \ion{H}{1} 
HVCs may be distant (greater than a few hundred kpc) analogs of nearer 
\ion{H}{1} HVCs (e.g., Braun \& Burton 2000;  Burton, Braun, \& 
Chengalur 2001),
but this hypothesis is not yet widely accepted.  The two metallicity 
measurements available for these clouds suggests that the gas has 
(O/H) $<0.5$~solar (Sembach et al. 2002), which is consistent with either
a low-metallicity  extragalactic origin or a Magellanic origin.

Two sight lines in our sample pass through or very near to isolated 
compact HVCs.  Ton\,S210 ($l=224.97\degr, b = -83.16\degr$) 
lies behind the outer regions of CHVC\,224.0--83.4--197.  
The properties of this cloud are considered by Sembach et al. (2000, 2002),
who find \ion{O}{6} absorption near the velocity of the \ion{H}{1}
21\,cm emission ($v_{LSR} \approx -200$ \kms).  They base their upper limit of 
(O/H) $<0.5$~solar ($3\sigma$ confidence) on the non-detection of the 
\ion{O}{1} $\lambda1039.230$ line in the $FUSE$ spectrum of this object.
The \ion{H}{1} column derived from the Lyman-series \ion{H}{1} absorption
is consistent with the \ion{H}{1} column from the 21\,cm measurements.
Sembach et al. (2002) also find
that the high velocity gas in this direction contains as much hot 
H$^+$, as 
inferred from the \ion{O}{6} absorption, as \ion{H}{1}, suggesting that the 
cloud has a substantial ionized outer boundary.  We discuss the 
production and significance of such a boundary in \S11.

The --200 \kms\ velocity of the CHVC\,224.0--83.4--197 \ion{O}{6} 
absorption in the Ton\,S210 spectrum stands out in our sample since it is 
the only feature that has a 
negative LSR velocity at Galactic longitudes $l > 180\degr$ (see \S 5.4 and 
Figure 8).  It also has the most extreme negative 
velocity in the GSR reference frame of any feature in the sample 
(Figure 9).  The peculiar velocity
of the cloud is substantially different from that of the bulk of the
Magellanic Stream in this direction ($v_{LSR} \sim -100$ \kms), 
but it is not yet possible to exclude the possibility that it could be 
a fragment of the tidal interaction between the Magellanic 
Clouds and the Galaxy (see Sembach et al. 2002).

The Mrk\,205 sight line ($l = 125.45\degr, b = +41.67\degr$)
passes through CHVC\,125+41--207, which is 
detected in \ion{H}{1} 21\,cm emission (Braun \& Burton 1999) 
and in \ion{Mg}{2} absorption (Bowen \& Blades 1993; Bowen, Blades, 
\& Pettini 1995). This cloud probably also has a sub-solar metallicity,
(Mg/H) $<0.2$ solar (see Gibson et al. 2001b; Wakker 2001), 
but additional measurements  are needed to confirm the low Mg 
abundance.  As noted previously, the data 
for Mrk\,205 are not optimal, and our limit of $\log N$(\ion{O}{6}) 
$< 13.79~(3\sigma)$ is not terribly strict.  Still, it appears that there is 
less \ion{O}{6} associated with the compact HVC along this sight
line than there is with the compact HVC along the Ton\,S210 sight line.

\subsubsection{Local Group Galaxies}
Several Local Group galaxies exhibit \ion{O}{6} absorption occurring at 
high velocities with respect to the LSR.  While not necessarily 
``high velocity'' in the rest frame of the galaxies, we mention them briefly
here for completeness.  Basic
results for the Magellanic Clouds have been described elsewhere (Hoopes 
et al. 2002; Howk et al. 2002b) and are not considered further here other than
to note that the data for both galaxies are consistent with the idea that
each is surrounded by a substantial halo of \ion{O}{6}.  The SMC also
exhibits \ion{O}{6} 
enhancements caused by local structures (e.g., 
supernova remnants, superbubbles, etc.). The \ion{O}{6} column densities
in the Magellanic Cloud halos are comparable to those found for the halo
of the Milky Way, despite the lower metallicities and lower masses
of the Clouds.  

A search for \ion{O}{6} in the 
halos of M\,31 and M\,33 in the directions of several objects yields mixed
results.    
Wannier\& Andersson (2002) detect \ion{O}{6} absorption 
toward QSO\,0045+3926 (a.k.a. RXJ0048.3+3941 or IO And) 
at an LSR velocity of --296 \kms. 
QSO\,0045+3926 is located at $5.3\times R_{25}$ from the center of M\,31. 
The measured velocity is within $\sim10$ \kms\ of that 
expected from an extension of a flat rotation curve for M\,31 at the 
location of the QSO.  A high negative velocity \ion{O}{6} component
is seen in many directions in this region of the sky ($l \sim 120\degr, 
b \sim -25\degr$), so it is difficult to know for certain whether the 
\ion{O}{6} feature toward QSO\,0045+3926
is unique to M\,31 (see Wakker et al. 2002).  High negative 
velocity absorption is also seen in the direction of several \ion{H}{2}
regions in M\,33 (e.g., NGC\,595 and probably NGC\,588).  However,
Wakker et al. (2002) find
no convincing evidence
for an extended, corotating \ion{O}{6}
halo around M\,33 at a level of $\log N$(\ion{O}{6}) $> 14$ based on the
non-detection of {\it positive} velocity gas at M\,33 velocities 
along several 
sight lines that pass near to M\,33.
Additional high quality $FUSE$  observations of other
sight lines in the directions of M\,31 and M\,33 would be useful.

Several sight lines in our sample pass within modest impact parameters 
($< 15$ kpc) of 
smaller Local Group galaxies (e.g., Draco, Leo~I, Ursa~Minor, Carina,
Sculptor).  None of these show unambiguous evidence of \ion{O}{6} 
absorption (see Wakker et al. 2002 for details).

\section{HIGH VELOCITY \ion{H}{1} 21\,CM EMISSION WITH NO CORRESPONDING \ion{O}{6} ABSORPTION}

In general, \ion{O}{6} absorption is seen at velocities at which 
high velocity \ion{H}{1} 21\,cm emission is detected.  However, there are 
a few notable exceptions to this rule.  We list the \ion{H}{1}
column densities and the \ion{O}{6} limits for these 6 cases in Table~9.
 Most of the \ion{O}{6} non-detections occur
for sight lines with low $FUSE$ data quality ($Q=1-2$ in 5/6 cases).
However, in all cases, the non-detection is significant.  One additional
case not listed in Table~9 is  the Complex~A \ion{H}{1} emission at 
--145 \kms\ toward PG\,0832+675;  since this star lies in 
front of Complex~A, no \ion{O}{6} absorption is expected.  

The 
absence of \ion{O}{6} along the sight lines listed in Table~9
may be due to small-scale structure within the 
high velocity gas sampled by the larger \ion{H}{1} beams.  With the 
exception of ESO\,265-G23, the absorption traced by \ion{C}{2} 
$\lambda1036.337$ at the \ion{H}{1} velocities noted is relatively weak, 
which supports the idea that these sight lines pass through low
density regions of the high velocity gas 
encompassed within the field of view of the \ion{H}{1} 21\,cm observations.
Furthermore, in three cases (Complex~A, Complex~C,
and the Outer Arm), other sight lines in the sample show both \ion{H}{1}
21\,cm emission and \ion{O}{6} absorption even though the sight lines 
in Table~9 do not.   

The Complex~M sight line
toward Ton\,1187 does not exhibit high velocity \ion{O}{6}, which may be
due to either lower ionization or small-scale structure.  
Complex~M has been detected in H$\alpha$ emission by Tufte et al. (1998),
which indicates that ionized gas is clearly present in some regions of the 
cloud.
Complex~M may contain  some small-scale 
structure since high velocity
absorption in lower ionization lines is present toward BD\,+38\,2182 
($d \sim 4.0$ kpc) but not 
toward HD\,93521 ($d \sim 1.9$ kpc), which is less than 30\arcmin\ away
(see Danly, Albert, \& Kuntz 1993; Wakker 2001). The smaller distance of 
HD\,93521 may also explain the absence of Complex~M absorption in this 
direction if Complex~M lies beyond z~$\sim 1.5$ kpc.

Either small-scale structure or ionization may be viable explanations for 
the absence of \ion{O}{6} in 
the remaining \ion{H}{1} HVCs in Table~9 (WD, WW478, WW84).  A possible
alternative explanation may be that the gas has a very low metallicity, which
would be very interesting since it would imply that the clouds are located at 
large distances. (Initial estimates indicate that clouds WD and WW84 have
metallicities $Z\sim0.1$.) This
hypothesis could be tested and the metallicities refined
by obtaining high angular resolution \ion{H}{1} 
data in conjunction with suitable STIS observations of the three
sight lines.

\section{NEW HIGH VELOCITY CLOUDS IDENTIFIED IN \ion{O}{6}} 

We observe high velocity \ion{O}{6} along many sight lines for which 
there are no previous detections of high velocity \ion{H}{1} 21\,cm
emission.  This suggests that much of the high velocity gas in these
directions is highly ionized.  
We discuss some of these features in this section.

\subsection{High Positive Velocity \ion{O}{6} ``Wings''}

Many of the sight lines in our sample exhibit broad, weak
\ion{O}{6} absorption features that extend asymmetrically 
from the Galactic thick disk/halo absorption velocities 
out to high positive velocities. Examples are shown in Figure~16,
where we have plotted the continuum-normalized  \ion{O}{6} $\lambda1031.926$ 
absorption profiles for 22
sight lines as a function of LSR velocity.  
In a few cases, the extensions (or wings)
comprise a significant fraction of the total 
\ion{O}{6} absorption along the sight line, but in most cases the positive
velocity wings extending beyond $v_{LSR} \approx +100$ \kms\ are much weaker 
than the thick disk/halo absorption along the same sight lines.  Two general
groupings of clouds listed in Table~1 other than the 
Magellanic Stream display high positive velocities.  These are the 
extreme positive velocity clouds in the north Galactic octant between
$180\degr < l < 360\degr$ (designated {\tt EPn} in Table~1), which have 
properties similar to those of identified 
\ion{H}{1} HVCs in the same general location (but not necessarily along the 
same sight lines), and the group of clouds designated {\tt Oth} that lie 
predominantly in the northern octant containing Complex~C and that have 
no identifiable counterpart in \ion{H}{1} 21\,cm emission. 

 All of 
the wing features carry a primary designation of {\tt EPn} (11/22 sight lines),
{\tt Oth} (10/22 sight lines) or {\tt MS} (1/22) in Table~1.  These features are 
denoted
with a dagger ($\dagger$) mark attached to their HVC identification. 
It is important to note that not all {\tt EPn} or {\tt Oth} features in 
Table~1 are broad absorption wings.
Eighteen of the 22 sight lines with high
positive velocity wings are located in the northern Galactic sky.  
The four sight lines in the southern sky (e.g.,  Mrk\,509, NGC\,1705,
PKS\,2005-489, PKS0405-12) are widely scattered in longitude and do not
appear to be associated in any obvious way with the 18 northern sight lines.
The absorption
wings can be seen quite clearly in the positive velocity channel maps
shown by Wakker et al. (2002).  We present a summary of the sky positions in
Figure~17, where we indicate column density of the wing features
in the same manner as in the  top panel of Figure~6.

The similarities 
in the general locations of the northern sight lines 
exhibiting absorption wings and in their velocities
suggest that many of the \ion{O}{6}
wing features could have a common origin.  One possibility is that the 
wings trace hot gas expelled from the Galactic disk.  
Sembach et al. (2001b) examined the 
absorption toward 3C\,273 and found that the high positive 
velocity \ion{O}{6} in
that direction has no detectable counterpart in other species, including 
\ion{C}{4}.  They interpreted the high velocity
\ion{O}{6} absorption as an indicator of 
hot gas flowing out of the Galactic disk.  The 3C\,273 sight line passes
through Radio Loops I and IV (Berkhuijsen, Haslam, \& Salter 1971) and 
through a region of 
enhanced soft X-ray emission associated with the North Polar Spur (Snowden
et al. 1995).  The radio loops are filled with X-ray emission and were probably
created by supernovae in the Galactic disk (Iwan 1980;
Snowden et al. 1995).  
The absence of lower ionization absorption and the
presence of 0.25 keV emission suggests that the \ion{O}{6}-bearing gas is 
hot ($T\sim 10^6$\,K).  Not all of the northern sight lines with high positive 
velocity wings lie in the direction of Loop I or the North Polar Spur, so 
it is not possible to associate all of these wings with these two large 
Galactic structures.  However, similar physical
processes may be at work along many
of these sight lines, and the wider-spread existence of the wings beyond the 
confines of Loop I may signal a larger scale production and acceleration
of hot gas in the northern Galactic sky than in the south.

An additional argument in favor of the idea that some of the positive velocity
wings trace gas expelled from the Galactic disk is
that hot gas expelled from the Galactic disk, perhaps as part of a 
``Galactic fountain'' (Shapiro \& Field 1976;
Bregman 1980; Norman \& Ikeuchi 1989), would likely contain \ion{O}{6}.  
Although most Galactic fountain models start with the assumption of a much
hotter flow ($T \sim 10^6$\,K), entrainment of cold gas in the flow or a 
lower temperature fountain would result in lower temperatures compatible with
the production of \ion{O}{6}.  Lower temperature fountain models can explain
the kinematics of some of the intermediate  velocity clouds 
(Houck \& Bregman 1990).
The thick disk and low halo of the Milky Way contain a substantial 
amount of \ion{O}{6}; the typical \ion{O}{6} column density for a sight line 
perpendicular to the Galactic plane is $\log N \sim 14.12$, with a $\sim0.25$
dex enhancement above this value in the northern Galactic hemisphere 
(Savage et al. 2002b).  The enhancement of the thick disk \ion{O}{6} in the 
north may indicate a generally  elevated level of hot gas production 
these latitudes.    For a few sight lines, like  ESO\,572-G34,
Mrk\,734, PG\,1001+291, or 
IRAS\,F1143-1810, the distinction between displaced thick disk/halo 
absorption and high velocity absorption is particularly nebulous.

Alternatively, some of the broad \ion{O}{6}
absorption wings may be entirely unrelated to 
the thick disk of the Milky Way.  Some of the gas could be tidal debris 
remaining from encounters of the Milky Way and the Magellanic Clouds or other 
small galaxies. Most of the wing sight lines lie near the orbit of the 
Magellanic system.  Another possibility might be that the wings are the 
positive velocity counterparts of the negative velocity {\tt LG} features 
in the southern Galactic sky.  The primary drawback to both 
of these hypotheses is that absorption wings are not discrete features
like those associated with the Magellanic Stream ({\tt MS}, {\tt MSe}) or Local Group
({\tt LG}) features in the south, and at least qualitatively, they appear to be 
a continuation of the thick disk/halo  velocity distribution.  
Additional observations of other species at 
these velocities, especially if abundances could be derived, would be 
particularly valuable in discriminating between the possible origins of the 
wings. 

Finally, we call attention to the Mrk\,509 sight line, which is one of the 
southern directions in the sample with a 
positive velocity wing.  Unlike the 3C\,273
sight line, this wing is detected in other species, including \ion{Si}{2},
\ion{Si}{4}, \ion{C}{4}, and \ion{N}{5} (see Sembach et al. 1999). In this 
case, the positive velocity wing is likely related to gas in the low
halo of the Galaxy.   The sight line passes within 5 kpc of the Galactic 
center, and several
other sight lines to halo stars near the Galactic center show similar 
features in \ion{C}{4} and \ion{Si}{4} (e.g., Savage, Massa, \& Sembach 1990;
Sembach, Savage, \& Massa 1991; Sembach, Savage, \& Lu 1995).

\subsection{High Velocity \ion{O}{6} in the Local Group?}

A subset of features listed in Table~1 have a category of {\tt LG} (Local
Group) assigned to them.  This label is suggestive of an origin and location
outside the Milky Way and was chosen because the absorption has some unusual
properties.  For most of these features, there is no obvious \ion{H}{1}
21\,cm emission at similar velocities directly along the sight lines. 
The most extensively
studied examples include the cloud toward Ton\,S210 (see \S6.5.2) and the
high negative velocity clouds toward Mrk\,509, which have 
ionization properties more indicative of clouds bathed in the extragalactic 
ionizing background field than of clouds in the halo of the Galaxy 
(see \S6.5.1).   All of the {\tt LG} clouds are located in the southern Galactic 
hemisphere, mostly in the $35\degr < l < 140\degr$ longitude range.  This is
also the same general area of sky containing many of the {\tt MSe}
sight lines.

Various authors have considered the possibility that some of the 
clouds in this general region of the sky (Local Group barycenter)
and in the opposite direction (Local Group anti-barycenter) belong to 
a similar population of clouds located outside the Milky Way (e.g., Blitz
et al. 1999).  
If we consider various standards of rest for the subsample of objects 
having {\tt LG}, {\tt MSe}, or {\tt EPn} classifications in Table~1, we find the 
average velocities and dispersions listed in Table~10.  The dispersion 
about the mean centroid velocity, $\sigma_{\langle \bar{v} \rangle}$, 
is a factor of two smaller
 in the LGSR reference frame than in the GSR reference frame, and a factor of
4 smaller than in the LSR reference frame.  A similar result can be seen
in Figure~7 for the entire ensemble of high velocity features in our 
sample, but restricting the sample to exclude points associated with 
known nearby features (e.g., Complexes A and C, the Outer Arm, the 
Magellanic Stream) results in a clearer difference between the rest frames.  

While
such a reduction is a necessary consequence for a distributed population
of Local Group clouds, it is not a sufficient requirement.   
The reduction
indicates that the clouds are not located nearby (i.e., $d < 1$ kpc),
but it cannot be used as the sole justification for a Local Group location
for several reasons.  A sample of clouds located within $\sim100$ kpc
of the Milky Way may have a significantly 
smaller dispersion in the GSR than in the LSR,
but the difference between the GSR and LGSR frames would be difficult to 
ascertain without a very large sample because of selection biases in the 
directions chosen for study.  An equally important bias is the 
strong selection against clouds with $|v_{LSR}| \lesssim  100$ \kms\ in
our sample, and thus 
the omission of clouds with both large and small values of $|v_{GSR}|$ 
and $|v_{LGSR}|$.  For example, a cloud with $v_{LSR} = 50$ \kms\ in the 
direction $b = 45\degr$, $l=(0\degr, 30\degr, 60\degr, 90\degr, 120\degr, 
150\degr,180\degr)$ has $v_{GSR} = (50, 128, 185, 206, 185, 128, 50)$ \kms\ 
and $v_{LGSR} = (-19, 79, 163, 209, 206, 155, 69)$ \kms.  Therefore, this 
sampling bias can cause the derived GSR and LGSR velocity distributions to be 
narrower (or perhaps even broader) than the actual distributions, 
depending on the velocities of the positions of the clouds omitted 
in the LSR velocity range $|v_{LSR}| \lesssim 100$ \kms.

\section{PRODUCTION OF \ion{O}{6}}

Conversion of \ion{O}{5} into \ion{O}{6} requires an energy of 114 eV,
which in principle can be supplied by either absorption of photons or by
collisions with other species (primarily electrons). We favor collisional
ionization as the primary production mechanism for most of the high 
velocity \ion{O}{6}
observed.  We discuss the reasons for this preference after considering
the collisional ionization and photoionization scenarios described below.

\subsection{Collisional Ionization in Hot Gas}
\subsubsection{General Considerations and Possible Relationship to X-ray 
Absorption Lines}

In collisional ionization equilibrium, the ionization fractions for 
each element depend on the abundances and temperature of the gas.  
For a plasma with solar abundances, \ion{O}{6} has a peak ionization
fraction f$_{\rm O\,VI}$ = (\ion{O}{6}/O) $\sim 0.22$ at $T \sim 2.8\times10^5$~K
(Sutherland \& Dopita 1993).  Although this is often the canonical
temperature quoted when referring to collisionally ionized \ion{O}{6},
the abundance of \ion{O}{6} remains high (f$_{\rm O\,VI} > 0.05$) at 
temperatures of $(2-4)\times10^5$~K.  In non-equilibrium, time-dependent
cooling situations, the temperature of peak \ion{O}{6} ionization fraction 
does not change significantly, but at lower temperatures the ionization 
fractions are larger than in the equilibrium case 
(see Shapiro \& Moore 1976; Sutherland \& Dopita 1993).  For example,
Shapiro \& Moore (1976) find that f$_{\rm O\,VI}> 0.03$ down to 
$T \sim 2\times10^4$~K for a gas cooling isochorically from an initial 
temperature of $10^6$~K.  The peak ionization fraction diminishes to account 
for this increase in f$_{\rm O\,VI}$ at lower temperatures, but f$_{\rm O\,VI}$ 
 at $T > T_{\rm peak}$ remains similar to its value under 
conditions of collisional ionization equilibrium.  

We compare the observed column densities of \ion{O}{6} and other well-observed
high ions to the column density predictions of various collisional 
ionization models in the next section.
Here, we note that recent X-ray absorption-line observations with the 
{\it Chandra X-ray Observatory} suggest that the zero-redshift
column densities of higher ionization species, such as \ion{O}{7} 
(and perhaps \ion{O}{8}), may be very large: $\log N$(\ion{O}{7}) 
$\gtrsim16$ (Fang \& Sembach 2002; Nicastro et al. 2002).  In collisional
ionization equilibrium at $\log T$(K)$  = (6.0,~6.2,~6.4)$, 
$N$(\ion{O}{7})/$N$(\ion{O}{6}) = (255,~233,~194), respectively 
(Sutherland \& Dopita 1993).  For $\log N$(\ion{O}{7}) = 16, this implies 
$\log N$(\ion{O}{6}) $\sim 13.6-13.7$ if the gas has $T \sim 10^6$~K.
Thus, the amount of \ion{O}{6} contained in the hot gas traced by the 
X-ray absorption could be roughly
comparable to the amounts of \ion{O}{6} 
found for the typical thick disk/halo absorption
or the high velocity gas. Even if the gas is much hotter, 
say $\sim10^7$~K, the amount of \ion{O}{6} predicted for such a large 
\ion{O}{7} column density [$\log N$(\ion{O}{6}) $\sim 13.3$]
would still be detectable with $FUSE$.  However, it is highly unlikely that 
all of the observed \ion{O}{6} in either the Galactic disk/halo or 
the high velocity 
gas is related exclusively to the hotter X-ray gas since lower ionization
species (e.g., \ion{H}{1}, \ion{C}{2}, \ion{C}{4}) are seen at similar
velocities and have miniscule ionization fractions in gas at temperatures
exceeding $10^6$~K.  It remains to be seen whether the velocity structure
of the higher temperature (\ion{O}{7}) gas is 
more consistent with disk/halo absorption or high velocity cloud 
absorption since the 
velocity resolution of the X-ray data is currently too poor
to discriminate between local Milky Way features and a more pervasive
Local Group medium. Comparisons of X-ray emission and absorption-line 
strengths along individual sight lines may help to discriminate between
these possibilities.

\subsubsection{Insights from Lower Ionization Stages}

Additional information about absorption by high ionization species is 
available for several of the sight lines in our sample.   We list the values 
of $N$(\ion{O}{6})/$N$(\ion{C}{4}) and $N$(\ion{O}{6})/$N$(\ion{N}{5})
for several of the high velocity \ion{O}{6} features in Table~11,
including the Magellanic Stream, Complex~C, and several Local Group clouds.  
The ratios are integrations over the velocity ranges covered by the high 
velocity \ion{O}{6}
features and thus are averages for the cases considered.   We also list
the ratios predicted by several by collisional ionization models and
the photoionization model discussed below.  The observed ratios
of \ion{O}{6}/\ion{C}{4} and \ion{O}{6}/\ion{N}{5} are typically greater
than unity, with the possible exception of the highly ionized clouds
toward Mrk\,509.

The  collisional ionization models considered in Table~11 include 
radiative cooling of $10^6$\,K gas (as in a Galactic fountain flow), thermal
conduction in the presence of magnetic fields, turbulent mixing of hot
($T\sim10^6$\,K) and warm ($T\sim10^4$\,K) gases, and time-averaged 
cooling of shock-heated gas in evolved supernova remnants.  The references 
and key assumptions
for each of the models are provided in the notes at the end of the table.  
Spitzer (1996)  summarizes these models and provides a brief discussion of the
ratios for the Galactic disk and halo.  Comparisons based on more
extensive information for the Galactic halo have been given elsewhere
(Sembach, Savage, \& Tripp. 1997; Savage, Sembach, \& Lu 1997, Savage et al.
2002b).  In addition
to the specific models listed in Table~11, we have also tabulated the 
predicted ratios if the gas is in collisional ionization equilibrium
using the ionization fractions calculated by Sutherland \& Dopita (1993)
and the solar abundances recommended by Holweger (2001); the
detailed dependences of the high ion ratios on temperature are shown
in Figure~18.  Incorporation of C, N, and O into dust should not affect 
these ratios significantly since all three elements are found predominantly
in the gas phase, but the ratios may be affected if there is a non-solar
relative abundance pattern (e.g., if N/O is sub-solar, as may be
the case for Complex~C; Richter et al. 2001).

\subsection{Photoionization}

\subsubsection{Extragalactic Radiation}
There is a considerable amount of information in the literature describing the 
conditions under which photoionization is relevant for highly ionized 
species such as \ion{O}{6}.  Here, we consider the simple case of a 
plane-parallel slab of gas  bathed in the extragalactic background 
produced by the integrated light of QSOs and AGNs.  Since all of the 
absorption features studied here occur within a few hundred \kms\
of the systemic velocity of the Milky Way, we consider only the ionization
of gas at zero redshift.  We used the ionization code {\tt CLOUDY} (v94.00;
Ferland 1996) to compute the ionization fraction of \ion{O}{6} 
(f$_{\rm O\,VI}$) in cases where the gas is optically thin to ionizing 
radiation ($\tau_{HI} < 1$; $\log N$(\ion{H}{1}) $ < 17.2$).  
We assumed the QSO spectral energy 
distribution given by Madau (1992)
normalized to a mean intensity at the Lyman limit 
J$_{\nu_0}$ = 1$\times$10$^{-23}$ erg~cm$^{-2}$~s$^{-1}$~Hz$^{-1}$~sr$^{-1}$
(Donahue, Aldering, \& Stocke 1995; Haardt \& Madau 1996; Shull et al. 1999). 
This procedure followed that outlined by Sembach et al. (1999)  

We summarize the results of these calculations in Figures~19 and 20, 
where we plot
f$_{\rm O\,VI}$ and $N$(\ion{O}{6}) as functions of ionization parameter
[ $U =(n_\gamma/n_H) \propto (J_\nu/n_H)$ ] for a 
metallicity $\log Z = -0.5$ and neutral hydrogen column densities
$\log N$(\ion{H}{1})$=14$ to 17.  The values 
of f$_{\rm O\,VI}$ do not depend strongly on the metallicity -- the curves
shown are similar to those for $\log Z$= --2~to~0.  
These calculations 
show that f$_{\rm O\,VI} < 0.25$ at most values of $U$, with the highest
values occurring for values of $\log U > -1.5$ .  For $\log U \lesssim -2$ 
(i.e., $n_H > 3\times10^{-5}$ cm$^{-2}$), f$_{\rm O\,VI} \lesssim 0.01$.  The values of 
f$_{\rm O\,VI}$ do not depend strongly on $N$(\ion{H}{1}) for 
$N$(\ion{H}{1}) $<10^{17}$ cm$^{-2}$.

In addition to the flux from AGNs and QSOs, the local extragalactic 
ionizing radiation field 
includes contributions from the integrated light of starburst galaxies
 and nearby galaxies (see Shull et al. 1999).
If the high velocity clouds are located within $\sim100$ kiloparsecs of 
the Milky Way, the detailed ionization properties of the clouds should 
depend primarily upon the escape
of ionizing photons from the Milky Way and LMC (Bland-Hawthorn \& 
Maloney 1999).  However, starlight should not affect  f$_{\rm O\,VI}$ or 
the observed \ion{O}{6} column densities.  Inclusion of an additional stellar 
spectrum in our photoionization models, like that produced by early-type OB 
stars, does not appreciably alter the predicted ionization fraction of 
\ion{O}{6}.  Furthermore, despite
intensive searches for stars associated with HVCs and the Magellanic 
Stream, none have yet been 
found (see Irwin, Demers, \& Kunkel 1990; Simon \& Blitz 2002), which 
indicates that there are few, if any,
{\it in~situ} stellar sources of ionizing photons.  The Magellanic
Bridge, a much younger portion of the Magellanic system than the Stream,
contains hot stars (Irwin et al. 1990) but is generally considered 
to be more closely associated with the Magellanic Clouds than with
the high velocity cloud system surrounding the Milky Way.  None of the sight 
lines in this \ion{O}{6} survey pass through the Magellanic Bridge
(but see Lehner 2002 for an example).

In the models shown in Figure~20, $N$(\ion{O}{6}) increases
with increasing hydrogen column density and scales approximately linearly
with metallicity.  The cloud size also scales roughly linearly with the 
total \ion{H}{1} column density and metallicity over the column density 
ranges considered.  A few simple scaling laws apply for fixed values of $U$:

\begin{center}
$N$(\ion{O}{6}) $\propto Z \times N$(\ion{H}{1}) \\
$n_{HI} \propto n_H \times (n_H/J_\nu)$ \\
$D = n_{HI}^{-1} \times N$(\ion{H}{1}) \\
\end{center}

\noindent
At a given ionization parameter, smaller sizes occur for clouds with lower
\ion{H}{1} column densities or higher metallicities.  To produce 
an \ion{O}{6} column density of $10^{13}$
cm$^{-2}$ requires a cloud size $D \gtrsim 40$ kpc for 
$n_H \ge 10^{-5}$ cm$^{-3}$, $\log N$(\ion{H}{1}) $> 14$, and 
$\log Z = -0.5$.  To produce 
an \ion{O}{6} column density of $10^{14}$ cm$^{-2}$ 
requires cloud sizes in excess
of several hundred kiloparsecs, even if the metallicity is as high as solar.
The assumption of a uniform, constant density cloud
may break down as the cloud size increases beyond these sizes.  
Cloud sizes  as large as these may be possible for pervasive media
outside galaxies (e.g., a Local Group medium) but are clearly too
large to be contained within the Milky Way).

We show the detailed
behavior of the \ion{O}{6}/\ion{C}{4}
and \ion{O}{6}/\ion{N}{5} ratios expected for the photoionization model in 
Figure~21, assuming relative abundances of C, N, and O equivalent to their 
solar proportions.  The ratios depend on the 
ionization parameter (or gas density 
for a fixed ionizing spectrum) and the \ion{H}{1} column density, so we show
ratios for values of $\log N$(\ion{H}{1}) = 14 and 17. The ratios are 
not  sensitive to metallicity in the range $Z=0.01-1.0$; the 
metallicity dependence is weaker than the \ion{H}{1} dependence.  The 
expected column density of \ion{O}{6} is greater than that of \ion{C}{4}
only for $U \gtrsim -1.5$, which corresponds to a density of 
$\lesssim10^{-5}$ cm$^{-3}$.  In Table~11 we list the high ion ratios 
for densities of $10^{-5}$ cm$^{-3}$ and $10^{-4}$ cm$^{-3}$.  In a few 
cases (e.g., Mrk~509, PKS~2155-304), the photoionization models can 
reproduce the observed ratios of \ion{O}{6}/\ion{C}{4} and 
\ion{O}{6}/\ion{N}{5} if the \ion{H}{1} column density is very low
[$N$(\ion{H}{1})~$\sim10^{14}$ cm$^{-2}$] and the gas density is low 
($n_H \lesssim 10^{-4}$ cm$^{-3}$).  However, the total observed
value of $N$(\ion{O}{6})
is difficult to reproduce unless the clouds are very large.

\subsubsection{Galactic Radiation}
Much of the discussion about photoionization in the preceding section is
relevant for discussions of photoionization by Galactic sources.  In both
the extragalactic and Galactic cases, 
the ionization parameter must be high for a substantial amount of 
\ion{O}{6} to be produced.  The production of \ion{O}{6} by ionizing
Galactic sources is restricted to special situations in which the 
interstellar gas has a very low density or lies in close proximity to the 
ionizing source.  The 
dilute ultraviolet radiation that leaks out of \ion{H}{2} regions and 
maintains the ionization of the 
warm ionized interstellar medium (WIM: $n_e \sim 0.08$ cm$^{-3}$,
$h_{\rm z}({\rm WIM}) \sim 1$ kpc; see Miller \& Cox 1993 and Reynolds 1993) 
is insufficient
to ionize \ion{O}{6} since the OB stars responsible for the WIM have 
strong \ion{He}{2} absorption edges at 54~eV.  

Several authors have considered the production of highly ionized atoms
by radiation from hot degenerate objects, such as He-poor white dwarf stars.
Dupree \& Raymond (1983) calculated the \ion{O}{6} column densities 
expected in the Str\"omgren spheres surrounding pure hydrogen white dwarfs
with surface temperatures $T \sim 60,000$\,K.  They found that the column
density of \ion{O}{6} depends upon both the size of the \ion{O}{6}
Str\"omgren region and the ambient ISM density; for ambient densities
of 0.01 and 0.1  cm$^{-2}$, the radii of the transition regions where the 
\ion{H}{1} fraction is 10\% are 6.8 and 2.1 pc, and the \ion{O}{6}
columns are $1.4\times10^{12}$ and $5.6\times10^{12}$, respectively.  
(The \ion{O}{6} Str\"omgren spheres are much smaller than the traditional
\ion{H}{2} Str\"omgren spheres.)
Thus, the \ion{O}{6} produced in such a region is both confined and 
weak compared to the observed \ion{O}{6} in the high velocity
gas.  Over a typical 1 kpc path through a 0.01 cm$^{-3}$ ISM, Dupree 
\& Raymond (1983) predict a total \ion{O}{6} column density of 
$(0.5-1.4)\times10^{13}$ cm$^{-2}$ might be intercepted by a sight line
given the (uncertain) space density of He-poor white dwarf stars.
This \ion{O}{6} column density is lower than  observed  for all 
features in Table~1, often by an order of magnitude or more.  Increasing
the path length beyond $\sim1$ kpc through the Galactic halo does not increase 
this prediction significantly since the  space density of white
dwarf stars is presumably highest near the Galactic plane.

Bregman \& Harrington (1986) have also considered the production of 
\ion{O}{6} in the Galactic halo, with a radiation field that includes 
both Galactic sources of radiation (hot stars, planetary nebulae, the 
Galactic soft X-ray background) and a contribution from the 
extragalactic background.  They find that a path length of $\sim10$ kpc through
a medium with $n_H = 0.001$ cm$^{-3}$ 
is needed to produce an \ion{O}{6} column density of $\sim1\times10^{13}$
cm$^{-2}$. In this regime, the conditions required
for the production of \ion{O}{6}
 essentially revert to the situation 
considered in \S9.2.1, only the problem of moving such extensive regions of 
the Galactic halo at high enough velocities to explain the observed 
\ion{O}{6} absorption features becomes untenable, even if the 
entire halo decouples from the underlying disk (see \S10.2).

\subsection{The Line Width - Column Density Relation}

There is an interesting correlation in the column densities and velocity 
widths of \ion{O}{6} absorbers found in different astrophysical environments.
This trend  was first noticed by Heckman et al. (2002) for starburst
systems and some of the Galactic and HVC absorption systems.  
In Figure~22  we plot \ion{O}{6} column density versus line width 
for \ion{O}{6} features in various environments, including 
the Galactic disk and halo, high velocity gas, and the Magellanic
Clouds.  The Galactic disk points are from the 
$Copernicus$ \ion{O}{6} survey (Jenkins 1978a, 1978b), 
which sampled gas within
$\sim1$ kpc of the Sun; only data points having \ion{O}{6}
$\lambda1031.926$ values of 
b = $\sqrt{2\langle (v - \bar{v})^2 \rangle}$ without substantial 
uncertainties due
to HD contamination are illustrated.
The remaining data points are based on $\lambda1031.926$ measurements from 
$FUSE$.  References to the sources for the data points
can be found in the figure caption.  

Production of \ion{O}{6} by photoionization should not produce the 
trend observed in Figure~22. Under conditions appropriate for 
photoionization, the line width should be dominated primarily by 
the temperature of the gas and the velocity separations of the 
different components.  Thus, one could expect to observe 
either very low or very high column densities at widths appropriate 
for photoionized gas (typically, $T \sim 10^4-10^5$\,K, or b $\sim2-10$\kms).
There is no reason to expect that all such components should have a similar 
column density, and therefore the distribution of data points would be 
more scattered than observed.

Collisional processes should produce a good correlation 
between $N$ and b since the column density scales linearly with
gas flow velocities in most collisional ionization situations.
Heckman et al. (2002) have described
these relationships and extend the result seen in Figure~22 to 
higher column density absorbers seen in starburst galaxies.  They also argue
that the \ion{O}{6} absorbers detected in the low-redshift IGM are 
collisionally ionized, with the exception of a few discrepant systems.  
Since most of the \ion{O}{6} IGM absorbers and the high velocity \ion{O}{6}
features occupy similar portions of Figure~22, understanding the ionization
of one type of system in more detail may lead to a better understanding of 
the other.

The data points for the high velocity \ion{O}{6} features follow
the trend defined by the Galactic 
disk and halo points in Figure~22.  The formal
slope of the high velocity \ion{O}{6} data points 
($\Delta \log N / \Delta \log $b) is 1.3, which is intermediate to the values
of 0.8 for the thick disk/halo and 1.7 for the nearby disk.  There is some
indication that the data points for the high velocity \ion{O}{6} features 
are slightly more scattered about the best fit line than the disk or 
halo points: $\sigma_{HV} \sim 0.36$ dex compared to $\sigma_{disk} \sim 0.29$
dex and $\sigma_{halo} \sim 0.15$ dex.  This may result from additional 
bulk motions, flows, and turbulence within these clouds compared to clouds
in the ISM of the Galaxy.

\subsection{Ionization Synopsis}
Given the above considerations, we consider it very likely that most of
the high velocity
\ion{O}{6} we observe is created primarily by non-equilibrium
collisional ionization processes.  This
conclusion is supported by several findings.  The amount of 
\ion{O}{6} produced by photoionization is smaller than the typical 
column densities observed unless the gas has a very low density and the 
clouds are extremely large -- so large in fact that the sizes are incompatible
with the distances to some HVCs 
(e.g., the Magellanic Stream and Complex~C).  Second, although 
the observed ionic ratios for some high velocity clouds  can occur through 
photoionization under special 
circumstances, the inferred cloud sizes required are again too large to 
allow a cloud location within the Milky Way - Magellanic Cloud system.
In other cases,
the ionic ratios are more reflective of collisional processes and are
similar to those found for Milky Way halo gas, which is even harder to 
photoionize than extragalactic clouds since the ionization sources (primarily
OB stars) produce a softer radiation field than the extragalactic ultraviolet
background. In the direction
of Mrk\,509, the ratios indicate that both photoionization and collisional
ionization processes may 
contribute since the amount of \ion{C}{4} relative to \ion{O}{6} is 
higher than the standard collisional ionization 
models predict.  The turbulent mixing
layer models can reproduce the observed $N$(\ion{O}{6})/$N$(\ion{C}{4}) 
ratio toward Mrk\,509, but cannot simultaneously
satisfy the constraints of the $N$(\ion{O}{6})/$N$(\ion{N}{5}) ratio.
Finally, the high velocity \ion{O}{6} features show a significant correlation
between \ion{O}{6} line width and column density, which is also observed
for other \ion{O}{6} systems that are predominantly collisionally ionized
(e.g., the Milky Way disk and halo, starburst outflows).

\section{KINEMATICS OF THE HIGH VELOCITY GAS}

\subsection{Expectations for a Corotating Halo}

By definition, the high velocity \ion{O}{6} absorption features considered
in this study generally have $|v_{LSR}| \gtrsim 100$ \kms\ with a few
exceptions.  This velocity cutoff provides isolation of the high velocity
absorption from the absorption produced by the thick disk/halo of the 
Galaxy.  The thick disk/halo \ion{O}{6} absorption is roughly characterized 
by a patchy plane-parallel slab of hot gas with an exponential scale height 
of 2.3
kiloparsecs in the direction perpendicular to the Galactic plane (see Savage 
et al. 2002b).  This scale height is comparable to the scale heights derived 
for other highly ionized species (e.g., \ion{C}{4}, \ion{N}{5}) using more 
limited datasets available with the $HST$
(Savage et al. 1997).  The thick disk 
high ions profiles can be modeled reasonably well 
assuming that the thick disk/halo gas follows essentially the same pattern of 
rotation around the Galactic center as the underlying disk gas for regions
within 1--2 kiloparsecs of the Galactic plane (Lu, Savage, \& Sembach 1994; 
Savage et al. 1997).  In some directions toward the Galactic center, this
coupling may break down at higher altitudes (Sembach et al. 1991, 1995a).  
It is therefore useful to quantify the 
magnitude of the deviations from corotation and to consider whether the 
high velocity \ion{O}{6} features can be described within the context of 
various rotation models for the gaseous Galactic halo.

To test whether the high velocity \ion{O}{6} we observe
can be modeled as a simple 
extension of the thick disk/halo distribution, we have calculated the 
\ion{O}{6} 
column density profiles expected for each sight line in the sample 
assuming that
the \ion{O}{6} has a vertical scale height of 2.3 kpc, a typical velocity
dispersion $\sigma = 42.5$ \kms\ (b = 60~\kms; Savage et al. 2002b), a 
mid-plane density $n_0$(\ion{O}{6})$ = 1.7\times10^{-8}$ cm$^{-3}$
(Jenkins, Bowen, \& Sembach 2002), and a halo rotation speed
identical to the rotation speed of the Galactic disk at the same Galactocentric
radii.  In this calculation we have used
the Galactic rotation curve derived by Clemens (1985) with 
a solar orbital speed of 220 \kms\ around the Galactic center. 
We terminated the calculation for each sight line at a distance of 10
kpc from the Galactic plane ($\approx 4$ scale heights).  
(See Savage et al. (1990) 
for specific details about the computation of the line profiles.)  
After converting each column density profile into a synthetic
absorption profile, we identified 
the extreme positive and negative velocities ($v_{pos}, v_{neg}$) at 
which each model absorption profile recovered to within 5\% of the 
continuum.  We chose the velocity interval between these cutoffs as a 
reasonable velocity range over which one would expect to detect 
weak high velocity
features that are extensions of the simple thick disk/halo distribution,
considering the quality of the $FUSE$ data.
Broadening this interval to accommodate slightly more sensitive measures of 
the absorption does not change our conclusions below appreciably.  Narrowing
it strengthens them.

The sight lines with the most negative predicted velocities are 
ESO\,141-G55 ($l=338.2\degr, b=-26.7\degr, v_{neg}=-169$ \kms),  
PKS\,2005-489 ($l=350.4\degr, b=-32.6\degr, v_{neg}=-164$ \kms), and
3C\,382.0 ($l=61.3\degr, b=17.4\degr, v_{neg}=-163$ \kms).  The sight lines
with the most positive predicted velocities are 
ESO\,265-G23  ($l=285.9\degr, b=16.6\degr, v_{pos}=187$ \kms),
Mrk\,509 ($l=36.0\degr, b=-29.9\degr, v_{pos}=153$ \kms), and 
PKS\,0558-504 ($l=258.0\degr, b=-28.6\degr, v_{pos}=150$ \kms),

We show the results of these calculations in 
Figure~23, where we plot the difference between the predicted cutoffs and 
the observed \ion{O}{6} velocity centroids 
as functions of Galactic longitude and latitude.  The figure is similar
to Figure~8, where we plot $v_{LSR}$ versus $l$ and $b$, except that 
the deviation velocity, 
$\Delta$$v=\bar{v}-v_{pred}$, is a measure of how far beyond the 
expected cutoffs each \ion{O}{6} feature lies.  Here, $v_{pred}$ is 
equal to $v_{neg}$ or $v_{pos}$, depending upon sign of the observed 
velocity $\bar{v}$.  The vertical bar on each 
point represents 
the width ($\pm$b in \kms) of the high velocity feature (see
Table~1) to give a sense of the velocity extent of the absorption.
The symbols are coded according to the classifications in Table~3 and are
the same as in Figure~3 to make comparisons of the figures
easier.
We plot horizontal dashed lines at $\Delta$$v = \pm30$ \kms\ to
illustrate that most of the observed velocities lie well outside those 
expected for corotation even after accounting for a substantial
amount of additional randomness to the velocity distribution of the smooth
distribution of components assumed in the model. 

There are several important things to note in Figure~23.  First,
in the Galactic longitude plot,  there is a distinct trend in the 
distribution of data points with longitude.  At $l<180\degr$, the values of 
$\Delta$$v$ are predominantly negative, whereas at $l>180\degr$ the 
values are predominantly positive.   Second, most of the data points 
deviate by at least 30 \kms\ from the predicted extrema; only 13 of the 85
points have $|\Delta$$v| \le 30$ \kms.  This clearly indicates
that the deviation velocities are not described well by a corotating gas layer.
(This is expected since the high velocity sample was selected in large part 
to avoid velocities associated with the Galactic thick disk and low halo.)
Third, in the Galactic latitude plot, the deviations in
the south ($\sigma_{\Delta v} \sim 113$ \kms) are on average a factor of 
$\sim1.4$ times
larger than those in the north  ($\sigma_{\Delta v} \sim 83$ \kms).  
A summary of the means and dispersions in $\Delta$$v$ for different Galactic 
quadrants is given in Table~12. 

\subsection{Expectations for a Non-Corotating Halo}

To see if the distribution of \ion{O}{6} velocities could be due to 
deviations from corotation in the halo gas layer, we have also calculated 
model \ion{O}{6} 
profiles assuming that the halo gas decouples from the disk gas.
In these models, the halo decoupling begins at an altitude 
z$_i$ and is complete at z$_f$;
the functional form of the decoupling is a linear decrease in rotation 
speed between z$_i$ and z$_f$.    We display the results for the model
with z$_i=1$ kpc and z$_f=3$ kpc in  Figure~24.  
This is equivalent
to having a static, non-rotating halo above z = 3 kpc, with the
predicted velocities beyond this distance simply being the projection of the 
Sun's motion in each direction.

There are several key differences in the velocities predicted by this model 
compared to the corotation model.
The trend in $\Delta$$v$ with Galactic longitude seen in Figure~23
has largely disappeared,
and the differences between north and south have decreased substantially
compared to the corotation model.  
The slight trend for negative values of $\Delta$$v$ in the south and 
positive values of $\Delta$$v$ in the north is weaker than before but
still present.
Inspection of Table~12 shows that the overall dispersion of $\Delta$$v$
for this model is 84 \kms\ compared to 115 \kms\ for the corotation model.

Twenty-three data points in the non-corotation model have 
$|\Delta$$v| \le 30$ \kms, considerably more than for the corotation case.
Still, this is only 23 out of the 85 cataloged high velocity features, 
implying that even this model does not account for most of the velocities.
Of these 23 features with $|\Delta$$v|$ $\le 30$ \kms, 9 are 
classified as {\tt Oth}. The remainder are divided roughly equally 
between the {\tt LG}, {\tt EPn}, and {\tt MSe} categories.
There is still a deviation
well above and beyond that expected, regardless of whether or not the halo
corotates (i.e., the \ion{O}{6} velocities do not obey any simple 
relation with the thick disk velocities.).

The smaller dispersion in the \ion{O}{6} deviation velocities in the
non-corotating model suggests that  much of the high velocity \ion{O}{6}
does not know about the rotation of the Galaxy, but it does not
imply that the Galaxy has 
a static halo of smoothly distributed \ion{O}{6}.
The model assumes a smooth gas distribution, but  the \ion{O}{6} in the 
thick disk/halo is known to be quite patchy in nature, even on small
angular scales (Howk et al. 2002a).  If there were a smooth gas distribution
and the halo were not rotating, we would expect to see absorption profiles
that had secondary absorption minima due to velocity crowding at the terminal 
velocities for some sight lines. We do not see such features.

\subsection{Searching for Outflows or a Hot Galactic Wind}

There is considerable interest in understanding whether galaxies have 
large-scale winds that are capable of altering the structure of the ISM
and expelling heavy elements into the IGM.  In all cases, the outflowing 
winds are expected to be hot, since it is typically supernovae and 
star-formation processes that power them.  As a result, the best evidence
for galactic winds often comes from X-ray observations and from ultraviolet
measurements of highly ionized species such as \ion{C}{4}, \ion{N}{5},
and \ion{O}{6} (see, e.g., Heckman 2002 and Martin 2002 for current
discussions of this topic and additional references).  An \ion{O}{6} 
outflow in the starburst galaxy NGC\,1705 has been observed by $FUSE$
(Heckman et al. 2001), and lower velocity hot gas flows may also be present 
in the LMC (Howk et al. 2002b).

In the Milky Way, there is evidence that a wind emanating from the Galactic 
center has resulted in the expulsion (or ionization) of much of the 
neutral ISM above z~$\sim1$ kpc in the inner 3 kpc of the Galaxy (Lockman 
1984).  The possibility of a strong Galactic wind has been considered
by Hirashita, Kamaya, \& Mineshige (1997), while a review of more 
quiescent wind activity and Galactic ``fountains'' is given by
Breitschwerdt \& Komossa (2000).  The detailed nature of a Galactic wind
depends upon many factors, including the energy input at the base of the wind,
the structure and density of the ISM, the gravitational potential of the 
Galaxy at the site of the wind flow, magnetic fields and cosmic ray 
support of material, and the efficiency of the entrainment of cooler material 
in the flow.  Few of these variables are known in detail, but models 
of winds in rotating galaxies like our own indicate that a Galactic 
wind could be driven
by cosmic rays and heated in part by Alfven waves, with the gas corotating 
with the underlying Galactic disk up to altitudes of several kiloparsecs 
(Zirakashvili et al. 1996).

Consider the  case of a spherically-symmetric unbound
wind that flows completely out of the Galaxy.  The simplest directions
to look for signs of an outflow would be directly toward or directly
away from the Galactic center, since the effects of Galactic rotation
would be minimized.  Toward $l \sim 180\degr$, we see high positive 
velocity \ion{O}{6}, as expected.  Toward $l \sim 0\degr$ we see both
positive and negative velocity features.  However, the number of sight 
lines in both directions is very limited.  At high latitudes 
($|b| \gtrsim 60\degr$), positive wind velocities should dominate, which is 
qualitatively consistent with the \ion{O}{6} velocities in Figure~8.
At lower latitudes, the expected wind velocity depends on Galactic longitude
and the distance of the wind particles from the Sun.  In the 
range $180\degr < l < 270\degr$, positive velocities are expected 
since the gas lies outside the solar circle.  Between $l=270\degr$ and 
$l=360\degr$, positive velocities are expected for the wind beyond
the Galactic center, while predominantly negative velocities are 
expected for gas within the solar
circle on the near side of the Galactic center.

The altitude at which a sight line intercepts the wind 
influences the observed wind velocity; the projected velocity for 
higher-z sight lines would be lower than for sight lines passing nearer
the disk of the Galaxy.  Furthermore, if the source of the wind was not a 
localized phenomenon but was instead spread over the disk of the Galaxy
(say, in spiral arm locations), then the gas motions may be dominated
by regional activity.  Indeed, evidence of localized outflow from the 
Scutum Supershell exists for several ionization stages, including \ion{O}{6}
(Sterling et al. 2002).  Some of \ion{O}{6} features we observe in our survey
(e.g., the positive velocity absorption wings)
are consistent with the presence of localized outflow from the Galactic 
disk.  
We conclude that the complexity of the high velocity \ion{O}{6} sky, 
selection effects, and the unknown extent of a Galactic wind make it
difficult to find conclusive evidence for a large-scale Galactic 
\ion{O}{6} outflow, nor can we rule out its presence without additional
information. 
Future observations targeted at addressing this issue in specific 
directions in both
the Milky Way and the Magellanic Clouds are needed.

\subsection{Expectations for \ion{O}{6} in the Local Group}

Several authors have argued convincingly against a distributed Local 
Group location for many of the \ion{H}{1} HVCs using statistical studies 
of the neutral gas in other galaxy groups (e.g., Charlton, Churchill, \& Rigby 2000; 
Zwaan \& Briggs 2000).  One
way to escape this argument is to require that the clouds be located 
close to the large galaxies within the groups ($d \lesssim100-200$ kpc).
For clouds in the Local Group, this makes it difficult to distinguish between
a cloud population related to the Milky Way and a truly Local Group
population based on kinematical information alone (i.e., the dispersions
in the cloud velocities in the GSR and LGSR reference frames are similar,
and strong observational selection effects apply - see \S8.2).
As a result, additional information is needed to determine if the 
clouds are located in the Local Group.

For the subsample of high velocity {\tt LG}, {\tt EPn}, and {\tt MSe} 
features  considered in Table~10, the dispersions of the velocity 
distributions in the GSR and LGSR reference frames are smaller than the
dispersion in the LSR reference frame.   Furthermore, in  the GSR and LGSR 
reference frames, the dispersions for this subsample are smaller than the 
dispersions listed in Table~6 for the entire high velocity \ion{O}{6} sample 
(Table~6).  The shaded regions of Figure~7 
highlight the velocity distributions of these 50 features.  
Several of the sight lines classified as {\tt LG} have ionization or
abundance information that support an extragalactic location (\S6.5).  
The {\tt  MSe} sight lines are included in this subsample, even though their 
velocities
are similar to those of the Magellanic Stream, since their positions on the sky
do not correspond to the locations of the bulk of the \ion{H}{1} 21\,cm emission
from the Stream.  Objects 
classified as {\tt LG}, {\tt EPn}, or {\tt MSe} in Table~1 would be excellent 
candidates for 
concerted efforts to determine the ionized gas content, 
gas-phase conditions, elemental abundances, and distances of the high velocity gas.

Increasing the number of \ion{O}{6} absorption-line measurements along
other sight lines in the regions of the sky
where  the {\tt LG}, {\tt EPn}, and {\tt MSe} features are found
is highly desirable.  
Unfortunately,
with current instrumentation, the number of additional background sources 
bright enough to observe in reasonable integration times is limited.  An
 alternate approach, that of imaging the \ion{O}{6} in emission at very
faint levels over 
large fields of view, may soon be possible with the NASA's Small Explorer 
mission for {\it Spectroscopy and Photometry of the Intergalactic Medium's 
Diffuse Radiation (SPIDR)}.  Such measurements would help to set the context 
for the absorption-line measurements in this study and perhaps reveal the 
association (if any) of these features with other high velocity gas.

\section{DISCUSSION}

High velocity \ion{O}{6} is both widespread and common 
along complete paths through the Galactic halo.  The high velocity \ion{O}{6} 
traces numerous phenomena, including tidal interactions with the 
Magellanic Clouds (via the Magellanic Stream), accretion of low metallicity
gas (e.g., Complex~C), highly ionized clouds (e.g., the Mrk~509 HVCs),
and the outflow of hot gas from the Galactic disk (e.g., the broad positive
velocity absorption features).  Distinguishing between all of the
possible phenomena occurring at large distances from the Galactic plane is not 
easy, and clearly any attempt to treat all detections of the high velocity
\ion{O}{6} as a single population of clouds would be a serious mistake.  
That being said, considerations of subsets of the high velocity \ion{O}{6} 
features suggest some intriguing possibilities.

Much of the high velocity \ion{O}{6} probably occurs at large distances
from the Galactic plane.  There have not yet been any detections of 
high velocity \ion{O}{6} at high latitudes toward halo stars out to distances
of several kiloparsecs.  The two stars in our sample 
are the most 
distant halo stars examined for high velocity \ion{O}{6} absorption
($d \sim 8-10$ kpc).
In a study of approximately two dozen halo star sight lines, Zsargo et al. 
(2002)
find low velocity \ion{O}{6} absorption associated with the Galactic thick 
disk/halo
but no high velocity \ion{O}{6} absorption at 
$|v_{LSR}| > 100$ \kms.  Despite the difficulties encountered in estimating
the stellar continua, this result is fairly robust, especially the
absence of discrete high velocity  features 
with modest optical depths.  Broad absorption 
wings, like those discussed in \S8.1, would be much more difficult to detect
against stellar 
continua with substantial curvature than against the
relatively  smooth spectra of QSOs and AGNs.  The  number of 
halo sight 
lines toward hot stars examined for high velocity \ion{O}{6} absorption
is small, so the absence of high velocity \ion{O}{6} in halo regions
within a few 
kiloparsecs of the Galactic disk is by no means ruled out.
Some high velocity \ion{O}{6}
is observed toward the Magellanic Clouds
 at velocities intermediate to those expected for the 
Galactic thick disk/halo and the Magellanic Clouds (Hoopes et al. 2002; 
Howk et al. 2002b).  The distance of the high velocity \ion{O}{6} in this
direction is unknown, but is likely to be more than a few kiloparsecs 
from the Galactic plane.

One possible explanation for much of the high velocity \ion{O}{6} is that
it arises at the boundaries between cool/warm 
clouds of gas and a very hot ($T > 10^6$\,K) Galactic corona or Local Group
medium.   A hot, highly extended 
corona much larger than the thick disk/halo region
might be left over from the formation of the 
Milky Way or Local Group, or may be the by-product of continuous accretion of 
smaller galaxies over time.  Interactions of this type have been proposed
to explain the observed H$\alpha$ emission from the Magellanic
Stream (Weiner \& Williams 1996), the shape and confinement 
of some Stream concentrations
(Stanimirovic et al. 2002), and the shapes of supergiant shells along
the outer edge of the LMC (de~Boer et al. 1998).
N-body simulations
of the tidal evolution and structure 
of the Magellanic Stream favor a low-density medium for imparting
weak drag forces to deflect some of the Stream gas and providing a possible
explanation for the absence of stars in the Stream (Gardiner \& Noguchi
1996; Gardiner 1999).  Moore \& Davis (1994) also postulated a hot,
low-density corona to provide ram pressure stripping of some of the Magellanic
Cloud gas and to explain the absence of gas in globular clusters and 
nearby dwarf spheroidal companions to the Milky Way (see also Blitz \&
Robishaw 2000).  Constraints on the density of the Galactic corona
from considerations of the survivability of the Magellanic Stream  range 
from $n \lesssim 10^{-4}$ cm$^{-3}$ based on dynamical arguments 
[e.g., drag (Moore \& Davis 1994)] to $n \lesssim 10^{-5}$ cm$^{-3}$
based on thermal considerations [e.g., heating and evaporation (Murali 2001)].
In either case, there would be little X-ray emission from the extended corona 
or Local Group medium at these 
densities, and no observational constraints on the 
extragalactic X-ray background observed by ROSAT
are violated by the presence of such a medium (see Rasmussen \& Pedersen 2001).

In such a scenario, one would expect the \ion{O}{6} to occur at the 
boundaries between neutral or weakly ionized gas clouds
and the hotter ($T \gtrsim10^6$ K) medium. 
The data for sight lines that pass through Complex~C are consistent with 
the idea that the Complex~C is falling into and interacting with a highly
extended corona.  The similarity in the general velocity structure of the 
\ion{H}{1} and 
\ion{O}{6} indicates that the spatial relationship is probably also close
(see Figure~14 and Table~8).  
The boundary between the \ion{H}{1} and the hot corona is probably thin 
enough that the negative velocity extension of the \ion{O}{6} is essentially 
determined by the kinematics at the boundary (i.e., by the motion of the 
\ion{H}{1}) rather than by larger scale thermal motions in hot gas or boundary
layer. The distance of Complex~C ($d > 5$ kpc, z~$> 3.5$ kpc) is sufficiently
large that any interactions must be occurring well above the Galactic plane,
either with the Galactic corona or with the outer regions of the thick 
disk/halo.
 
Hydrodynamical simulations of clouds moving through a hot, low-density 
medium show that weak bow shocks develop on the leading edges of the 
clouds as the gas is compressed and heated (Quilis \& Moore 2001).  Although
the simulations were motivated in an attempt to explain the 
``head-tail'' structures seen for some HVCs as they pass through the 
lower halo (Br\"uns, et al. 2000; Br\"uns, Kerp, \& Pagels 2001), 
they also provide some guidance for lower density situations.
The exact nature of the interaction depends on the relative speeds and
densities of the two gases.   The high velocity \ion{O}{6} features
generally have velocities comparable to the adiabatic sound speed in a 
medium of the type described above, and both projection effects and 
motion of the medium itself (e.g., a Galactic wind) would serve to increase
the Mach number of the \ion{O}{6} clouds.  However, even if the clouds are 
not moving at 
supersonic speeds relative to the ambient medium, some viscous or turbulent
stripping of the cooler gas occurs (see Quilis \& Moore 2001).  This
type of turbulent mixing of hot and warm gases has been described in detail
by Slavin, Shull, \& Begelman (1993).  Thermal conduction may also be efficient,
depending upon the orientation of the magnetic field relative to the 
conduction front (Borkowski, Balbus, \& Fristrom
1990). Fast ionization of cold gas may also occur through the ``critical
velocity effect'' if the coronal plasma itself
is magnetized (see Konz et al. 2001).  
Thus, the exact details of the mixing and heating
of the gas depend on the geometry of the clouds, the entrainment velocity of 
the gas, the presence of magnetic fields, and several other physical
quantities.

The presence or absence of neutral gas associated with the high velocity
\ion{O}{6} in the interacting corona scenario would depend on several
factors, most notably the mass of the cool/warm gas cloud passing through the
corona and the heating efficiency.  The good correspondence of 
the \ion{O}{6} absorption velocities and the \ion{H}{1} 21\,cm emission 
velocities in the directions of substantial \ion{H}{1} concentrations (e.g.,
Complex~C, the Magellanic Stream - see Figures~14 and 15 and Table~8) argue
that the interface region between the corona and the cooler material must be 
relatively compact compared to the overall size of the cloud.  The larger
line widths seen for \ion{O}{6} are consistent with additional sources of 
turbulence expected for mixing zones in the interface region. 
Photoionization of \ion{O}{6} is very unlikely in such a region since low
gas density and large path lengths are required if the photoionization is due
to the extragalactic background.

Most conductive interface
models and the turbulent mixing layer models predict \ion{O}{6} column 
densities of $\sim10^{13}$ cm$^{-2}$ per conduction front and even less
for mixing layers (if the mixing is slow - i.e., no shocks).  
Magnetic fields parallel to conduction fronts inhibit conduction and result
in lower \ion{O}{6} column densities.  These single-interface
column densities are approximately
an order of magnitude less than those observed (see Table~1).  There are 
several possible explanations for this.  First, there may be small scale 
macro-structure (clumps) within the clouds that have similar
enough velocities that the absorption features from the fragments blend 
together and cannot be distinguished.  The boundary of each fragment would
contribute to the total column density, and multiple ($\gtrsim10$) 
boundaries would be required to explain the observed values of $N$(\ion{O}{6}).
The \ion{O}{6} profiles are generally quite broad and are consistent with the
presence of multiple, blended components.  Second, turbulent mixing in eddy 
layers in the conduction zone increases the likelihood of more efficient 
mixing 
and greater \ion{O}{6} columns.  The hydrodynamical simulations show that
the transition-temperature gas layers where \ion{O}{6} should exist are 
turbulent in nature (see, e.g., Figure~2 in Quilis \& Moore  2001).  Third, there
may be an external source of heat that increases the depth of the 
mixing layer.  The column densities in mixing layers depend on the temperature
of the post-mixed gas and on 
the product of the velocity of the hot gas that enters the mixing layers and 
the fraction of mass deposited by the hot gas in the 
turbulent flow (Slavin et al. 1993).
In the case of a cloud moving through a low density medium, the source of 
heat could be weak shocks or the ram pressure heating itself.

There are numerous examples of stellar tidal streams in the 
Galactic halo (see Majewski, Hawley, \& Munn 1996; Mart\'inez-Delgado et al.
2001; Newberg et al. 2002).
Sources of the high velocity material entering the Galactic corona
might include infalling or tidally 
disturbed galaxies, such as the Sagittarius dwarf (Ibata et al. 2001).  
Tidal interactions of the fragments
and  galaxies in the Local Group could drive dynamical instabilities in
some of the smaller galaxies and strip interstellar material from gas-rich 
irregular galaxies (Mayer et al. 2001).  Since viscous processes
affect the gas but not the stars in interacting systems, tracing the 
original source of the high velocity gas may prove impossible.  However,
the fact that no pristine (zero-metallicity) HVCs have yet been identified
indicates that some previous metal enrichment inside galaxies has probably
occurred (see \S7 and Table~9 for some potentially low metallicity clouds
 that require further study).  In the case of Complex~C, 
the low metallicity 
($Z\sim0.1-0.2$)
indicates a history different from that of the interstellar material in the 
Galactic disk.  

The column density of \ion{O}{6}  within the hot ($ T > 10^6$\,K) 
gas of the corona or 
Local Group gas itself
is given by $N$(\ion{O}{6}) = (O/H)$_\odot$ $Z$ f$_{\rm O\,VI}$ $nL$,
where $L$ is the  path length through the absorbing gas.  
At $T \ge 10^6$\,K, f$_{\rm O\,VI}$ $\le 3.8\times10^{-3}$ 
(Sutherland \& Dopita 1993). For $n
 = 10^{-5}$ cm$^{-3}$,  $N$(\ion{O}{6})\,[cm$^{-2}$] $\le 6.4\times10^{10}
Z~L$\,[kpc].  A hot corona with an extent of 1 Mpc and a metallicity 
$Z=0.1$ would produce an \ion{O}{6} column density of less than $10^{13}$ 
cm$^{-2}$.  Thus, the corona would have an \ion{O}{6} column density 
significantly less than that of the Milky Way thick disk/halo and would 
be difficult to detect directly through its \ion{O}{6} absorption.  However,
if the gas had a higher metallicity (e.g., $Z=0.3$), and a higher density
(e.g., $n = 10^{-4}$ cm$^{-3}$), then the amount of \ion{O}{6} expected
would be substantial [$N$(\ion{O}{6})~$> 3\times10^{13}$ cm$^{-2}$]
even if the corona extended only $\sim 100$ kpc.  Such a corona should 
show X-ray \ion{O}{7} absorption features that may also be detectable 
(see \S9.1).    A possible manifestation of the coronal gas might be very 
broad, shallow absorption wings like those discussed in \S8.1. 
Indeed, preliminary results 
from several authors (Fang \& Sembach 2002; Nicastro et al. 2002) demonstrate
that \ion{O}{7} absorption is present near zero velocity along at least two 
high latitude sight lines, which is consistent with the presence of a 
large, nearby reservoir of hot gas.  

Assuming spherical geometry, a radius $R \ge 70$ kpc, and density
$n_{\rm H} \sim 10^{-5}$, the mass of the hot tenuous gas in the corona
is $M \gtrsim 4\times10^8 M_\odot$.  
If the hot corona is a prolate spheroid
with an axial ratio $a/b$ = 2:1 and a semi-minor axis of $b \ge 70$ kpc, then
$M \gtrsim 10^9 M_\odot$.  If the hot medium fills the Local Group,
the mass would exceed $10^{11} M_\odot$.

An alternative source for the high velocity \ion{O}{6} 
may be that the clouds and any associated \ion{H}{1} fragments are simply 
condensations within
large remnant Local Group gas structures  
falling in on the Galaxy (Oort 1970).
Cosmological structure formation models predict large numbers of cooling 
fragments embedded in dark matter, and some of these structures should be 
observable in X-ray absorption (e.g., Fang et al. 2002) or in
\ion{O}{6} absorption as the gas cools through the 
$T=10^5-10^6$\,K
temperature regime.  Estimates of the number density of these structures 
are consistent with the observed IGM \ion{O}{6} detection rate (Fang \&
Bryan 2001).  This situation is in many ways analogous to the coronal
model developed above because only about 30\% of the hot gas is detectable 
in \ion{O}{6} absorption, while the remaining $\sim70\%$ is too hot to 
observe  (Dav\'e et al. 2001). However, unlike the coronal model in which the 
cooler gas may be tidal fragments interacting with a Galactic corona
or Local Group medium, 
in this scenario the 
\ion{O}{6}-bearing gas and hotter gas share a common evolution within 
isolated evolving structures with no external baryonic medium required
(though dark matter confinement would likely be necessary -- see, e.g.,
 Kepner et al. 1999).

Distinguishing between these two models will be difficult, but there are 
several tests that may yield some additional insight.  Metallicity 
determinations for a substantial number of the high velocity \ion{H}{1} and 
\ion{O}{6} features would help to distinguish between Magellanic debris
and lower metallicity ($Z\lesssim0.1$) gas expected for a cooling gas
structure or debris from smaller Local Group galaxies.  High-resolution 
imaging of the H$\alpha$ emission from the high velocity \ion{O}{6} 
absorbers would constrain their relationship with the \ion{H}{1} and 
perhaps reveal additional signatures of interactions with an extended Galactic 
corona (e.g., bow shock structures or turbulently mixed layers of ionized
gas between the \ion{H}{1} and hot corona).  Some high velocity clouds
have already been detected in H$\alpha$ emission at modest spatial
resolution:  the 
Magellanic Stream (Weiner \& Williams 1996), Complexes~A, C, and
M (Tufte et al. 1998), and several compact \ion{H}{1} HVCs in the 
$100\degr < l < 160\degr$, $-60\degr < b < -30\degr$ region of the sky
(Tufte et al. 2002).  While it is often assumed that all of the H$\alpha$
emission is caused by photoionization, the results of our study indicate
that collisional ionization processes should not be ignored.
$HST$ observations of 
\ion{C}{4} and \ion{N}{5} associated with the high velocity \ion{O}{6} 
would help to constrain the ionization properties of the clouds.  Finally,
imaging of the high velocity gas in \ion{O}{6} and X-ray transitions
at sufficient spectral resolution to isolate individual spectral lines
would provide much needed insight into the structure of the hot gas and its
relationship with cooler material.
This is currently difficult with existing instrumentation but should
be possible with proposed ultraviolet and X-ray observatories (e.g., 
$SPIDR$ and $Constellation-X$).

\section{CONCLUSIONS}
In this paper we report the results of an extensive  $FUSE$ study of
high velocity \ion{O}{6} absorption along complete sight lines through 
the Galactic halo in the directions toward 100 AGNs/QSOs and two 
distant halo stars.  Companion 
studies describe the observations and data (Wakker et al. 2002) and the 
results for \ion{O}{6} absorption
produced by the thick disk and halo of the Milky Way (Savage et al. 2002b).  
In this study, the cutoff
between high velocity gas and Galactic thick disk/halo absorption was generally
near $|v_{LSR}| \sim 100$ \kms.  Our main 
conclusions regarding the high velocity gas are as follows:

\noindent
1) We identify 85 individual 
high velocity \ion{O}{6} features along the 102 sight lines in our sample.
A critical part of this identification process involved detailed consideration
of the absorption produced by \ion{O}{6} and other species 
(primarily H$_2$) in the thick disk and halo of the Galaxy, as well as the 
absorption produced by low-redshift intergalactic absorption lines of 
\ion{H}{1} and ionized metal species.

\noindent 
2)  We searched for absorption in a velocity range of $-1200 < v_{LSR} < +1200$
\kms\ around the \ion{O}{6} $\lambda1031.926$ line.  
With few exceptions, the high velocity \ion{O}{6}
absorption is confined to $|v_{LSR}| < 400$ \kms,  
indicating that the \ion{O}{6} features observed are either associated with 
the Milky Way or nearby clouds within the Local Group.
The 85 high velocity \ion{O}{6} features have velocity centroids ranging 
from $-372 \lesssim \bar{v}_{LSR} \lesssim -90$ \kms\ to 
$+93 \lesssim \bar{v}_{LSR}
\lesssim +385$ \kms. There are an additional 6  
confirmed or very likely ($>90$\% confidence) detections and 2 tentative 
detections of \ion{O}{6} 
between $v_{LSR} = +500$ and +1200 \kms; these very high velocity
features probably trace intergalactic gas beyond the Local Group.

\noindent
3) The 85 high velocity \ion{O}{6} features have logarithmic column densities
of 13.06 to 14.59, with an average of $\langle \log N \rangle = 13.95\pm0.34$ and a median of 13.97.  
The average \ion{O}{6}
column density is a factor of 2.7 times lower than the typical column
density for a sight line through the \ion{O}{6} layer in the thick disk/halo
of the Galaxy.

\noindent
4) The line widths of the 85 high velocity \ion{O}{6} features range from
$\sim 16$ \kms\ to 81\kms, with an average of 
$\langle {\rm b} \rangle = 40\pm14$ \kms.
The lowest values of b are close to the thermal width expected for \ion{O}{6}
at $T \sim 3\times10^5$\,K, while the higher values of b indicate that 
additional non-thermal broadening mechanisms are common.

\noindent
5) We detect high velocity \ion{O}{6} $\lambda1031.926$ absorption 
with integrated (total) values of $W_\lambda \gtrsim 30$ m\AA\ at 
$\ge 3\sigma$ 
confidence along 59 of the 102 sight lines surveyed.  For the highest 
quality sub-sample of the dataset, the high velocity detection frequency 
increases to 22 of 26 sight lines.  Forty of the 59 
sight lines have high velocity \ion{O}{6} $\lambda1031.926$ absorption
with  $W_\lambda > 100$ m\AA, and 27 have $W_\lambda > 150$ m\AA.  
Converting these \ion{O}{6} detections into estimates of $N$(H$^+$) in the 
hot gas indicates
that $\sim60$\% of the sky (and perhaps as much as $\sim85$\%) 
is covered by hot ionized hydrogen at a level of 
$N({\rm H}^+) \gtrsim4\times10^{16}$ cm$^{-2}$ if the high velocity gas has 
a metallicity similar to that of the Magellanic Stream ($Z\sim0.2-0.3$), and  
$\sim30$\% of the sky is covered at a level of 
$N({\rm H}^+)\gtrsim4\times10^{17}$ cm$^{-2}$.  The covering factor of the 
hot, high velocity H$^+$ associated with the  \ion{O}{6}
is similar to that found for high velocity 
\ion{H}{1} 21\,cm emission of warm neutral
gas at a comparable column density level.  

\noindent
6) High velocity 
\ion{O}{6} absorption is observed in almost all cases where high velocity 
\ion{H}{1} 21\,cm emission is observed.  The 6 sight lines for which this 
is not true may contain small-scale structure within the field of view
of the 21\,cm observations that is not traced by the \ion{O}{6} absorption
measures.  In a few cases, ionization effects may also explain
the absence of \ion{O}{6}.  Three of the six cases are good candidates 
for follow-up STIS observations to determine if low metallicity may be the
reason for the absence of \ion{O}{6} absorption.

\noindent
7) Some of the high velocity \ion{O}{6} is associated with well-known 
high velocity structures.  These include the Magellanic Stream,
Complex~A, Complex~C, the Outer Arm, 
and several discrete \ion{H}{1} HVCs.

\noindent
8) Some of the high velocity \ion{O}{6} features have no counterpart 
in \ion{H}{1} 21\,cm emission.  These include discrete high velocity
features as well as broad positive velocity
\ion{O}{6} absorption wings that blend with lower velocity absorption
features produced by the Galactic thick disk/halo. The 
discrete features may typify clouds located in the Local Group.

\noindent
9) The broad, high velocity \ion{O}{6} absorption wings are concentrated
mainly in the northern Galactic hemisphere (18/22 sight lines) 
and may trace either tidal debris or thick disk/halo gas 
that has been accelerated to high velocities by star-formation activity in the 
Galactic disk.  If the latter interpretation is correct, the
 gas may be related to the northern hemisphere 
thick disk/halo \ion{O}{6} enhancement observed by Savage et al. (2002b).

\noindent
10) Most of the high velocity \ion{O}{6} features have velocities incompatible 
with those of Galactic rotation (by definition).  The kinematics of the 
high velocity \ion{O}{6} are not described adequately by models in which the 
Galactic halo decouples from the underlying disk.  There is also no obvious
signature of a hot Galactic wind emanating from the Galactic center, but 
selection
effects preclude a definitive statement about its existence.

\noindent 
11) The dispersion about the mean of the high velocity \ion{O}{6}
centroids  decreases when the velocities are converted from the
LSR to the GSR and LGSR reference frames.   While this 
reduction is expected if the 
\ion{O}{6} is associated with gas in a highly extended Galactic corona or 
in the Local Group,  it does not stand alone as sufficient proof of an
extragalactic location.   The clear 
separation of  the various phenomena  producing high velocity \ion{O}{6}
in and near the Galaxy will require continuing studies of the distances,
kinematics,  elemental abundances,  and physical states of the different 
types of \ion{O}{6} HVCs. 

\noindent
12) We find it unlikely that many of the observed \ion{O}{6} features are 
produced by photoionization, even
if the gas is irradiated by extragalactic ultraviolet background radiation.
Rather, several observational constraints indicate that collisional
ionization in hot ($T\sim10^5-10^6$\,K) 
gas is likely the dominant ionization process for most of 
the \ion{O}{6}.  The constraints include the amount of \ion{O}{6} observed, 
the ratios of \ion{O}{6} column densities to those of other highly
ionized species (\ion{C}{4}, \ion{N}{5}), and the strong correlation between
$N$(\ion{O}{6}) and \ion{O}{6} line width.

\noindent
13) Consideration of the possible sources of collisional ionization 
favors production of some of the high velocity  \ion{O}{6} at the turbulent
boundaries between cool/warm 
clouds of gas and a highly-extended, hot ($T > 10^6$\,K) Galactic corona or 
Local Group.   The corona must have a low density 
($n \lesssim 10^{-4}-10^{-5}$ cm$^{-3}$) and be very large ($R \gtrsim70$ kpc) 
to explain the \ion{O}{6} 
observed in the Magellanic Stream and other putative Local Group clouds.
This corona is much more extensive than the Galactic thick disk/halo region
considered in previous hot gas investigations.

\noindent
14) The existence of a hot, highly extended Galactic corona
or Local Group medium
and the prevalence of high velocity \ion{O}{6} are consistent with predictions
that there should be a considerable amount of hot gas  left over
from the formation of the Local Group.  
Descriptions of galaxy evolution will need to account for highly ionized
gas of the type observed in this study, and future X-ray studies of 
hot gas in the Local Group
 will need to consider  carefully the relationship of the 
X-ray absorption/emission to the complex high velocity 
absorption observed in \ion{O}{6}.

\acknowledgements
This work is based on data obtained for the Guaranteed Time Team by the
NASA-CNES-CSA $FUSE$ mission operated by the Johns Hopkins University.
We thank Chris Howk for assistance with implementation of the {\tt CLOUDY}
ionization code, and members of the $FUSE$ \ion{O}{6} working group 
for encouragement and helpful discussions.
This research has made use of the NASA/IPAC Extragalactic Database (NED),
which is operated by the Jet Propulsion Laboratory,
California Institute of Technology, under contract with the National 
Aeronautics and Space Administration. 
KRS acknowledges financial support through NASA contract NAS5-32985
and Long Term Space Astrophysics grant NAG5-3485.  BPW acknowledges
financial support through NASA grants  NAG5-8967, NAG5-9024, and NAG5-9179.

\newpage
\noindent
\begin{center}
{\bf FIGURE CAPTIONS}  \\
\end{center}
Unfortunately, all figures are in PNG format due to size restrictions
imposed by astro-ph.   These figures look ``best'' displayed on a computer
screeen or printed as gray-scale images.  Higher quality (postscript)
versions are available upon request.

\begin{figure}[h!]
\caption{{\bf(See accompanying PNG file.)}
$FUSE$ LiF1A observations of the 1015--1040\,\AA\ spectral region of 
three AGN/QSOs in the high velocity \ion{O}{6} survey.  The data have 
a velocity resolution of $\sim20$ \kms\ (FWHM) and are binned to $\sim10$ \kms\
($\sim0.033$\,\AA) samples.  Prominent interstellar lines, including the 
two lines of the \ion{O}{6} doublet at 1031.926\,\AA\ and 1037.617\,\AA,
are identified above each spectrum at their rest (laboratory) wavelengths.  
The wavelengths of strong molecular hydrogen lines in the $J=0-4$ rotational 
levels are indicated in the top panel; the H$_2$ lines decrease in strength
from the top to bottom panels. Additional \ion{H}{1} and metal lines from 
intervening intergalactic clouds are present in the middle and bottom panels.
Crossed circles mark the locations of terrestrial airglow lines of \ion{H}{1} 
and \ion{O}{1}.
High velocity \ion{O}{6} $\lambda1031.926$
absorption is present along all three sight lines.}
\end{figure}

\begin{figure}[h!]
\caption{{\bf(See accompanying PNG file.)}
{\it Upper panels:} Normalized intensity profiles for the \ion{O}{6} 
$\lambda\lambda1031.926, 1037.617$ lines toward Mrk\,509, ESO\,572-G34, and
PG\,1116+215.
The \ion{O}{6} $\lambda$1031.926 line is shown at the top of each panel, 
together with a 
model of the H$_2$ (6--0) P(3) $\lambda1031.191$  and (6--0) R(4) 
$\lambda1032.349$ absorption features (thin solid line).  The \ion{O}{6} 
$\lambda$1037.617 
line is shown below the 1031.926\,\AA\ line, together with horizontal bars
indicating the integration ranges of the high velocity absorption. 
The high velocity portions of 
the $\lambda1037.617$ absorption are sometimes blended 
with other Galactic ISM features [\ion{C}{2} $\lambda1036.337$, \ion{C}{2}$^*$
$\lambda1037.012$, H$_2$ (5--0) R(1) $\lambda1037.149$ and P(1) $1038.157$].  
{\it Bottom panels:} Apparent column density profiles
for the \ion{O}{6}
$\lambda$1031.926 absorption (heavy line) and $\lambda$1037.617
 absorption (light line).
The values of $N_a(v)$ for the two lines agree well when a direct comparison
is possible, 
indicating that there are no unresolved saturated structures within the lines 
at these velocities.}

\end{figure}

\begin{figure}[h!]
\caption{{\bf(See accompanying PNG files.)}
({\it Top panel:}) All-sky Hammer-Aitoff projection of the 
102 sight lines in the 
$FUSE$ \ion{O}{6} survey.  In this projection,  the Galactic anti-center 
is at the center of the figure, and Galactic longitude increases to the left.  
The locations and names of the 100 QSOs/AGNs and 2 halo stars are 
indicated. In a few
cases, the names have been abbreviated for clarity.
({\it Bottom panel:}) High velocity \ion{O}{6} identifications.  
The ``type'' of high velocity gas (if any) observed 
along each sight line is coded according to the categories 
described in Table~3.  For some sight lines, multiple high velocity 
features are present.  Plus symbols indicate the positions of the 
 high velocity gas non-detections listed in Table~2.}
\end{figure}

\begin{figure}[ht!]
\caption{{\bf(See accompanying PNG file.)}
Logarithmic \ion{O}{6} column densities [log $N$(\ion{O}{6}) 
(cm$^{-2}$)]
for the Milky Way thick disk/halo (dashed line) and high velocity  
(solid line) absorption features listed in Table~1.  The bin size 
for these distributions is 0.10 dex.}
\end{figure}

\begin{figure}[h!]
\caption{{\bf(See accompanying PNG file.)}
({\it Top panel:})
A comparison of the \ion{O}{6} column densities of the 
thick disk/halo and high velocity absorption along the 59 sight lines
where high velocity \ion{O}{6} absorption is observed.   Open 
circles indicate sight 
lines with $b < 0\degr$; filled circles indicate sight lines with $b > 0\degr$.
 A modest anti-correlation of the two quantities exists (dashed line).  
({\it Bottom panel:}) A comparison of the 
high velocity \ion{O}{6} column densities and sight-line-averaged ROSAT 
0.25~keV X-ray intensities ($I(X)$, in units of $10^{-6}$ cnt~s$^{-1}$~arcmin$^{-2}$). 
A modest anti-correlation of
the two quantities exists (dashed line).}  
\end{figure}

\begin{figure}[h!]
\caption{{\bf(See accompanying PNG files.)}
All-sky Hammer-Aitoff projections of the high velocity \ion{O}{6} 
column densities listed in Table~1.  In the top panel, the logarithmic
column density is coded according to symbol size (see legend), 
with the symbol split in 
in half if two features are present.  In the bottom panel, the column density 
is color coded. Detections are 
plotted as colored circles that are 12\degr\ in radius.  When two features
were detected within 12\degr\ of each other (either along the same sight line 
or along adjoining sight lines), the shaded area size is adjusted accordingly.
The mottled appearance of some regions indicates lower reliability 
measurements. Null detections are indicated as 
either downward-pointing triangles with a size proportional to the upper
limit on the column density
(top panel) or as small  ($2.5\degr$ radius) 
color-coded regions indicating the upper limit
(bottom panel).}
\end{figure}

\begin{figure}[h!]
\caption{{\bf(See accompanying PNG file.)}
Distributions of high velocity \ion{O}{6} profile centroids in the 
Local Standard
of Rest, Galactic Standard of Rest, and the Local Group Standard of Rest
reference frames
for the 85 features listed in Table~1.  The distributions for the subsample
of 50 {\tt LG}, {\tt EPn}, and {\tt MSe} features considered in \S\S8.2 and 10.4
are shown as shaded regions.
By definition,
the absence of features at $-100 \lesssim v_{LSR} \lesssim 100$ \kms\ in
the top panel occurs because of blending with the 
Galactic thick disk/halo \ion{O}{6} distribution.
The bin size for these distributions is 20 \kms.  The 
dashed Gaussian profiles are fits to the distributions containing all the high 
velocity features; in the top panel, the fit is performed only for $|v_{LSR}| \ge
100$ \kms.
The ensemble velocity centroids (~$\langle \bar{v} \rangle$~) and widths
($\sigma$ and FWHM)
are given in the upper left corner for each distribution (see also Table~6). }
\end{figure}

\begin{figure}[h!]
\caption{{\bf(See accompanying PNG file.)}
High velocity \ion{O}{6} centroids in the Local Standard of 
Rest reference frame.  In the top panel, filled circles denote sight lines 
with $b>0\degr$, and open circles denote sight lines with $b<0\degr$.  
In the bottom panel, filled circles denote sight lines 
with $180\degr < l < 360\degr$, and open circles denote sight lines with 
$0\degr < l < 180\degr$.
 The vertical bar on each point is the velocity width
($\pm$b) measured for each feature.  By definition, few points fall in the 
region with $-100 \lesssim \bar{v}_{LSR} \lesssim 100$ \kms.}
\end{figure}

\begin{figure}[h!]
\caption{{\bf(See accompanying PNG file.)}
High velocity \ion{O}{6} centroids in the Galactic Standard of 
Rest reference frame.  Filled circles denote sight lines with $b>0\degr$,
and open circles denote sight lines with $b<0\degr$.  
 The vertical bar on each point is the velocity width
($\pm$b) measured for each feature.  The dashed curves bound the region of the
plot that would be occupied by gas in the Galactic disk ($|b| \le 20\degr$)
with $|\bar{v}_{LSR}| \le 100$ \kms; 
by definition of the sample, few points fall in this region.}
\end{figure}

\begin{figure}[h!]
\caption{{\bf(See accompanying PNG files.)}
All-sky Hammer-Aitoff projections of the \ion{O}{6} velocities 
in the LSR and GSR reference frames.  These maps include features listed in
Table~1 and do not include gas at velocities attributed to the thick
disk/halo 
of the Galaxy. The velocities of the \ion{O}{6} features are color coded
and displayed as filled regions of radius 12$\degr$
for each direction where high velocity gas is detected. 
When two features
were detected within 12\degr\ of each other (either along the same sight line 
or along adjoining sight lines), the shaded area size is adjusted accordingly.
Points with no colors indicate null detections of high velocity \ion{O}{6}.}
\end{figure}

\begin{figure}[h!]
\caption{{\bf(See accompanying PNG files.)}
(a) Hammer-Aitoff projection of the high velocity \ion{H}{1} sky based on 
21\,cm emission measurements (adapted from Wakker et al. 2002).  
The \ion{H}{1} data has a spatial resolution of  approximately
36\arcmin\ and is representative of gas with 
$N$(\ion{H}{1}) $> 2\times10^{18}$ cm$^{-2}$.  Data for $|b| < 20\degr$
have been omitted for clarity.  The positions of the 
high velocity \ion{O}{6} features listed in Table~1 are denoted by the 
large circles,  
with the fill color indicating velocity on the same color scale used for the 
\ion{H}{1} emission.  If more than one high velocity \ion{O}{6}
feature is present, the circle is split and the velocities are color-coded
in each section of the circle.  Open circles indicate null detections 
(Table~2). ``X'' marks indicate the locations of the two stellar sight lines
in the sample. 
(b) Azimuthal equal-area projection 
of the data shown in Figure~11a, looking up toward the north Galactic pole.
(c) Azimuthal equal-area projection of the data shown in Figure~11a, 
looking down toward the south Galactic pole.}
\end{figure}

\begin{figure}[h!]
\caption{{\bf(See accompanying PNG file.)}
\ion{O}{6} profile widths for the Milky Way thick disk/halo
(dashed line) and high velocity  (solid line) absorption features. These
observed widths have not been corrected for the small amount of 
instrumental line broadening; this broadening (b$_{inst} \sim 12-15$ \kms)
is important only for the narrowest lines observed.
The bin size for these distributions 
is 5 \kms.
The arrow indicates the thermal width (b = 17.6 \kms) for a single 
component at $T = 3\times10^5$\,K.}
\end{figure}

\newpage
\clearpage
\begin{figure}[h!]
\caption{{\bf(See accompanying PNG files.)} A Leiden-Dwingeloo survey
 grey-scale map of the high velocity \ion{H}{1} 
21\,cm emission in the region of sky near Complex~C.  The contours denote
an \ion{H}{1} column density of $5\times10^{18}$ cm$^{-2}$.  The top panel 
shows the emission over the velocity range $-200 \le v_{\rm LSR} \le -150$
\kms.  The bottom panel 
shows the emission over the velocity range $-150 \le v_{\rm LSR} \le -100$
\kms.  
 The $FUSE$ sight lines are marked in red. Additional features in this 
region of the sky (Complex~A, IV~Arch, Complex~K) are also indicated in the 
lower panel.}
\end{figure}

\begin{figure}[h!]
\caption{{\bf(See accompanying PNG file.)} \ion{H}{1} 21\,cm emission, \ion{O}{6} $\lambda1031.926$, and 
\ion{C}{2} $\lambda 1036.337$ absorption profiles for nine sight lines
in the general direction of Complex~C.  The NRAO 140-foot \ion{H}{1} 
spectra have a beam size of 21\arcmin\ (FWHM).  The solid line overplotted on the \ion{O}{6}
profile is a model of the interstellar H$_2$ absorption in the (6--0) P(3) and 
R(4) lines.  The horizontal bar below the \ion{O}{6} profile shows the 
integration range for the high velocity \ion{O}{6}.  
\ion{C}{2}$^*$ $\lambda1037.018$ 
absorption is seen at a velocity of $\sim197$ \kms\ relative to the 
\ion{C}{2} line.}
\end{figure}

\begin{figure}[h!]
\caption{{\bf(See accompanying PNG file.)} \ion{H}{1} 21\,cm emission, \ion{O}{6} $\lambda1031.926$, and 
\ion{C}{2} $\lambda 1036.337$ absorption profiles for six directions
identified in Table~1 as Magellanic Stream ({\tt MS/MSe}) sight lines.  
The top set of panels 
shows sight lines containing high negative velocity gas, and the bottom set
of panels shows sight lines containing high positive velocity gas.  
 The  \ion{H}{1} 
spectra come from a variety of sources (Green\,Bank/NRAO, Parkes, Villa-Elisa). 
The values of $T_b$ have been scaled in the plots by the 
factors indicated to show the HVC emission.  
The solid line overplotted on the \ion{O}{6}
profile is a model of the interstellar H$_2$ absorption in the (6--0) P(3) and 
R(4) lines.  The horizontal bar above the \ion{C}{2} profile shows the 
integration range for the high velocity gas.  \ion{C}{2}$^*$ $\lambda1037.018$ 
absorption is seen at a velocity of $\sim197$ \kms\ relative to the 
\ion{C}{2} line.}
\end{figure}

\begin{figure}[h!]
\caption{{\bf(See accompanying PNG file.)} 
Normalized intensity versus LSR velocity for 22 objects in our 
sample
with  high positive velocity \ion{O}{6} absorption wings.  These
components are identified in Table~1 with a dagger ($\dagger$) symbol 
attached to the 
feature identification. Identifications of all high velocity \ion{O}{6}
features are provided above each spectrum.
In many cases, the absorption wings join smoothly with the low velocity 
absorption produced by the Galactic thick disk/halo.   The smooth solid 
curves overplotted on some of the spectra are models of the H$_2$ (6--0)
P(3) and R(4) absorption along the sight lines.
The IGM absorption feature in the 3C\,273 spectrum is \ion{H}{1} Ly$\beta$ 
at a redshift $z\approx0.0053$.}
\end{figure}

\begin{figure}[h!]
\caption{{\bf(See accompanying PNG file.)} 
All-sky distribution of 22 sight lines containing high positive
velocity \ion{O}{6} absorption wings.  This map includes those sight lines
for which a dagger ($\dagger$) symbol is attached to the high velocity feature 
identifications in Table~1.  The symbol size indicates the \ion{O}{6}
column density of the absorption feature. Note that most of these sight lines
are located in the northern Galactic hemisphere.}
\end{figure}

\begin{figure}[h!]
\caption{{\bf(See accompanying PNG file.)} 
Ratios of $N$(\ion{O}{6})/$N$(\ion{C}{4}) and 
$N$(\ion{O}{6})/$N$(\ion{N}{5}) as a function of temperature (in K) 
under conditions of collisional 
ionization equilibrium (Sutherland \& Dopita 1993).  The calculations 
assume intrinsic solar abundance ratios, (O/C) = 1.4 and (O/N) = 6.4
(Holweger 2001), and
identical gas-phase depletions onto dust grains, 
$\delta$(C) = $\delta$(N) = $\delta$(O). Differential depletion of these 
elements onto dust grains will raise or lower the resulting curves by less 
than $\sim0.3$ dex since none of these elements is strongly depleted from
the gas-phase in diffuse cloud environments (see Savage \& Sembach 1996).}
\end{figure}

\begin{figure}[h!]
\caption{{\bf(See accompanying PNG file.)} 
Ionization fractions of \ion{O}{6}, \ion{N}{5}, \ion{C}{4}, 
and \ion{H}{1}  versus ionization parameter, 
$U = \langle n_\gamma/n_H \rangle$, and gas density, $n_H$, 
for  uniform dust-free gas clouds with 1/3 solar metallicity
subjected to the extragalactic ionizing background at $z = 0$.  Models
for two values of total \ion{H}{1} column density are plotted.  The 
models assume an incident AGN/QSO spectrum with an intensity 
$J_{\nu_0} = 10^{-23}$ erg~cm$^{-2}$~s$^{-1}$~Hz$^{-1}$~sr$^{-1}$ at 
the Lyman limit.  The curves shown do not vary strongly as a function of 
metallicity for $Z = 0.01 - 1.0$.}

\end{figure}

\begin{figure}[h!]
\caption{{\bf(See accompanying PNG file.)} 
\ion{O}{6} column density as a function  of ionization 
parameter and gas density for extragalactic clouds similar to those
 considered 
in Figure~19.  $N$(\ion{O}{6}) is shown for four 
values of the assumed \ion{H}{1} column density.  $N$(\ion{O}{6}) 
scales roughly linearly with metallicity for the range of $N$(\ion{H}{1})
values shown.  At a given ionization parameter (or density), the cloud 
size, $D$, increases with the \ion{H}{1} column density; values
of $D$ (in kpc) are given immediately above several of the points.}
\end{figure}

\begin{figure}[h!]
\caption{{\bf(See accompanying PNG file.)} 
Column density ratios  $N$(\ion{O}{6})/$N$(\ion{C}{4})
 and $N$(\ion{O}{6})/$N$(\ion{N}{5})
as a function of ionization 
parameter and gas density for the extragalactic cloud considered 
in Figure~19, assuming solar relative abundances for C, N, and O.  Models
for two values of total \ion{H}{1} column density are plotted. 
The ratios do not change substantially for metallicities in the range 
0.01--1.0 solar. }
\end{figure}

\begin{figure}[h!]
\caption{{\bf(See accompanying PNG file.)} 
\ion{O}{6} column density [$N$(\ion{O}{6}) cm$^{-2}$] versus line width 
[b (\kms)] for a variety of 
\ion{O}{6} absorption systems.  Data points include absorption observed by
$FUSE$ in the Galactic halo (Savage et al. 2002), high velocity clouds
(this paper), the SMC (Hoopes et al. 2002), and the LMC (Howk et al. 2002).
Data points for absorption in the Galactic disk are from the $Copernicus$
\ion{O}{6} survey (Jenkins 1978a, 1978b). 
The straight diagonal line indicates the approximate detection limit 
for $FUSE$ data if $W_{1032} = 30$ m\,\AA.
Note the good correlation between $N$ and b.  The thermal widths 
of single components at various temperatures are indicated at the top of 
the figure.}
\end{figure}

\begin{figure}[h!]
\caption{{\bf(See accompanying PNG file.)} 
Deviations of the \ion{O}{6} velocities from those predicted by
the corotating halo model described in the text versus Galactic longitude
(upper panel) and latitude (lower panel).  The deviation velocity,
$\Delta v = \bar{v} - v_{\rm pred}$, is defined
as the difference between the measured 
velocity centroid of the high velocity \ion{O}{6}
feature and the velocity extremum of the $\lambda1031.926$ 
absorption expected for a smooth, corotating, exponentially-stratified 
\ion{O}{6} gas layer with a scale height of 2.3 kpc, and an intrinsic gas
velocity dispersion (b = $\sqrt{2}\sigma$) of 60~\kms. The different symbols
denote the various high velocity feature designations according to the 
code given to the right of the lower panel.}
\end{figure}

\begin{figure}[h!]
\caption{{\bf(See accompanying PNG file.)} 
Same as Figure~23, except for the non-corotating gas layer described
in the text.}
\end{figure}

\clearpage
\newpage


\begin{deluxetable}{lrrclrrcccr}
\tabletypesize{\small}
\tablewidth{510pt} 
\tablecaption{High Velocity O\,{\sc vi} Detections}
\tablehead{Name & \multicolumn{1}{c}{\small $l$} 
                & \multicolumn{1}{c}{\small $b$} 
		& Q\tablenotemark{a} 
                & \multicolumn{1}{c}{\small ID\tablenotemark{b}}
                & \multicolumn{2}{c}{\small $v_{min,max}$} 
                & \multicolumn{1}{c}{\small $\bar{v}$} 
                & \multicolumn{1}{c}{\small b\tablenotemark{c}} 
                & \multicolumn{1}{c}{\small $\log N \pm \sigma_{sc} \pm \sigma_{sys}$ }
                & \multicolumn{1}{c}{\small $\frac{W_\lambda}{\sigma_{W}}$ } \\
		& \multicolumn{1}{c}{\small (\degr)} 
		& \multicolumn{1}{c}{\small (\degr)}
		& &
		& \multicolumn{2}{c}{\footnotesize (\kms)}
		& \multicolumn{1}{c}{\footnotesize (\kms)}
		& \multicolumn{1}{c}{\footnotesize (\kms)}
}
\startdata
PKS\,2155--304 &  17.73 & $-52.25$ &  4 & LG & $-180$ & $ -85$ & --129 & 35$\pm$3 & 13.80$\pm$0.03$\pm$0.11 & 17.5 \\
 &  &  &  & LG & $-280$ & $-180$ & --232 & 42$\pm$4 & 13.57$\pm$0.05$\pm$0.14 &  8.6 \\
Mrk\,509 &  35.97 & $-29.86$ &  4 & LG & $-345$ & $-180$ & --247 & (57$\pm$4) & 14.24$\pm$0.02$\pm$0.08 & 25.0 \\
 &  &  &  & LG & $-180$ & $-100$ & --143 & ((31$\pm$3)) & 13.76$\pm$0.04$\pm$0.15 & 12.6 \\
 &  &  &  & Oth$^\dagger$ & $ 115$ & $ 200$ & \phn152 & ((34$\pm$4)) & 13.55$\pm$0.06$\pm$0.22 &  7.0 \\
MRC\,2251--178 &  46.20 & $-61.33$ &  3 & LG & $-340$ & $-180$ & --258 & 65$\pm$4 & 13.91$\pm$0.11$\pm$0.05 &  3.7 \\
 &  &  &  & LG & $-145$ & $ -65$ & \phn--95 & 31$\pm$4 & 14.06$\pm$0.06$\pm$0.10 &  8.4 \\
PHL\,1811 &  47.46 & $-44.81$ &  3 & LG & $-360$ & $-295$ & --322 & 21$\pm$3 & 13.85$\pm$0.09$\pm$0.09 &  5.2 \\
 &  &  &  & LG & $-200$ & $ -65$ & --142 & (42$\pm$3) & 14.38$\pm$0.04$\pm$0.08 & 13.0 \\
Mrk\,506 &  54.18 & $ 31.50$ &  1 & C & $-180$ & $-100$ & --144 & (33$\pm$4) & 14.05$\pm$0.11$\pm$0.14 &  4.5 \\
Mrk\,478 &  59.24 & $ 65.03$ &  3 & Oth & $ 340$ & $ 435$ & \phn385 & 35$\pm$4 & 13.83$\pm$0.10$\pm$0.05 &  4.4 \\
3C\,382.0 &  61.31 & $ 17.45$ &  1 & OA & $-130$ & $ -50$ & \phn--90 & ((30$\pm$3)) & 14.14$\pm$0.10$\pm$0.09 &  5.1 \\
Mrk\,501 &  63.60 & $ 38.86$ &  2 & C & $-150$ & $-100$ & --122 & ((22$\pm$3)) & 13.81$\pm$0.11$\pm$0.19 &  4.3 \\
Mrk\,1513 &  63.67 & $-29.07$ &  3 & MSe & $-390$ & $-220$ & --293 & 49$\pm$4 & 14.29$\pm$0.05$\pm$0.04 &  9.0 \\
Mrk\,926 &  64.09 & $-58.76$ &  1 & LG & $-220$ & $ -65$ & --125 & (60$\pm$5) & 14.44$\pm$0.12$\pm$0.05 &  4.5 \\
 &  &  &  & MSe & $-395$ & $-220$ & --295 & (57$\pm$5) & 14.41$\pm$0.12$\pm$0.04 &  3.8 \\
Mrk\,304 &  75.99 & $-34.22$ &  3 & MSe & $-355$ & $-240$ & --304 & 40$\pm$3 & 14.47$\pm$0.06$\pm$0.04 & 11.4 \\
NGC\,7469 &  83.10 & $-45.47$ &  4 & MS & $-370$ & $-235$ & --304 & (54$\pm$5) & 14.18$\pm$0.04$\pm$0.11 & 11.9 \\
 &  &  &  & LG & $-235$ & $-120$ & --185 & ((43$\pm$4)) & 14.22$\pm$0.03$\pm$0.16 & 14.2 \\
PG\,1626+554 &  84.51 & $ 42.19$ &  2 & Oth$^\dagger$ & $  85$ & $ 210$ & \phn132 & (48$\pm$4) & 14.08$\pm$0.11$\pm$0.07 &  4.1 \\
 &  &  &  & C & $-260$ & $ -65$ & --152 & ((81$\pm$6)) & 14.35$\pm$0.08$\pm$0.05 &  5.9 \\
NGC\,7714 &  88.22 & $-55.56$ &  2 & MS & $-310$ & $-230$ & --259 & (26$\pm$3) & 14.13$\pm$0.09$\pm$0.09 &  5.6 \\
Mrk\,290 &  91.49 & $ 47.95$ &  1 & C & $-190$ & $ -70$ & --132 & ((52$\pm$5)) & 14.20$\pm$0.09$\pm$0.07 &  4.7 \\
PG\,2349--014 &  91.66 & $-60.36$ &  1 & LG & $-205$ & $ -90$ & --142 & (51$\pm$5) & 14.19$\pm$0.09$\pm$0.09 &  5.5 \\
 &  &  &  & MS & $-415$ & $-270$ & --324 & (43$\pm$8) & 14.00$\pm$0.14$\pm$0.11 &  3.0 \\
UGC\,12163 &  92.14 & $-25.34$ &  2 & MSe & $-345$ & $-210$ & --274 & (41$\pm$5) & 14.16$\pm$0.08$\pm$0.05 &  6.0 \\
H\,1821+643 &  94.00 & $ 27.42$ &  4 & OA & $-160$ & $-100$ & --122 & ((22$\pm$5)) & 13.87$\pm$0.04$\pm$0.18 & 12.0 \\
 &  &  &  & Oth & $-225$ & $-160$ & --192 & (24$\pm$5) & 13.72$\pm$0.05$\pm$0.13 &  8.0 \\
 &  &  &  & Oth & $-285$ & $-235$ & --258 & 16$\pm$5 & 13.24$\pm$0.13$\pm$0.20 &  3.0 \\
Mrk\,876 &  98.27 & $ 40.38$ &  4 & Oth$^\dagger$ & $  85$ & $ 140$ & \phn107 & ((20$\pm$5)) & 13.23$\pm$0.13$\pm$0.31 &  3.0 \\
 &  &  &  & C & $-200$ & $-100$ & --142 & ((38$\pm$5)) & 14.05$\pm$0.04$\pm$0.13 & 14.1 \\
Mrk\,817 & 100.30 & $ 53.48$ &  4 & Oth$^\dagger$ & $  60$ & $ 140$ & \phn+93 & ((35$\pm$5)) & 13.30$\pm$0.08$\pm$0.21 &  4.8 \\
 &  &  &  & C & $-165$ & $ -90$ & --116 & ((27$\pm$4)) & 13.88$\pm$0.03$\pm$0.17 & 19.2 \\
Mrk\,335 & 108.76 & $-41.42$ &  4 & MSe & $-375$ & $-235$ & --305 & (40$\pm$4) & 13.87$\pm$0.05$\pm$0.14 &  9.0 \\
 &  &  &  & LG & $-225$ & $ -75$ & --178 & ((54$\pm$4)) & 14.14$\pm$0.03$\pm$0.10 & 16.9 \\
PG\,1351+640 & 111.89 & $ 52.02$ &  3 & C & $-160$ & $-100$ & --124 & ((25$\pm$3)) & 13.67$\pm$0.08$\pm$0.23 &  4.9 \\
 &  &  &  & Oth$^\dagger$ & $ 100$ & $ 160$ & \phn125 & ((27$\pm$4)) & 13.44$\pm$0.13$\pm$0.22 &  3.1 \\
Mrk\,279 & 115.04 & $ 46.86$ &  4 & C & $-210$ & $-115$ & --154 & (34$\pm$3) & 13.67$\pm$0.05$\pm$0.13 &  8.8 \\
PG\,1259+593 & 120.56 & $ 58.05$ &  4 & Oth$^\dagger$ & $ 100$ & $ 185$ & \phn139 & 34$\pm$7 & 13.06$\pm$0.15$\pm$0.27 &  2.3 \\
 &  &  &  & C & $-150$ & $ -80$ & --110 & 26$\pm$3 & 13.72$\pm$0.04$\pm$0.13 & 14.2 \\
Mrk\,1502 & 123.75 & $-50.18$ &  2 & LG & $-210$ & $ -95$ & --159 & (38$\pm$3) & 14.28$\pm$0.07$\pm$0.06 &  7.0 \\
PG\,0052+251 & 123.91 & $-37.44$ &  2 & LG & $-240$ & $-155$ & --193 & ((21$\pm$3)) & 13.92$\pm$0.11$\pm$0.17 &  3.5 \\
 &  &  &  & MSe & $-415$ & $-270$ & --334 & (59$\pm$4) & 14.17$\pm$0.09$\pm$0.05 &  4.6 \\
Mrk\,352 & 125.03 & $-31.01$ &  2 & MSe & $-365$ & $-250$ & --300 & (42$\pm$5) & 13.96$\pm$0.14$\pm$0.08 &  3.0 \\
 &  &  &  & LG & $-250$ & $-115$ & --185 & ((48$\pm$6)) & 14.14$\pm$0.10$\pm$0.08 &  4.6 \\
Mrk\,357 & 132.20 & $-39.14$ &  2 & LG & $-220$ & $-140$ & --183 & ((34$\pm$4)) & 13.88$\pm$0.11$\pm$0.24 &  3.8 \\
 &  &  &  & MSe & $-340$ & $-220$ & --279 & ((44$\pm$4)) & 14.12$\pm$0.09$\pm$0.16 &  5.5 \\
NGC\,588 & 133.34 & $-31.39$ &  2 & LG\tablenotemark{d} & $-310$ & $-125$ & --212 & 56$\pm$4 & 14.59$\pm$0.05$\pm$0.03 &  9.7 \\
 &  &  &  & MSe & $-420$ & $-310$ & --372 & 43$\pm$4 & 14.08$\pm$0.10$\pm$0.07 &  4.5 \\
NGC\,595 & 133.53 & $-31.31$ &  2 & M33 & $-215$ & $ -80$ & --149 & (46$\pm$4) & 14.35$\pm$0.07$\pm$0.04 &  7.1 \\
 &  &  &  & MSe & $-365$ & $-215$ & --284 & (56$\pm$4) & 14.28$\pm$0.09$\pm$0.04 &  4.9 \\
Ton\,S180 & 139.00 & $-85.07$ &  3 & LG & $-265$ & $-100$ & --156 & 46$\pm$6 & 14.18$\pm$0.05$\pm$0.04 &  8.1 \\
 &  &  &  & Oth & $ 220$ & $ 280$ & \phn251 & 23$\pm$6 & 13.64$\pm$0.09$\pm$0.08 &  4.7 \\
NGC\,3310 & 156.60 & $ 54.06$ &  4 & Oth & $-195$ & $-135$ & --162 & (23$\pm$4) & 13.58$\pm$0.09$\pm$0.19 &  4.7 \\
Mrk\,106 & 161.14 & $ 42.88$ &  1 & A & $-150$ & $-100$ & --125 & ((22$\pm$4)) & 13.81$\pm$0.14$\pm$0.17 &  3.0 \\
PG\,0953+414 & 179.79 & $ 51.71$ &  4 & EPn$^\dagger$ & $ 100$ & $ 235$ & \phn146 & ((53$\pm$4)) & 13.83$\pm$0.06$\pm$0.13 &  6.8 \\
Mrk\,421 & 179.83 & $ 65.03$ &  4 & EPn$^\dagger$ & $ 100$ & $ 185$ & \phn135 & ((34$\pm$4)) & 13.51$\pm$0.11$\pm$0.19 &  3.4 \\
PG\,0947+396 & 182.85 & $ 50.75$ &  2 & EPn$^\dagger$ & $ 100$ & $ 175$ & \phn136 & ((30$\pm$4)) & 14.19$\pm$0.10$\pm$0.11 &  5.6 \\
NGC\,3991 & 185.68 & $ 77.20$ &  1 & EPn & $ 145$ & $ 320$ & \phn229 & (63$\pm$5) & 14.28$\pm$0.13$\pm$0.02 &  3.2 \\
HS\,1102+3441 & 188.56 & $ 66.22$ &  1 & EPn$^\dagger$ & $  95$ & $ 210$ & \phn144 & ((48$\pm$7)) & 14.30$\pm$0.10$\pm$0.07 &  4.3 \\
PG\,0844+349 & 188.56 & $ 37.97$ &  4 & EPn$^\dagger$ & $ 120$ & $ 250$ & \phn179 & (48$\pm$5) & 13.75$\pm$0.08$\pm$0.14 &  4.8 \\
PG\,1001+291 & 200.09 & $ 53.20$ &  1 & EPn$^\dagger$ & $ 100$ & $ 200$ & \phn131 & ((40$\pm$5)) & 13.97$\pm$0.13$\pm$0.16 &  3.1 \\
PKS\,0405--12 & 204.93 & $-41.76$ &  4 & Oth$^\dagger$ & $ 100$ & $ 210$ & \phn143 & (41$\pm$4) & 13.75$\pm$0.07$\pm$0.09 &  6.1 \\
NGC\,4670 & 212.69 & $ 88.63$ &  3 & Oth & $ 320$ & $ 420$ & \phn363 & 29$\pm$4 & 13.96$\pm$0.05$\pm$0.04 &  8.1 \\
PG\,1116+215 & 223.36 & $ 68.21$ &  4 & EPn & $ 230$ & $ 310$ & \phn259 & (28$\pm$4) & 13.28$\pm$0.13$\pm$0.30 &  3.1 \\
 &  &  &  & EPn & $ 115$ & $ 230$ & \phn181 & 42$\pm$4 & 14.12$\pm$0.03$\pm$0.07 & 19.0 \\
Ton\,S210 & 224.97 & $-83.16$ &  4 & LG & $-250$ & $-120$ & --187 & 53$\pm$4 & 13.97$\pm$0.05$\pm$0.11 &  8.3 \\
Mrk\,734 & 244.75 & $ 63.94$ &  2 & EPn$^\dagger$ & $ 140$ & $ 275$ & \phn195 & ((45$\pm$4)) & 14.10$\pm$0.10$\pm$0.16 &  4.1 \\
HE\,0226--4110 & 253.94 & $-65.77$ &  4 & MSe\tablenotemark{e} & $  80$ & $ 260$ & \phn164 & 59$\pm$4 & 13.98$\pm$0.05$\pm$0.03 &  8.8 \\
PKS\,0558--504 & 257.96 & $-28.57$ &  3 & MSe\tablenotemark{e} & $ 210$ & $ 315$ & \phn258 & 36$\pm$5 & 13.68$\pm$0.09$\pm$0.07 &  4.5 \\
NGC\,1705 & 261.08 & $-38.74$ &  4 & MS$^\dagger$ & $ 120$ & $ 245$ & \phn183 & ((53$\pm$5)) & 13.78$\pm$0.04$\pm$0.41 &  6.8 \\
 &  &  &  & MS & $ 245$ & $ 405$ & \phn330 & 55$\pm$4 & 14.31$\pm$0.02$\pm$0.14 & 22.2 \\
HE\,1115--1735 & 273.65 & $ 39.64$ &  1 & EPn$^\dagger$ & $  90$ & $ 175$ & \phn131 & 28$\pm$3 & 14.13$\pm$0.10$\pm$0.06 &  5.0 \\
{\scriptsize IRAS\,F11431--1810} & 281.85 & $ 41.71$ &  3 & EPn$^\dagger$ & $ 210$ & $ 320$ & \phn254 & 37$\pm$5 & 13.85$\pm$0.11$\pm$0.08 &  3.6 \\
 &  &  &  & EPn$^\dagger$ & $ 100$ & $ 210$ & \phn139 & ((41$\pm$4)) & 14.25$\pm$0.06$\pm$0.12 &  8.9 \\
ESO\,265--G23 & 285.91 & $ 16.59$ &  1 & EPn & $ 200$ & $ 310$ & \phn262 & (39$\pm$5) & 14.44$\pm$0.09$\pm$0.07 &  6.1 \\
ESO\,572--G34 & 286.12 & $ 42.12$ &  3 & EPn$^\dagger$ & $ 100$ & $ 275$ & \phn171 & ((65$\pm$4)) & 14.47$\pm$0.03$\pm$0.06 & 16.6 \\
3C\,273.0 & 289.95 & $ 64.36$ &  4 & EPn$^\dagger$ & $ 105$ & $ 160$ & \phn125 & (16$\pm$4) & 13.17$\pm$0.11$\pm$0.33 &  3.4 \\
 &  &  &  & EPn$^\dagger$ & $ 160$ & $ 260$ & \phn210 & 31$\pm$4 & 13.52$\pm$0.07$\pm$0.08 &  6.3 \\
Fairall\,9 & 295.07 & $-57.83$ &  2 & MS & $ 100$ & $ 275$ & \phn183 & (72$\pm$4) & 14.33$\pm$0.07$\pm$0.07 &  7.4 \\
Tol\,1247--232 & 302.60 & $ 39.30$ &  2 & EPn$^\dagger$ & $ 100$ & $ 255$ & \phn168 & ((66$\pm$6)) & 14.18$\pm$0.10$\pm$0.07 &  4.4 \\
PG\,1302--102 & 308.59 & $ 52.16$ &  4 & EPn & $ 190$ & $ 340$ & \phn256 & 51$\pm$4 & 13.95$\pm$0.06$\pm$0.04 &  7.5 \\
ESO\,141--G55 & 338.18 & $-26.71$ &  4 & Oth & $ 140$ & $ 225$ & \phn176 & (28$\pm$3) & 13.45$\pm$0.11$\pm$0.14 &  3.6 \\
Mrk\,1383 & 349.22 & $ 55.12$ &  4 & Oth$^\dagger$ & $ 100$ & $ 160$ & \phn125 & (23$\pm$3) & 13.42$\pm$0.09$\pm$0.18 &  5.0 \\
PKS\,2005--489 & 350.37 & $-32.60$ &  3 & Oth$^\dagger$ & $ 120$ & $ 225$ & \phn156 & ((42$\pm$5)) & 13.92$\pm$0.06$\pm$0.17 &  7.9 \\
\enddata
\tablenotetext{a}{Data quality ($S/N$ per 20 \kms\ resolution element):   
$Q = 1$ ($S/N = 3-5$),
$Q = 2$ ($S/N = 5-9$), $Q = 3$ ($S/N = 9-14$), $Q = 4$ ($S/N >14$).}
\tablenotetext{b}{HVC identification:  {\tt A} = Complex~A, {\tt C} = Complex~C,
{\tt EPn} = Extreme Positive (North), {\tt LG} = Local Group, {\tt MS} = Magellanic Stream,
{\tt MSe} = Magellanic Stream extension, {\tt OA} = Outer Arm, {\tt Oth} = Other.  A dagger
symbol ($\dagger$) indicates that the feature is also considered a high
velocity ``wing'' feature, as described in \S8.1}
\tablenotetext{c}{Observed line width as defined in \S3.3.  These widths have 
not been corrected for the $FUSE$ instrumental broadening (b$_{inst} \sim
12-15$ \kms).  Values in parentheses 
are somewhat uncertain because they depend upon the integration
limits adopted.  Double parentheses 
indicate that the velocity cutoffs used 
for the integration are not well constrained by the data and were chosen
according to the rules outlined in the text.} 
\tablenotetext{d}{This NGC\,588 component may contain a contribution from M\,33
(see Wakker et al. 2002).}
\tablenotetext{e}{These features toward HE\,0226-4410 and PKS\,0558-504 
are probably closely related to the {\tt MS}
components (see \S6.3).}
\end{deluxetable}

\clearpage
\newpage

\begin{deluxetable}{lcccccc}
\tablecolumns{5}
\tablewidth{0pt} 
\tablecaption{Sight Lines with No High Velocity \ion{O}{6} Detected by $FUSE$}
\tablehead{Name & $l$ & $b$ & Q\tablenotemark{a} & $\acute{\sigma_{W}}$\tablenotemark{b}& $\log N$(\ion{O}{6})\tablenotemark{c} & Note\tablenotemark{d} \\
& (\degr) & (\degr) & & (m\AA) & ($3\sigma$)}
\startdata
PG\,1352+183 & \phn\phn4.37 & \phn72.87 & 1 &   51.6 & $<$14.09 \\
PG\,1404+226 & \phn21.48 & \phn72.37 & 1 &   33.6 & $<$13.91 \\
NGC\,5548 & \phn31.96 & \phn70.50 & 2 &   29.1 & $<$13.84 \\
PG\,1402+261 & \phn32.96 & \phn73.46 & 2 &   24.3 & $<$13.76 & 1\\
vZ\,1128\tablenotemark{e} & \phn42.50 & \phn78.68 & 4 &    \phn4.5 & $<$13.04 \\
Mrk\,829 & \phn58.76 & \phn63.25 & 2 &   21.1 & $<$13.70 & 5 \\
PG\,1444+407 & \phn69.90 & \phn62.72 & 2 &   30.0 & $<$13.86 \\
SBS\,1415+437 & \phn81.96 & \phn66.20 & 2 &   25.7 & $<$13.79 \\
PG\,1411+442 & \phn83.83 & \phn66.35 & 2 &   42.4 & $<$14.01 & 3\\
PG\,1415+451 & \phn84.72 & \phn65.32 & 1 &   28.6 & $<$13.84 & 4\\
Mrk\,487 & \phn87.84 & \phn49.03 & 2 &   45.1 & $<$14.03 \\
Mrk\,59 & 111.54 & \phn82.12 & 3 &    \phn9.5 & $<$13.36 & 5\\
Mrk\,205 & 125.45 & \phn41.67 & 1 &   52.2 & $<$14.10 \\
3C\,249.1 & 130.39 & \phn38.55 & 2 &   28.2 & $<$13.83 & 1,2\\
Mrk\,209 & 134.15 & \phn68.08 & 3 &   14.7 & $<$13.55 & 4\\
PG\,0804+761 & 138.28 & \phn31.03 & 4 &    \phn7.1 & $<$13.23 & 1,2\\
HS\,0624+6907 & 145.71 & \phn23.35 & 2 &   19.7 & $<$13.67 & 1,2\\
PG\,0832+675\tablenotemark{e} & 147.75 & \phn35.01 & 3 &   18.3 & $<$13.64 & 1,3,4\\
V\,II\,Zw\,118 & 151.36 & \phn25.99 & 4 &    8.4 & $<$13.30 & 1,2\\
MS\,0700.7+6338 & 152.47 & \phn25.63 & 1 &   42.0 & $<$14.00 & 1\\
NGC\,4151 & 155.08 & \phn75.06 & 3 &   11.1 & $<$13.42 & 1,4\\
Mrk\,9 & 158.36 & \phn28.75 & 3 &   16.1 & $<$13.59 & 1,2\\
Mrk\,116 & 160.53 & \phn44.84 & 3 &   19.3 & $<$13.67 & 1,2\\
Mrk\,79 & 168.60 & \phn28.38 & 2 &   21.2 & $<$13.70 & 1,2\\
NGC\,1068 & 172.10 & --51.93 & 4 &   10.5 & $<$13.40 & 1,4\\
NGC\,985 & 180.84 & --59.49 & 3 &   14.9 & $<$13.55 & 1\\
Ton\,1187 & 188.33 & \phn55.38 & 2 &   36.9 & $<$13.95 \\
SBS\,0335--052 & 191.34 & --44.69 & 2 &   36.0 & $<$13.94 \\
PG\,0832+251 & 199.49 & \phn33.15 & 1 &   44.4 & $<$14.03 & 1\\
HE\,0238--1904 & 200.48 & --63.63 & 1 &   47.3 & $<$14.05 & 1,4\\
Mrk\,1095 & 201.69 & --21.13 & 3 &    \phn9.2 & $<$13.34 & 1,2\\
Mrk\,36 & 201.76 & \phn66.49 & 1 &   57.0 & $<$14.14 & 1\\
NGC\,3504 & 204.60 & \phn66.04 & 1 &   42.2 & $<$14.00 & 5\\
Mrk\,618 & 206.72 & --34.66 & 2 &   22.4 & $<$13.73 & 1,2\\
PG\,1004+130 & 225.12 & \phn49.12 & 3 &   16.3 & $<$13.59 & 1,2,4\\
HE\,0450--2958 & 231.13 & --37.59 & 2 &   54.2 & $<$14.11 \\
NGC\,1399 & 236.72 & --53.63 & 2 &   50.3 & $<$14.08 \\
PG\,1211+143 & 267.55 & \phn74.32 & 3 &   16.9 & $<$13.61 & 1,4\\
Mrk\,771 & 269.44 & \phn81.74 & 1 &   24.1 & $<$13.76 \\
NGC\,4649 & 295.88 & \phn74.34 & 1 &   47.4 & $<$14.05 \\
PG\,1307+085 & 316.79 & \phn70.71 & 3 &   18.6 & $<$13.65 & 1\\
HE\,1326--0516 & 320.07 & \phn56.07 & 1 &   45.2 & $<$14.03 \\
Tol\,1924--416 & 356.94 & --24.10 & 2 &   19.3 & $<$13.67 & 1\\
\enddata
\tablenotetext{a}{Data quality ($S/N$ per 20 \kms\ resolution element):   
$Q$~=~1 ($S/N = 3-5$),
$Q$~=~2 ($S/N = 5-9$), $Q$~=~3 ($S/N = 9-14$), $Q$~=~4 ($S/N >14$).}
\tablenotetext{b}{Scaled equivalent width error ($1\sigma$).  This error was
derived from the equivalent width error found for the Galactic 
thick disk/halo \ion{O}{6}
absorption along the sight line (Wakker et al. 2002; Savage et al. 2002b)
and is appropriate for a velocity integration range spanning 100 \kms.  It includes
statistical and continuum placement uncertainties.}
\tablenotetext{c}{These are $3\sigma$ limits set by the listed value of 
$\acute{\sigma_{W}}$ and the assumption of a linear curve of growth (see \S3.2).}
\tablenotetext{d}{Notes: \\
\noindent(1) Sight line contains H$_2$ (6--0) P(3) $\lambda1031.191$
absorption.  Actual limit may be higher if considering absorption near 
$\approx-214$ \kms. 
\\
\noindent(2) Sight line contains H$_2$ (6--0) R(4) $\lambda1032.349$
absorption.  Actual limit may be higher if considering absorption near 
$\approx+123$ \kms.
\\
\noindent(3) Due to other absorption features, actual limit may be higher if 
considering absorption at 
high negative velocities. 
\\
\noindent(4) Due to other absorption features, actual limit may be higher if 
considering absorption at high positive velocities.
\\
\noindent(5) Actual limit may be higher, depending on velocity range chosen
 because continuum placement is difficult at high velocities.}
\tablenotetext{e}{Stellar sight line with z = 9.8 kpc for vZ\,1128 and 
z = 4.7 kpc for PG\,0832+675.}
\end{deluxetable}

\clearpage
\newpage

\begin{deluxetable}{lccccccc}
\tabletypesize{\small}
\tablewidth{500pt}
\tablecaption{\ion{O}{6} High Velocity Classifications}
\tablehead{Classification & ID & \# & $\langle~\log N~\rangle$ & $\langle~\bar{v}~\rangle$ 
& $\langle$ b $\rangle$ & $l$ & $b$  \\
& & &  & (\kms) & (\kms) & (\degr) & (\degr)}
\startdata
Complex A 		& A & \phn1 & 13.81  & --125 & 22  & 161.1 & 42.9 \\
\\
Complex C 		& C & \phn9 & 13.93 (0.24) & --132 (\phn15) & 37 (18)  & \phn54.2 to 120.6 & \phn31.5 to \phn58.0 \\
\\
Extreme Positive North	& EPn & 19 & 13.96 (0.37) & \phn181 (\phn49) & 42 (13) & 179.8 to 308.6 & \phn16.6 to \phn77.2\\
\\
Magellanic Stream \\
\,\,\,\,\,($v_{LSR} > 0$ \kms)& MS   & \phn3 & 14.14 (0.31) & \phn232 (\phn84) & 60 (10) & 261.1 to 295.1 & --57.8 to --38.7\\
\,\,\,\,\,($v_{LSR} < 0$ \kms)& MS   & \phn3 & 14.10 (0.10) & --296 (\phn34) & 40 (13) & \phn83.1 to \phn91.7 & --60.4 to --45.5\\
\\
Magellanic Stream Ext.	 \\
\,\,\,\,\,($v_{LSR} > 0$ \kms)&  MSe  & \phn2 & 13.83 (0.21) & \phn211 (\phn66) & 47 (16) & 253.9 to 258.0 & --65.8 to --28.6\\
\,\,\,\,\,($v_{LSR} < 0$ \kms)&  MSe  & 10    & 14.18 (0.19) & --304 (\phn29) & 47 (\phn7) & \phn63.7 to 133.5 & --58.8 to --25.3\\
\\
Outer Arm 		& OA & \phn2 & 14.01 (0.19) & --106 (\phn22) & 26 (\phn6) & \phn61.3 to \phn94.0 & \phn17.5 to \phn27.4\\
\\
Local Group 		& LG & 19 & 14.08 (0.26) & --182 (\phn54) & 43 (12) & \phn17.7 to 225.0 & --85.1 to --29.9 \\
\\
Other 		 \\
\,\,\,\,\,($v_{LSR} > 0$ \kms)&  Oth   & 13 & 13.59 (0.31) & \phn180 (\phn94) & 32 (\phn8) & \phn36.0 to 350.4 & --85.1 to \phn88.6 \\
\,\,\,\,\,($v_{LSR} < 0$ \kms)&  Oth   & \phn3 & 13.51 (0.25) & --204 (\phn49) & 21 (\phn4) & \phn94.0 to 156.6 & \phn\phn27.4 to \phn54.1 \
\enddata

\end{deluxetable}
\clearpage
\newpage

\begin{deluxetable}{cccccc}
\tablewidth{0pt}
\tablecaption{High Velocity \ion{O}{6} Detection Summary}
\tablehead{$\Sigma W_\lambda$ Threshold & \multicolumn{5}{c}{Number of Sight Lines with [$\frac{W_\lambda}{\sigma_{W}}]_{_{tot}} > 3.0$ \tablenotemark{a,b}} \\
(m\AA) & All $Q$  & $Q=1$ & $Q=2$ & $Q=3$ & $Q=4$} 
\startdata
\phn30 & 59 (58\%)\tablenotemark{c} & 11 (48\%)\tablenotemark{c} & 14 (47\%)\tablenotemark{c} & 12 (52\%)\tablenotemark{c} & 22 (85\%)\\
\phn50 & 55 (54\%)\tablenotemark{c} & 11 (48\%)\tablenotemark{c} & 14 (47\%)\tablenotemark{c} & 12 (52\%)\tablenotemark{c} & 18 (69\%)\\
100    & 40 (39\%) & \phn9 (39\%) & 13 (43\%) & \phn7 (30\%) & 11 (42\%)\\
150    & 27 (26\%) & \phn6 (26\%) & \phn9 (30\%)& \phn7 (30\%) & \phn5 (19\%)\\
200    & 19 (19\%) & \phn3 (13\%)& \phn7 (23\%) & \phn5 (22\%) & \phn4 (15\%)\\
\enddata
\tablenotetext{a}{Number of sight lines (and percentages) with 
total integrated high velocity \ion{O}{6} $\lambda1031.926$
equivalent widths, $\Sigma W_{1032}$, greater than the listed equivalent width 
threshold.}
\tablenotetext{b}{There are 102 sight lines in the sample: 23 with $Q=1$, 30 
with $Q=2$, 23 with $Q=3$, and 26 with $Q=4$.}
\tablenotetext{c}{These $Q=1-3$ 
values should be treated with caution since the data quality is not 
sufficient to reliably exclude weak features.  These values are likely to 
be lower
limits if the $Q=4$ values are representative of the true detection
frequency (see text).}
\end{deluxetable}

\clearpage
\newpage

\begin{deluxetable}{cccccccc}
\tablewidth{0pt} 
\tablecaption{High Velocity \ion{O}{6} Sky Covering Percentages -- Column Densities\tablenotemark{a}}
\tablehead{Galactic& Sight Lines & \multicolumn{6}{c}{\ion{O}{6} Column Density Threshold, $N_0$ (cm$^{-2}$)} \\
Region & in Sample & $1.0\times10^{13}$ & $2.5\times10^{13}$ & $5.0\times10^{13}$   
& $7.5\times10^{13}$ & $1.0\times10^{14}$ & $2.5\times10^{14}$  } 
\startdata
All Sky & 102 & 59 (58\%) & 55 (54\%) & 48 (47\%) & 38 (37\%) & 31 (30\%) & 6 (6\%)\\
\cutinhead{$0\degr < l < 180\degr$}
$b < 0\degr$ & \phn20  & 19 (95\%) & 18 (90\%) & 15 (75\%) & 11 (55\%) & \phn9 (45\%) & 1 (5\%) \\
$b > 0\degr$ &  \phn40  & 16 (40\%)  & 16 (40\%) & 15 (38\%) & 12 (30\%) & 11 (28\%) & 2 (5\%) \\
\cutinhead{$180\degr < l < 360\degr$}
$b < 0\degr$ &  \phn16 & \phn8 (50\%) & \phn7 (44\%) & \phn6 (38\%) & \phn6 (38\%) & \phn4 (35\%) & 1 (6\%) \\
$b > 0\degr$ &  \phn26 & 16 (62\%) & 14 (54\%) & 12 (46\%) & \phn9 (35\%) & \phn7 (27\%) & 2 (8\%)
\enddata
\tablenotetext{a}{Number (percentage) of sight lines for which 
$\sum$$N$(\ion{O}{6})$_{\rm HV}~\ge~N_0$.  Entries for $N_0 < 10^{14}$ cm$^{-2}$ 
should be treated as provisional since data quality may prevent detection of 
weaker features along some sight lines.}
\end{deluxetable}

\clearpage
\newpage

\begin{deluxetable}{lcccc}
\tablewidth{0pt} 
\tablecaption{Velocity Centroid Means and Standard Deviations\tablenotemark{a}}
\tablehead{Reference Frame & Longitude Range & $b<0\degr$ & $b>0\degr$ & All $b$}
\startdata
Local Standard of Rest	
                & $0\degr < l<180\degr$ & $-205\pm127$ & $-38\pm158$ & $-138\pm162$\\
	   	& $180\degr < l < 360\degr$ & $\phn156\pm142$ & $193\pm64$ & $\phn181\pm95$ \\
	   	& All $l$ & $-129\pm196$ & $66\pm170$ &  $-33\pm207$  \\
\\
Galactic Standard of Rest	
                & $0\degr<l<180\degr$ & $-85\pm111$ &  $88\pm142$ & $-15\pm150$ \\
	   	& $180\degr<l<360\degr$ & $55\pm106$ & $108\pm92$ & $91\pm98$ \\
	   	& All $l$ & $-56\pm123$ & $97\pm121$ & $20\pm144$  \\
\\
Local Group Standard of Rest	
                & $0\degr<l<180\degr$ & $-40\pm110$ &  $94\pm136$ & $14\pm137$ \\
	   	& $180\degr<l<360\degr$ & $56\pm94$ & $69\pm108$ & $65\pm102$ \\
	   	& All $l$ & $-19\pm113$ & $82\pm124$ & $31\pm128$  \\
\\
& & \multicolumn{3}{c}{Number of High Velocity \ion{O}{6} Features} \\
& $0\degr < l<180\degr$      & [34] & [23] & [57] \\
& $180\degr < l < 360\degr$  & [\phn9] & [19] & [28] \\
& All $l$                    & [43] & [42] & [85] \\
\enddata
\tablenotetext{a}{Velocity centroid means, 
$\langle \bar{v} \rangle$, and standard deviations, $\sigma_{\langle \bar{v} \rangle}$, of the high velocity \ion{O}{6} features.  
All velocities are in \kms.  
The number of  features  
in each longitude-latitude interval is listed
in square brackets below the velocity entries.}
\end{deluxetable}

\clearpage
\newpage


\clearpage
\newpage

\begin{deluxetable}{ccccccc}
\tablewidth{0pt}
\tablecaption{High Velocity \ion{O}{6} Sky Covering Percentages -- Centroid Velocities\tablenotemark{a}}
\tablehead{Velocity Threshold & All Sky & \multicolumn{2}{c}{\underline{$0\degr < l < 180\degr$}} 
&  \multicolumn{2}{c}{\underline{$180\degr < l < 360\degr$}}\\
$v_0$ (\kms) & & $b<0\degr$ & $b>0\degr$ & $b<0\degr$ & $b>0\degr$} 
\startdata
--300 & \phn8 (\phn8\%) [\phn8\%]   & \phn8 (40\%)  & \phn0 (\phn0\%) & 0 (\phn0\%) & \phn0 (\phn0\%)\\
--250 & 16 (16\%) [12\%] & 15 (75\%) & \phn1 (\phn3\%) & 0 (\phn0\%) & \phn0 (\phn0\%)\\
--200 & 18 (18\%) [19\%] & 17 (85\%) & \phn1 (\phn3\%) & 0 (\phn0\%) & \phn0 (\phn0\%)\\
--150 & 24 (24\%) [31\%] & 19 (95\%) & \phn4 (10\%) & 1 (\phn6\%) & \phn0 (\phn0\%)\\
\phn--90 & 33 (32\%) [42\%] & 19 (95\%) & 13 (33\%) & 1 (\phn6\%) & \phn0 (\phn0\%)\\
\phantom{--0}90  & 33 (32\%) [58\%] & \phn2 (10\%) & \phn8 (20\%) & 7 (44\%) & 16 (62\%)\\
\phn150 & 20 (20\%) [31\%] & \phn2 (10\%) & \phn1 (\phn3\%) & 6 (38\%) & 11 (42\%)\\
\phn200 & 11 (11\%) [15\%] & \phn1 (\phn5\%) & \phn1 (\phn3\%) & 2 (13\%) & \phn7 (27\%)\\
\phn250 & \phn9 (\phn9\%) [12\%]& \phn1 (\phn5\%) & \phn1 (\phn3\%) & 2 (13\%) & \phn5 (19\%)\\
\phn300 & \phn3 (\phn3\%) [\phn4\%]& \phn0 (\phn0\%) & \phn1 (\phn3\%) & 1 (\phn6\%) & \phn1 (\phn4\%)\\
\tableline
\# Sight Lines\tablenotemark{b} & 102 [26] & 20 & 40 & 16 & 26
\enddata
\tablenotetext{a}{Number (percentage) of sight lines in the sample 
that contain 
at least one high velocity \ion{O}{6} feature with either $\bar{v} \le v_0$ (for $v_0 < 0$ \kms) or
$\bar{v} \ge v_0$ (for $v_0 > 0$ \kms).  Data quality affects these 
percentages since weak features are below the detection limits for some  
sight lines (i.e., these values should 
probably be treated as lower limits -- see text).  Values listed in square brackets [~]
are for the highest quality ($Q=4$) sight lines only.}
\tablenotetext{b}{Number of sight lines in each longitude-latitude region.}
\end{deluxetable}

\clearpage
\newpage

\begin{deluxetable}{lccccccccc}
\tablewidth{0pt} 
\tabletypesize{\small}
\tablecaption{\ion{H}{1} - \ion{O}{6} Comparison for Complex~C Sight Lines}
\tablehead{Sight Line & $l$ & $b$ & $\bar{v}_{\rm HI}$\tablenotemark{a} & $N$(\ion{H}{1})\tablenotemark{a} & $\bar{v}_{\rm OVI}$ & $N$(\ion{O}{6}) & $\bar{v}_{\rm HI}$--$\bar{v}_{\rm OVI}$ & $\frac{N(H\,I)}{N(O\,VI)}$ & Note\\
& {\scriptsize (\degr)} & {\scriptsize(\degr)} & {\scriptsize(\kms)} & {\scriptsize(10$^{18}$ cm$^{-2}$)} & {\scriptsize(\kms)} & {\scriptsize(10$^{13}$ cm$^{-2}$)} & {\scriptsize(\kms)} & {\scriptsize(x10$^{5}$)}}
\startdata
Mrk\,279 & 115.04 & 46.86 & --137/[--102] & 20.4/[12.3] & --154 & 4.68 & --17 & 4.4 & 1\\
\\
Mrk\,290 & 91.49 & 47.95  & --134/[--105] & 94.7/[38.5] & --132 & 15.8 & +2 & 6.0 & 1\\
\\
Mrk\,501 & 63.60 & 36.86 & --116 & 10.7 & --122 & 6.46 & --6 & 1.7\\
\\
Mrk\,506 & 54.18 & 31.50 & --155 & 4.4 & --144 & 11.2 & +11 & 0.4\\
\\
Mrk\,817 & 100.30 & 53.48 & --109 & 32.4 & --116 & 7.59 & (--7) & 4.3 & 2\\
\\
Mrk\,876 & 98.27 & 40.38 & [--173]/--133 & [4.7]/19.9 & --142 & 11.2 & (--9)  & 1.8 & 1,2\\
\\
PG\,1259+593 & 120.56 & 58.05 & --128 & 89.5 & --110 & 5.25 & +18 & 17.\\
\\
PG\,1351+640 & 111.89 & 52.02 & --154/[--115] & 59.6/[7.7] & --124 & 5.68 & (+30) & 13. & 1,2\\
\\
PG\,1626+554 & 84.51 & 42.19 & --131 & 26.9 & --152 & 22.4 & --21 & 1.2\\
\\
Averages: & \nodata & \nodata & \nodata & \nodata & \nodata & \nodata & $-2\pm15$ & $5.5\pm5.7$ & 3,4
\enddata
\tablenotetext{a}{Values derived from Effelsberg \ion{H}{1} 21\,cm spectra.
Values in brackets [~] indicate secondary features.}
\tablenotetext{b}{Notes: \\
\noindent
(1) Values for 
$\bar{v}_{\rm HI}-\bar{v}_{\rm OVI}$ and 
$N$(\ion{H}{1})/$N$(\ion{O}{6}) do not include the \ion{H}{1} 
components listed in brackets. \\
\noindent
(2) Velocity difference is uncertain because
$\bar{v}_{\rm OVI}$ relies on integration limits that are based on 
something other than the \ion{O}{6} data (i.e., the \ion{H}{1} velocity
structure). \\
\noindent
(3) If the uncertain
values listed in parentheses are included, the average of 
$\bar{v}_{\rm HI}-\bar{v}_{\rm OVI}$ becomes $0\pm17$ \kms. 
\\
\noindent
(4) The quoted uncertainties are standard deviations of the mean values.}
\end{deluxetable}

\clearpage
\newpage
\begin{deluxetable}{lccccccl}
\tablewidth{0pt} 
\tablecaption{High Velocity \ion{H}{1} 21\,cm Emission Features with No \ion{O}{6}}
\tablehead{Object & $l$ & $b$ & $\bar{v}_{21\,cm}$ & $\log N$(\ion{H}{1})\tablenotemark{a} & $\log N$(\ion{O}{6})\tablenotemark{b} & 
$Q\tablenotemark{c}$ & \multicolumn{1}{c}{Comment} \\
& (\degr) & (\degr) & (\kms) & ($3\sigma$)}
\startdata
PG\,0052+251	& 123.91 & --37.44 & $-121$ & 18.60 &$<13.72$ & 2 & WW478  \\
Mrk\,205	& 125.45 & \phn41.67 & $-202$ & 19.21 &$<13.79$ & 1 & WW84 \\
HS 0624+6907	& 145.71 & \phn22.35 & $-100$ & 19.32 &$<13.61$ & 2 & Outer Arm \\
Mrk\,116	& 160.53 & \phn44.84 & $-170$ & 19.33 &$<13.49$ & 3 & Complex A \\
Ton\,1187	& 188.33 & \phn55.38 & $-104$ & 19.23 &$<13.65$ & 2 & Complex M \\
ESO\,265-G23	& 285.91 & \phn16.59 & +117  & 19.71 &$<13.86$\tablenotemark{d} & 1 & Complex WD\\
\enddata
\tablenotetext{a}{The \ion{H}{1} 21\,cm emission profiles are shown in Figure~1
of Wakker et al. (2002).}
\tablenotetext{b}{\ion{O}{6} column density limit over an integration range of
$\approx50$ \kms\ centered on the observed velocity of the \ion{H}{1} 
21\,cm emission.}
\tablenotetext{c}{Data quality of the $FUSE$ observation.}
\tablenotetext{d}{A higher quality $FUSE$ Guest Investigator observation 
obtained recently by B.K.~Gibson and colleagues confirms this non-detection.}
\end{deluxetable}

\clearpage
\newpage
\begin{deluxetable}{lcc}
\tablewidth{0pt} 
\tablecaption{Velocity Summary of Possible Local Group Features\tablenotemark{a}}
\tablehead{Reference Frame & $\langle \bar{v} \rangle$ & $\sigma_{\langle \bar{v} \rangle}$ \\
& (\kms) & (\kms)}
\startdata
Local Standard of Rest		& --53 & 214
\\
Galactic Standard of Rest	& --22 & 123	
\\
Local Group Standard of Rest	& --21 & \phn63
\enddata
\tablenotetext{a}{The 50 \ion{O}{6} absorption features considered have classifications of 
{\tt EPn}, {\tt MSe}, or {\tt LG} as defined in Table~3.}
\end{deluxetable}

\clearpage
\newpage

\begin{deluxetable}{lcccccc}
\tablewidth{0pt} 
\tablecaption{High Ion Column Density Ratios}
\tablehead{Sight Line & $\bar{v}$ & ID & $\log N$(\ion{O}{6}) & $\frac{N{\rm(O\,VI)}}{N{\rm (C\,IV)}}$ & $\frac{N{\rm (O\,VI)}}{N{\rm (N\,V)}}$ & Notes\tablenotemark{b}\\
& (\kms) }
\startdata
Mrk\,509 	& --247 & LG & 14.24 & $<0.3$ & $<0.2$ & 1\\
\\
PKS\,2155-304 	& --232 & LG & 13.57 & $\sim1$ & $>3$ & 2\\
	 	& --129 & LG &  13.80 & $\sim2$ & $>4.6$ & 2\\ 
\\
NGC\,1705	& +330 & MS & 14.31 & $>3$ & \nodata & 3 \\
\\
Fairall~9	& +183 & MS & 14.33 & $\sim7$ & $>3$ & 4\\
\\
PG\,1259+593	& --110	& C & 13.72 & $\sim3.5$ & $>3.8$ & 5 \\
\\
$\langle$Galactic Halo$\rangle$ & $\sim0$ & \nodata &  $\sim14.37\pm0.18$ & $\sim1.5$ & $\sim5$ & 6\\
\cutinhead{Model Predictions\tablenotemark{a}}
Turbulent Mixing 	& \nodata & \nodata & \nodata & 0.1-0.8 & 2.5-8 & 7\\
\\
Radiative Cooling 	& \nodata & \nodata & \nodata  & 2-10 & 7--14 & 8\\
\\
Conduction 		& \nodata & \nodata & \nodata & 1.6--4.5 & 4--7 & 9\\
\\
Supernova Remnants	& \nodata & \nodata & \nodata & 5-10 & 12--16 & 10\\
\\
C.I.E. ($T<2\times10^5$\,K) & \nodata & \nodata & \nodata & $<2.6$ & $<0.7$ & 11\\
C.I.E. ($T=3\times10^5$\,K) & \nodata & \nodata & \nodata & 134 & 50 & 11\\
C.I.E. ($T=5\times10^5$\,K) & \nodata & \nodata & \nodata & 50 & 47 & 11 \\
\\
Photo.	($n_H=10^{-5}$ cm$^{-3}$)	& \nodata & \nodata & \nodata & 1.0 & 3.6 & 12\\
Photo.	($n_H=10^{-4}$ cm$^{-3}$)	& \nodata & \nodata & \nodata & 0.01& 0.5 & 12
\enddata
\tablenotetext{a}{Model predictions assume solar abundance ratios unless
otherwise noted.}
\tablenotetext{b}{Notes: (see next page)}
\end{deluxetable}

\clearpage
\newpage

\begin{center}
Table~11 (continued)\\
\end{center}
\noindent Notes: \\
\noindent (1) \ion{C}{4} and \ion{N}{5} from
Sembach et al. (1999).  The values listed are the sums for the 
two \ion{C}{4}-HVCs along the sight line at $-283$ and $-228$ \kms.
An additional high velocity \ion{O}{6} feature at $\bar{v} = -143$ \kms\
(see Table~1) has no obvious counterpart in \ion{C}{4} or \ion{N}{5}. \\
\noindent
(2) \ion{C}{4} and \ion{N}{5} from
Sembach et al. (1999).  \\
\noindent
(3) \ion{C}{4} from Sembach  (2002). Continuum placement for
\ion{C}{4} is difficult. \\
\noindent
(4) \ion{C}{4} from Lu, Savage, \& Sembach (1994).\\
\noindent
(5) \ion{C}{4} and \ion{N}{5} from extant STIS data.  See Fox et al. (2002).\\
\noindent
(6) Galactic halo averages from data compiled by Sembach, Savage, \& Tripp (1997),
Savage, {\mbox Sembach}, \& Lu (1997), and Savage et al. (2002b).\\
\noindent
(7) Turbulent mixing layer results from Slavin, Shull, \& Begelman (1993)
for TMLs with post-mixed temperatures $T = (1-3) \times 10^5$\,K and hot gas
entrainment velocities of $25-100$ \kms.\\
\noindent
(8) Radiative cooling results from Edgar \& Chevalier (1986) for a range of 
models with varying assumptions about the transition of the flow from 
isochoric to isobaric cooling.\\
(9) Magnetized thermal conduction front results from 
Borkowski, Balbus, \& Fristrom (1990) for a range of magnetic field 
orientations inclined between $0\degr$ to $60\degr$ to the face of the cloud 
at an age of $2.5\times10^5$ years.\\
\noindent
(10) Evolved supernova remnant results from Slavin \& Cox (1992) for isolated
explosions in a medium with densities of $0.1-0.3$ cm$^{-3}$ and 
$B \approx 1\mu$G.  Values are for remnants with ages of $4\times10^5$ to 
$5\times10^6$ years.\\
\noindent
(11) Collisional ionization equilibrium ratios assuming solar abundance
ratios (Sutherland \& Dopita 1993; Holweger 2001). \\
\noindent
(12) Photoionization model results shown in Figure~19.  The values quoted
are appropriate for models with $N$(\ion{H}{1}) = $10^{14}$ cm$^{-2}$, 
a metallicity of $\sim1/3$ solar, and a solar relative abundance pattern.
If the relative abundances are not in their solar proportions (e.g., if 
N/O is sub-solar as appears to be the case for Complex~C), the ratios 
may differ from those plotted.
Higher values of $N$(\ion{H}{1}) result in lower ratios.  
See \S9.2
for additional descriptions of the model.

\clearpage
\newpage

\begin{deluxetable}{lcccc}
\tablewidth{0pt} 
\tablecaption{Means and Standard Deviations of $\Delta$$v$ \tablenotemark{a,b}}
\tablehead{Rotation Model & Longitude Range  & $b<0\degr$ & $b>0\degr$ & All $b$}
\startdata
Corotating Halo & $l<180\degr$ & $-110\pm88$ & $\phn8\pm76$ & $-62\pm101$\\
	   	& $l>180\degr$ & $\phantom{-0}71\pm81$ & $94\pm65$ & $\phantom{-}86\pm\phn70$ \\
	   	& All $l$ & $-72\pm113$ & $47\pm83$ & $-13\pm115$  \\
\\
Decoupled Halo	& $l<180\degr$ & $-27\pm76$ & $70\pm71$ & $11\pm87$ \\
	   	& $l>180\degr$ & $\phantom{-0}0\pm56$ & $33\pm89$ & $23\pm80$ \\
	   	& All $l$ & $-22\pm73$ & $52\pm81$ & $15\pm84$  \\
\enddata
\tablenotetext{a}{Values listed are $\langle\Delta v\rangle \pm \sigma_{\langle\Delta v\rangle}$.
All velocities are in \kms.}
\tablenotetext{b}{$\Delta$$v$ = $\bar{v} - v_{pred}$.  
See Table~1 for values of $\bar{v}$.}
\end{deluxetable}

\end{document}